\documentclass[pdflatex,iicol,sn-mathphys-num]{sn-jnl}
\usepackage{soul}
\usepackage[utf8]{inputenc}
\usepackage[T1]{fontenc}
\usepackage{amsfonts}
\usepackage{tikz}
\usepackage{graphicx}
\usepackage[caption=false]{subfig}
\usepackage{tensor} 
\usepackage{mathtools} 
\AtBeginDocument{\allowdisplaybreaks[0]}
\usepackage{amssymb}
\usepackage{enumerate}
\usepackage{euscript}
\usepackage{color}
\usepackage{braket}
\usepackage{booktabs}
\usepackage{flushend}
\makeatletter
\newcommand{\widefiguretype}{\def\@captype{figure}}
\newcommand{\compactappendixheadings}{%
  \renewcommand{\sectionfont}{\normalfont\fontsize{11bp}{13bp}\selectfont\bfseries\raggedright\boldmath}%
  \renewcommand{\@hangfrom}[1]{\noindent##1}%
}
\makeatother
\DeclareMathOperator{\sech}{sech}

\begin{document}

\title{Holographic Entanglement Entropy of~Network~CFT}
\author[1,2,3]{\fnm{Zhi-Rong} \sur{Deng}}
\email{dengzhr2024@lzu.edu.cn}
\author*[1,2,3]{\fnm{Jie} \sur{Yang}}
\email{yangjiev@lzu.edu.cn}

\affil[1]{%
  \orgdiv{Key Laboratory for Quantum Theory and Applications of MoE},
  \orgname{Lanzhou University},
  \orgaddress{\city{Lanzhou}, \postcode{730000}, \country{China}}}
\affil[2]{%
  \orgdiv{Lanzhou Center for Theoretical Physics,
    Key Laboratory of Theoretical Physics of Gansu Province},
  \orgname{Lanzhou University},
  \orgaddress{\city{Lanzhou}, \postcode{730000}, \country{China}}}
\affil[3]{%
  \orgdiv{Institute of Theoretical Physics and
    Research Center of Gravitation},
  \orgname{Lanzhou University},
  \orgaddress{\city{Lanzhou}, \postcode{730000}, \country{China}}}

\abstract{
	Miao et al. \cite{Guo:2025sbm, Guo:2026prm} proposed the AdS/NCFT correspondence, which explores the gravity dual of conformal field theories defined on networks. While previous studies investigated this model within Einstein and Gauss-Bonnet gravity (first- and second-order Lovelock gravity), in this paper, we generalize the framework to higher-order Lovelock gravity and more complex network structures. We derive the extremal and junction conditions for generalized Ryu-Takayanagi (RT) surfaces in third-order Lovelock gravity, which can be naturally extended to arbitrary higher-order Lovelock theories. Focusing on a network model comprising one node and three branches, we numerically investigate the relationship between the network holographic entanglement entropy (HEE) and gravitational coupling coefficients. Furthermore, by analyzing the HEE across various network configurations, we find that the primary structural feature affecting the network HEE is the presence of internal nodes within line-shaped networks. Our results indicate that the network HEE decreases as the number or depth of internal branches increases.
}


\AddToHook{cmd/printkeywords/after}[epjc-ack]{%
  \par\addvspace{10pt}%
  {\small\noindent\textbf{Acknowledgements}\par
   \noindent We are grateful to Rong-Xin Miao and Yu Guo for useful discussions.\par}%
}
\maketitle
\RemoveFromHook{cmd/printkeywords/after}[epjc-ack]
\raggedbottom

	
	\section{Introduction}

	The AdS/CFT correspondence provides a holographic framework to study gravity theories, proposing that quantum gravity in the bulk is dual to a conformal field theory (CFT) on the boundary \cite{Maldacena:1997re}. Entanglement entropy provides a probe of quantum correlations in field theory. In two-dimensional conformal field theories, its scaling also encodes information about the central charge and the presence of boundaries \cite{Calabrese:2004eu}. The Ryu-Takayanagi (RT) proposal \cite{Ryu:2006bv} is the most celebrated formula within the AdS/CFT correspondence; it establishes an equality between the entanglement entropy of a subregion in the CFT living on the asymptotic boundary of AdS and the area of the RT surface, which is the extremal surface homologous to that subregion. The gravitational replica construction provides a derivational basis for the minimal-area prescription \cite{Lewkowycz:2013nqa}. Originally, this proposal was formulated for AdS spacetimes with Einstein gravity, but it has since been generalized to models involving higher-order gravity, including higher-order Lovelock gravity \cite{Hung:2011xb,deBoer:2011wk,Dong:2013qoa}. In these higher-order gravity models, the holographic entanglement entropy (HEE) is not merely the area of the generalized RT surface; rather, it is given by an integral over the generalized RT surface, whose integrand contains terms that depend on the intrinsic curvature.
	
	Lovelock gravity \cite{Lovelock1971TheET} is a gravity theory in which the relative coefficients of the curvature terms at each order are precisely chosen to yield only second-order derivatives in the equations of motion. Its action is expressed as \cite{Myers:1987yn}
	\begin{equation}
	\begin{aligned}
	I_{\text{Lovelock}} &= \frac{1}{16\pi G}
	 \sum_{p=0}^{\left[\frac{d}{2}\right]}\alpha_p
	 \biggl(\int_{\text{Bulk}}d^{d+1}x\sqrt{-g}\,\mathcal{L}^{(p)}\\
	 &\qquad+\int_{\text{bdy}}d^dy\sqrt{-h}\,\mathcal{Q}^{(p)}\biggr),
	\end{aligned}
	\label{eq:Lovelock action}
	\end{equation}
	where $p$ is the order of the Lovelock gravity, while $\alpha_p$, $\mathcal{L}^{(p)}$, and $\mathcal{Q}^{(p)}$ denote the corresponding gravitational coupling coefficient, bulk term, and boundary term in the action, respectively. The notation $\left[\frac{d}{2}\right]$ represents the integer part of $\frac{d}{2}$. If the bulk spacetime has an even dimension $(d+1)$, an extra topological term of order $p=\frac{d+1}{2}$ can be added to the action, which does not contribute to the equations of motion. The explicit expressions for $\mathcal{L}^{(p)}$ and $\mathcal{Q}^{(p)}$ are given by Eqs. (\ref{eq:Lp}) and (\ref{eq:Qp}) in Appendix \ref{Explicit Tensor Expressions in Lovelock Theories}, respectively.

	The Lovelock combinations retain second-order equations of motion, while the corresponding boundary terms supply the boundary completion needed for the variational problem.

	The HEE in higher-order Lovelock gravity is given by \cite{Jacobson:1993xs}
	\begin{equation}
	\begin{aligned}
	S_{\text{Lovelock}} &= \frac{1}{4G}
	 \sum_{p=1}^{\left[\frac{d}{2}\right]}\alpha_p
	 \biggl(\int_\Gamma d^{d-1}\xi\sqrt{h_\Gamma}\,\tilde{\mathcal{L}}^{(p)}\\
	 &\qquad+\int_\gamma d^{d-2}\eta\sqrt{h_\gamma}\,\tilde{\mathcal{Q}}^{(p)}\biggr),
	\end{aligned}
	\label{eq:Lovelock HEE}
	\end{equation}
	where $\Gamma$ and $\gamma$ denote the generalized RT surface and its intersection with the boundary, respectively. A formal structural correspondence exists between the entanglement entropy and the gravity action, as explicitly shown by the relations $\tilde{\mathcal{L}}^{(p)}=p\mathcal{L}^{(p-1)}$ and $\tilde{\mathcal{Q}}^{(p)}=p\mathcal{Q}^{(p-1)}$. However, this precise mapping requires that the generalized Kronecker delta, the Riemann curvature, and both the intrinsic and extrinsic curvatures be replaced by their respective induced counterparts on the surface. The explicit expressions for $\tilde{\mathcal{L}}^{(p)}$ and  $\tilde{\mathcal{Q}}^{(p)}$ are given by Eqs. (\ref{eq:tLp}) and (\ref{eq:tQp}) in Appendix \ref{Explicit Tensor Expressions in Lovelock Theories}, respectively.

	In free-field models of wire and CFT junctions, entanglement between branches depends on transmission through the node~\cite{CalabreseMintchevVicari2012,GutperleMiller2017}.

\begin{figure}[!htbp]
    		\centering
    		\includegraphics[width=0.9\linewidth]{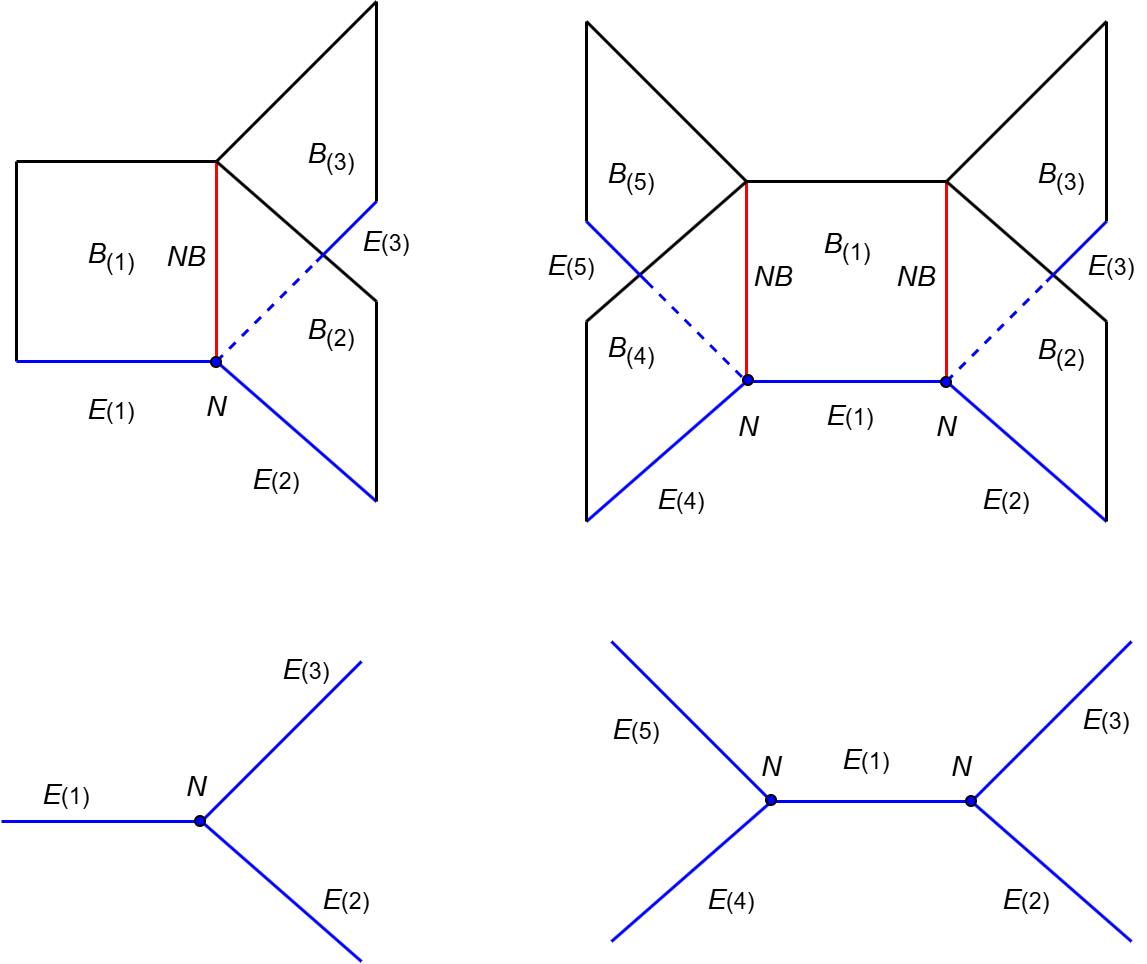}
    		\caption{Geometry of holographic networks. The bulk geometry for a one-node, three-branch network model (top left), where $B_{(m)}$ denotes the distinct bulk spacetimes, $E_{(m)}$ represents the boundary edges, and $NB$ is the Net-brane connecting them at the node $N$. The corresponding network conformal field theory (NCFT) defined on the boundary (bottom left). The bulk configuration of a two-node, five-branch network model (top right). The dual NCFT boundary structure for the two-node configuration (bottom right)}
    		\label{fig:AdS/NCFT}
	\end{figure}

	Recently, Miao et al. proposed the AdS/NCFT (network CFT) model \cite{Guo:2025sbm, Guo:2026prm}, which serves as a multi-branch generalization of AdS/BCFT (boundary CFT) \cite{Takayanagi:2011zk,Fujita:2011fp} and AdS/ICFT (interface CFT) \cite{Bachas:2001vj}. As illustrated in Fig. \ref{fig:AdS/NCFT}, $E_{(m)}$ and $N$ denote the edges and nodes of the network, respectively. Different CFTs reside on the edges, and the nodes link them together. The gravity dual of a network node is a brane, referred to as a Net-brane (NB). This Net-brane connects distinct AdS bulk regions $B_{(m)}$, which may be governed by different gravity theories. Both the total action and the corresponding HEE functional comprise contributions from all bulk regions and the Net-brane. 

	By varying the total action, the resulting boundary term yields the junction condition at the Net-brane, which gives the equation satisfied by the NB tension (detailed in Section \ref{Tension}). Separately, by varying the entropy functional, the internal and boundary terms of the variation provide the extremal and junction conditions for the generalized RT surface, respectively (detailed in Section \ref{RT and HEE}). For the numerical computations in this paper, we focus on a network model consisting of $n$ edges $E_{(m)}$ connected by a single node $N$. In Section \ref{Numerical Results}, we first verify that our numerical results perfectly match the analytical expressions; we then investigate the holographic network for Lovelock gravity up to the third order, exploring the effects of varying gravitational coupling coefficients across different bulk branches. Section \ref{HEE in Generalized Network} analyzes how the configuration the network influence the network HEE. Finally, we summarize our results and draw conclusions in Section \ref{Conclusion}.

	\section{Tension on Net-brane}\label{Tension}
	The total action of our model is expressed as
	\begin{align}
		I &= \sum_{m}^{n}\frac{1}{16\pi G_{(m)}}\int_{B_{(m)}}
		    d^{d+1}x_{(m)}\sqrt{-g_{(m)}}\nonumber\\*
		  &\quad \times\bigl(\alpha_{0(m)}\mathcal{L}^{(0)}_{(m)}
		    +\alpha_{1(m)}\mathcal{L}^{(1)}_{(m)}\nonumber\\*
		  &\qquad\quad+\alpha_{2(m)}\mathcal{L}^{(2)}_{(m)}
		    +\alpha_{3(m)}\mathcal{L}^{(3)}_{(m)}\bigr)\nonumber\\*
		  &\quad+\int_{NB}d^d y\sqrt{-h}\biggl[
		    \sum_{m}^{n}\frac{1}{16\pi G_{(m)}}\nonumber\\*
		  &\qquad \times\bigl(\alpha_{0(m)}\mathcal{Q}^{(0)}_{(m)}
		    +\alpha_{1(m)}\mathcal{Q}^{(1)}_{(m)}\nonumber\\*
		  &\qquad\quad+\alpha_{2(m)}\mathcal{Q}^{(2)}_{(m)}
		    +\alpha_{3(m)}\mathcal{Q}^{(3)}_{(m)}\bigr)\nonumber\\*
		  &\qquad+\frac{1}{16\pi G}\times\left(-2T\right)\biggr],
	\label{eq:action}
	\end{align}
	\begin{equation*}
	{\setlength{\arraycolsep}{2pt}
	\begin{array}{@{}lll@{}}
		\mathcal{L}_{(m)}^{(0)} = 1 & \mathcal{Q}_{(m)}^{(0)} = 0
		  & \alpha_{0(m)} = \frac{d(d-1)}{L_{(m)}} \\[1ex]
		\mathcal{L}_{(m)}^{(1)} = R_{(m)} &
		  \mathcal{Q}_{(m)}^{(1)} = 2(\prescript{NB}{}{K})_{(m)}
		  & \alpha_{1(m)} = 1\\
		\cdots
	\end{array}}
	\end{equation*}
	where $m$ indexes the different branches. The subscript $(m)$ indicates that the respective quantities belong to the $m$-th branch, allowing us to independently vary Newton's constant, the bare AdS radius $L_{(m)}$, and the higher-order couplings $\alpha_{2(m)}$ and $\alpha_{3(m)}$ across different branches. Here, $n$ represents the total number of branches, which we set to $n=3$ in our numerical computations. Furthermore, $T$ denotes the tension of the Net-brane, and $G$ (without a subscript) is the effective Newton's constant, defined as
	\begin{equation}
		\frac{n}{G}=\sum_m^n\frac{1}{G_{(m)}}.
	\label{eq:effective Newton's constant}
	\end{equation}

	The explicit expressions for the higher-order terms $\mathcal{L}^{(2)}_{(m)}$, $\mathcal{L}^{(3)}_{(m)}$, $\mathcal{Q}^{(2)}_{(m)}$, and $\mathcal{Q}^{(3)}_{(m)}$ can be derived from Eqs. (\ref{eq:Lp}) and (\ref{eq:Qp}) by substituting the global quantities with their corresponding branch-specific counterparts. Specifically, this involves the replacements $g\mapsto g_{(m)}$, $R\mapsto R_{(m)}$, and $(\prescript{NB}{}{K})\mapsto (\prescript{NB}{}{K})_{(m)}$. (Recall that we denote the boundary extrinsic and intrinsic curvatures $(\prescript{b}{}{K})$ and $(\prescript{b}{}{R})$ as $(\prescript{NB}{}{K})$ and $(\prescript{NB}{}{R})$, respectively).

	Following the variational approach to brane and multiway junctions~\cite{Davis2003,ShenPengLi2024}, we vary the cubic Lovelock action to obtain
	\begin{align}
		&\delta I = \sum_{m}^{n}\frac{1}{16\pi G_{(m)}}\int_{B_{(m)}}
		  d^{d+1}x_{(m)}\sqrt{-g_{(m)}}\nonumber\\*
		&\quad\times\bigl(\alpha_{0(m)}\mathcal{E}^{(0)}_{(m)\mu\nu}
		  +\alpha_{1(m)}\mathcal{E}^{(1)}_{(m)\mu\nu}\nonumber\\*
		&\quad+\alpha_{2(m)}\mathcal{E}^{(2)}_{(m)\mu\nu}
		  +\alpha_{3(m)}\mathcal{E}^{(3)}_{(m)\mu\nu}\bigr)
		  \delta g^{\mu\nu}_{(m)}\nonumber\\*
		&\quad+\int_{NB}d^d y\sqrt{-h}\biggl[\nonumber\\*
		&\quad\sum_{m}^{n}\frac{1}{16\pi G_{(m)}}\bigl(
		  \alpha_{0(m)}\mathcal{J}^{(0)}_{(m)ij}
		  +\alpha_{1(m)}\mathcal{J}^{(1)}_{(m)ij}\nonumber\\*
		&\qquad+\alpha_{2(m)}\mathcal{J}^{(2)}_{(m)ij}
		  +\alpha_{3(m)}\mathcal{J}^{(3)}_{(m)ij}\bigr)\nonumber\\*
		&\qquad+\frac{1}{16\pi G}\times\left(Th_{ij}\right)
		  \biggr]\delta h^{ij}.
	\label{eq:action variation}
	\end{align}
	 The explicit expressions for $\mathcal{E}^{(p)}_{\mu\nu}$ and $\mathcal{J}^{(p)}_{ij}$ are given by Eqs. (\ref{eq:Ep}) and (\ref{eq:Jp}) in Appendix \ref{Explicit Tensor Expressions in Lovelock Theories}, respectively. Similarly, for each specific branch, these quantities must be substituted with their local equivalents: $g\mapsto g_{(m)}$, $R\mapsto R_{(m)}$ and $(\prescript{NB}{}{K})\mapsto (\prescript{NB}{}{K})_{(m)}$.

	Note that $\mathcal{J}^{(0)}_{(m)ij}=0$. Therefore, the tension $T$ on the Net-brane satisfies
	\begin{equation}
		\begin{aligned}
		&Th_{ij} = -\sum_m^n\frac{G}{G_{(m)}}\\
		&\quad\times\bigl(\alpha_{1(m)}\mathcal{J}^{(1)}_{(m)ij}
		  +\alpha_{2(m)}\mathcal{J}^{(2)}_{(m)ij}
		  +\alpha_{3(m)}\mathcal{J}^{(3)}_{(m)ij}\bigr).
		\end{aligned}
	\label{eq:raw tension}
	\end{equation}

	We use the bulk metric in branch $B_{(m)}$ as follows (note that the subscript $(m)$ is omitted for $s$, $l$, $z$, and $x$ for brevity):
	\begin{equation}
	\begin{gathered}
	\text{Poincar\'e AdS:}\\
	ds^2=\frac{l^2}{z^2}\left(-dt^2+dz^2+dx^2+\delta_{ab}d\eta^a d\eta^b\right),
	\end{gathered}
	\label{eq:Poincare AdS}
	\end{equation}
	where we consistently set the effective AdS radius to $l_{(m)}=l=1$. In these coordinates, the network edge is situated at $z=0$, the node is at $z=x=0$, and the Net-brane is located at
	\begin{equation}
		\text{Net-brane:}\quad x=-\sinh(\rho)z.
	\label{eq:NB}
	\end{equation}
	The position of the Net-brane can be adjusted by varying the parameter $\rho$.

	Evaluating the junction condition in these coordinates, the tension takes the explicit form
	\begin{equation}
		\begin{aligned}
		&T = \sum_m^n\frac{G}{G_{(m)}}\\
		&\quad\times\bigl(\alpha_{1(m)}T^{(1)}
		  +\alpha_{2(m)}T^{(2)}+\alpha_{3(m)}T^{(3)}\bigr),
		\end{aligned}
	\label{eq:tension}
	\end{equation}
	\begin{align}
		T^{(1)} &= \frac{d-1}{l}\tanh(\rho)\nonumber\\
		T^{(2)} &= -\frac{(d-1)(d-2)(d-3)}{6l^3}\nonumber\\*
		  &\quad\times\left[9\sinh(\rho)+\sinh(3\rho)\right]\sech^3(\rho)\nonumber\\
		T^{(3)} &= \frac{(d-1)(d-2)(d-3)(d-4)(d-5)}{40l^5}\nonumber\\*
		  &\quad\times\left[89+28\cosh(2\rho)+3\cosh(4\rho)\right]\nonumber\\*
		  &\quad\times\sech^4(\rho)\tanh(\rho)\nonumber
	\end{align}
where $T^{(1)}$, $T^{(2)}$, and $T^{(3)}$ are the tension contributions corresponding to the Einstein, Gauss-Bonnet, and cubic Lovelock orders, respectively. The expression for $T^{(1)}$ precisely matches that given in \cite{Guo:2025sbm}.
	
	\section{Genernalized RT Surface and HEE}\label{RT and HEE}
	
	In this section, we derive the extremal and junction conditions for the generalized RT surface and calculate the HEE by evaluating the on-shell entropy functional over this surface.

	Extending the Gauss--Bonnet network HEE functional~\cite{Guo:2026prm} to cubic Lovelock order gives
	\begin{align}
		&S = \sum_{m}^{n}\frac{1}{4 G_{(m)}}\int_{\Gamma_{(m)}}
		  d^{d-1}\xi_{(m)}\sqrt{h_{\Gamma_{(m)}}}\nonumber\\*
		  &\quad\times\bigl(\alpha_{1(m)}\tilde{\mathcal{L}}^{(1)}_{(m)}
		  +\alpha_{2(m)}\tilde{\mathcal{L}}^{(2)}_{(m)}
		  +\alpha_{3(m)}\tilde{\mathcal{L}}^{(3)}_{(m)}\bigr)\nonumber\\*
		  &\quad+\int_{\gamma}d^{d-2}\eta\sqrt{h_{\gamma}}
		  \biggl[\sum_{m}^{n}\frac{1}{4 G_{(m)}}\nonumber\\*
		  &\times\bigl(\alpha_{1(m)}\tilde{\mathcal{Q}}^{(1)}_{(m)}
		  +\alpha_{2(m)}\tilde{\mathcal{Q}}^{(2)}_{(m)}
		  +\alpha_{3(m)}\tilde{\mathcal{Q}}^{(3)}_{(m)}\bigr)\biggr],
	\label{eq:HEE}
	\end{align}
where $\Gamma_{(m)}$ represent the generalized RT surfaces in different branches, which intersect the Net-brane at a common surface $\gamma$. The explicit expressions of $\tilde{\mathcal{L}}^{(p)}_{(m)}$ and $\tilde{\mathcal{Q}}^{(p)}_{(m)}$ are given by Eqs. (\ref{eq:tLp}) and (\ref{eq:tQp}), with the global quantities replaced by their branch-specific counterparts. Specifically, this involves the replacements $h_{\Gamma}\mapsto h_{\Gamma_{(m)}}$, $(\prescript{\Gamma}{}{R})\mapsto (\prescript{\Gamma_{(m)}}{}{R})$, and $(\prescript{\gamma}{}{K})\mapsto (\prescript{\gamma}{}{K})_{(m)}$. Recall that $(\prescript{\gamma}{}{K})_{(m)}$ represents the extrinsic curvature evaluated from the RT surface $\Gamma_{(m)}$ to the intersection $\gamma=\Gamma_{(m)}\cap NB$. Note that $\tilde{\mathcal{Q}}^{(1)}_{(m)}=0$.

	Taking the variation of the entropy functional (\ref{eq:HEE}), we obtain
	\begin{align}
		&\delta S = \sum_{m}^{n}\frac{1}{4 G_{(m)}}\int_{\Gamma_{(m)}}
		  d^{d-1}\xi_{(m)}\sqrt{h_{\Gamma_{(m)}}}\nonumber\\*
		  &\quad\times\bigl(\alpha_{1(m)}\tilde{\mathcal{E}}^{(1)}_{(m)\alpha\beta}
		  +\alpha_{2(m)}\tilde{\mathcal{E}}^{(2)}_{(m)\alpha\beta}
		  \nonumber\\*
		&\qquad+\alpha_{3(m)}\tilde{\mathcal{E}}^{(3)}_{(m)\alpha\beta}\bigr)
		  \delta (h_{\Gamma_{(m)}})^{\alpha\beta}\nonumber\\*
		  &\quad+\int_{\gamma}d^{d-2}\eta\sqrt{h_{\gamma}}
		  \biggl[\sum_{m}^{n}\frac{1}{4 G_{(m)}}\nonumber\\*
		  &\scalebox{0.86}{$\displaystyle
		  {}\times\bigl(\alpha_{1(m)}\tilde{\mathcal{J}}^{(1)}_{(m)ab}
		  +\alpha_{2(m)}\tilde{\mathcal{J}}^{(2)}_{(m)ab}
		  +\alpha_{3(m)}\tilde{\mathcal{J}}^{(3)}_{(m)ab}\bigr)
		  \biggr]\delta (h_{\gamma})^{ab}.$}
	\label{eq:HEE variation}
	\end{align}
	
	Given the identities $\tilde{\mathcal{L}}^{(p)}=p\mathcal{L}^{(p-1)}$ and $\tilde{\mathcal{Q}}^{(p)}=p\mathcal{Q}^{(p-1)}$, the relations $\tilde{\mathcal{E}}^{(p)}=p\mathcal{E}^{(p-1)}$ and $\tilde{\mathcal{J}}^{(p)}=p\mathcal{J}^{(p-1)}$ also naturally hold, provided that the generalized Kronecker delta and curvature tensors are replaced with their induced counterparts on the respective surfaces. The explicit expressions for $\tilde{\mathcal{E}}^{(p)}_{\alpha\beta}$ and $\tilde{\mathcal{J}}^{(p)}_{ab}$ are given by Eqs. (\ref{eq:tEp}) and (\ref{eq:tJp}) in Appendix \ref{Explicit Tensor Expressions in Lovelock Theories}, respectively. 

	Similarly, for each specific branch, these quantities must be substituted with their local equivalents: $h_{\Gamma}\mapsto h_{\Gamma_{(m)}}$, $(\prescript{\Gamma}{}{R})\mapsto (\prescript{\Gamma_{(m)}}{}{R})$ and $(\prescript{\gamma}{}{K})\mapsto (\prescript{\gamma}{}{K})_{(m)}$. Note that $\tilde{\mathcal{J}}^{(1)}_{(m)ab}=0$.
	
	Next, we present the extremal and junction conditions for the generalized RT surfaces. Hereafter, we omit the prefix "generalized" for brevity. The detailed derivation is deferred to Appendix \ref{Derivation of Extremal and Junction Conditions for RT Surfaces}, and the results are given as follows:

	\begin{equation}
		\text{Extremal Condition:}\quad\mathcal{A}_{(m)}^{\alpha\beta}(K_\Gamma)_{(m)\mu\alpha\beta}=0
	\label{eq:raw extremal condition}
	\end{equation}
	\begin{align}
		&\text{Junction Condition:}\nonumber\\
		&\sum_m^n\frac{1}{G_{(m)}}\biggl(
		 n_{{(m)}\alpha}\mathcal{A}_{(m)}^{\alpha\beta}
		 \partial_{(m)\beta}x_{(m)}^\nu g_{(m)\mu\nu}\nonumber\\*
		&\qquad-\hat{D}_a\mathcal{B}_{(m)}^{ab}
		 \partial_b x_{(m)}^\nu g_{(m)\mu\nu}\nonumber\\*
		&\qquad-\mathcal{B}_{(m)}^{ab}(K_\gamma)_{(m)\mu ab}
		 \biggr)\frac{\partial x_{(m)}^\mu}{\partial y^i(\eta^a)}=0.
	\label{eq:raw junction condition}
	\end{align}
	
	We assume the bulk branches $B_{(m)}$ are described by the Poincar\'e coordinates given in Eq. (\ref{eq:Poincare AdS}). We consider strip subregions in the NCFT branches, which correspond to spatial regions extending infinitely along all $\eta^a$ directions. For the RT surfaces, we adopt the following parametric coordinates (note that the branch index $(m)$ for $(\prescript{\Gamma}{}{s})$, $z$, and $s$ will be omitted from here on until the final results for brevity):
	\begin{equation}
		d(\prescript{\Gamma}{}{s})^2=\frac{l^2}{z(s)^2}\left(ds^2+\delta_{ab}d\eta^a d\eta^b\right).
	\label{eq:RT paramter coordinate}
	\end{equation}

	Due to the translational symmetry along all $\eta^a$ directions, the problem simplifies to studying the 1D projected curves of the RT surfaces onto the $xz$-plane. To parameterize these curves smoothly, we use $s$, which strictly denotes the arc length parameter of these projected curves. The intersection $\gamma$ is defined to be at $s=0$. To reduce the order of the resulting differential equations, we introduce the following substitutions:
	\begin{align}
		&x'(s)=\sin(\theta(s))\qquad x''(s)=\cos(\theta(s))\theta'(s)\nonumber\\
		&z'(s)=\cos(\theta(s))\qquad z''(s)=-\sin(\theta(s))\theta'(s),
	\label{eq:xpzp to theta}
	\end{align}

	where the prime ($'$) denotes the derivative with respect to $s$, and $\theta$ is the angle measured from the positive $z$-axis to the tangent vector of the projected RT curve. Upon making these substitutions, Eq. (\ref{eq:raw extremal condition}) yields two non-trivial components, both of which reduce to the identical differential equation for the extremal condition. On the other hand, Eq. (\ref{eq:raw junction condition}) yields only one non-trivial component, which provides the specific junction condition at the Net-brane. The explicit forms of specific extremal and junction conditions are given by Eqs. (\ref{eq:specific extremal condition}) and (\ref{eq:specific NCFT junction condition}).
	
	We denote the length of the subregion along the $x$-direction as $x_{(m)L}$ and introduce a UV cut-off at $z = \varepsilon_z$. In the numerical calculations, we always set $\varepsilon_z=10^{-5}$. In summary, the complete set of differential equations and boundary conditions is given by:
	\begin{equation}
	\begin{cases}
    \text{Eq. (\ref{eq:specific extremal condition})} &m=1,\,\cdots,\,n\\
		z_{(m)}'(s_{(m)})=\cos(\theta_{(m)}(s_{(m)})) &m=1,\,\cdots,\,n\\
    x_{(m)}'(s_{(m)})=\sin(\theta_{(m)}(s_{(m)})) &m=1,\,\cdots,\,n
	\end{cases}
	\label{eq:NCFT full differential equation}
	\end{equation}
	\begin{equation}
	\begin{cases}
    \text{Eq. (\ref{eq:specific NCFT junction condition})} \\
	z_{(m)}(s_{(m)}=0)=z_b &m=1,\,\cdots,\,n\\
		x_{(m)}(s_{(m)}=0)=-\sinh(\rho)z_b &m=1,\,\cdots,\,n\\
	z_{(m)}(s_{(m)E})=\varepsilon_z &m=1,\,\cdots,\,n\\
	x_{(m)}(s_{(m)E})=x_{(m)L} &m=1,\,\cdots,\,n
	\end{cases}
	\label{eq:NCFT full boundary condition}
	\end{equation}
	
	In our numerical calculations, we employ the shooting method to find valid solutions. We first guess a set of initial values for $z_b$ and $\theta_{(m)0}$ that satisfy the junction condition (\ref{eq:specific NCFT junction condition}). We then integrate the extremal equations (\ref{eq:NCFT full differential equation}) inward until the coordinate $z_{(m)}(s_{(m)})$ reaches the UV cut-off $\varepsilon_z$; we define this endpoint as $s_{(m)E}$. Next, we verify whether the corresponding $x$-coordinate satisfies the boundary condition $x_{(m)}(s_{(m)E}) = x_{(m)L}$ for all branches. A configuration that satisfies this matching condition across all branches constitutes a valid geometric solution. If multiple valid configurations are found, the physical RT surface is the one that minimizes the HEE. The specific expressions for the boundary and bulk HEE integrals in Eq. (\ref{eq:HEE}) are given by Eqs. (\ref{eq:SBdy}) and (\ref{eq:SBulk}), respectively.

	\section{Numerical Results}\label{Numerical Results}
	
	In this section, we employ numerical calculations to demonstrate the relationship between the HEE and the couplings. Specifically, we vary $\rho$, which alters the location of the Net-brane, to explore the relationship between the HEE and $\rho$ for a given set of fixed couplings.

	We adopt the same notations and concepts for the HEE as in \cite{Guo:2025sbm, Guo:2026prm}. Here, $S_{\text{NCFT}}(\rho)$ denotes the HEE of the subregions in the NCFT, and $S_{\text{BCFT}}(\rho)$ denotes the total HEE of the subregions of the isolated $n$-branch BCFTs. The computational distinction between $S_{\text{NCFT}}(\rho)$ and $S_{\text{BCFT}}(\rho)$ lies in the fact that, to evaluate the latter, we replace Eq.~(\ref{eq:specific NCFT junction condition}) with the $n$-branch junction condition (\ref{eq:specific BCFT junction condition}), and replace $z_b$ with $z_{(m)b}$ in the boundary conditions (\ref{eq:NCFT full boundary condition}). Our actual interest lies not in the raw values of $S_{\text{NCFT}}(\rho)$ or $S_{\text{BCFT}}(\rho)$—which vary with the UV cutoff $\varepsilon_z^S$ of the HEE integral—but rather in a quantity that represents the entanglement entropy of the network and is independent of $\varepsilon_z^S$. We use three candidates to describe the network entanglement entropy, which are defined as
	\begin{align}
 		&S_{\text{I}}(\rho)=S_{\text{NCFT}}(\rho)-S_{\text{BCFT}}(\rho=0)\nonumber\\
		&S_{\text{II}}(\rho)=S_{\text{NCFT}}(\rho)-S_{\text{NCFT}}(\rho=0)\nonumber\\
		&S_{\text{III}}(\rho)=S_{\text{NCFT}}(\rho)-S_{\text{BCFT}}(\rho).
	\label{eq:network ee}
	\end{align}

	From Eq. (\ref{eq:network ee}), the three types of network HEE share underlying connections—namely, a constant difference between $S_{\text{I}}(\rho)$ and $S_{\text{II}}(\rho)$, and the relations $S_{\text{I}}(\rho=0)=S_{\text{III}}(\rho=0)$ and $S_{\text{II}}(\rho=0)=0$. 

	As the numerical results below demonstrate, these three types of network HEE remain independent of $\varepsilon_z^S$ within numerical error tolerances. Because they are free from UV cutoff dependence, they genuinely reflect the intrinsic entanglement nature of the network. Therefore, we ultimately use the three types of network HEE, $S_{\text{I}}(\rho)$, $S_{\text{II}}(\rho)$, and $S_{\text{III}}(\rho)$ to define the calculated network HEE in this paper.
	
	Notably, we set a UV cutoff $\varepsilon_z^S$ for the HEE integral that differs from the UV cutoff $\varepsilon_z$ used in (\ref{eq:NCFT full boundary condition}), as an overly small $\varepsilon_z^S$ introduces significant numerical deviations in the HEE integral, especially for large $d$. Therefore, we set relatively larger values of $\varepsilon_z^S$ for larger dimensions $d$, as follows:
	\begin{align}
 		&\varepsilon_z^S=10^{-5} &\text{for}\,d=2\nonumber\\
		&\varepsilon_z^S=5\times10^{-3} &\text{for}\,d=4\nonumber\\
		&\varepsilon_z^S=2\times10^{-1} &\text{for}\,d=6 &.
	\label{eq:ezS}
	\end{align}
	
	\subsection{Recovering previous results}

\begin{figure*}[!tp]
		\centering
		\subfloat[one-branch $S^1_{\text{BCFT}}(\rho)$]{%
			\includegraphics[width=0.35\linewidth]{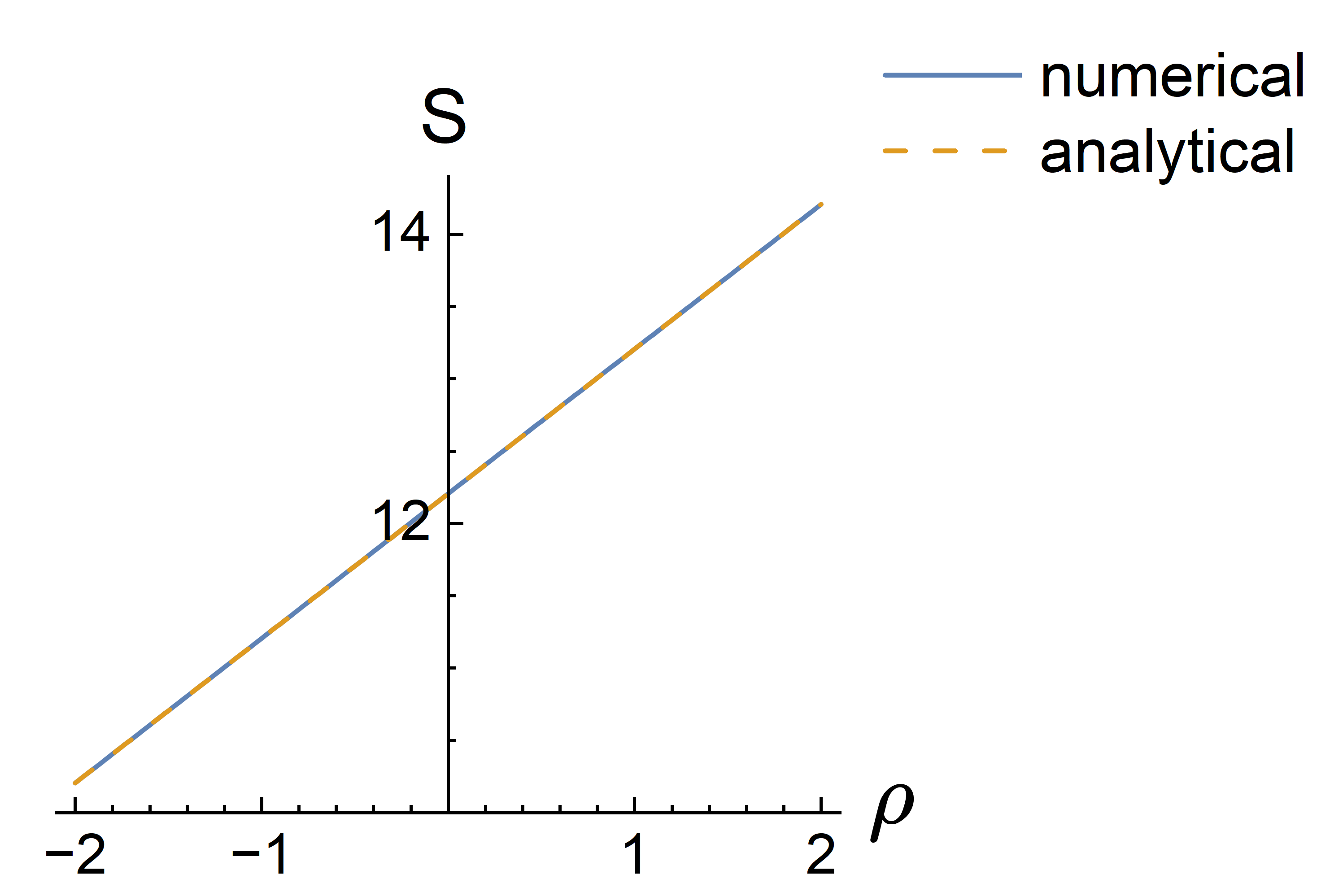}%
			\label{fig:Einstein_d=2_BCFT_analytical_vs_numerical}}\hspace{0.06\linewidth}
		\subfloat[RT surfaces]{%
			\includegraphics[width=0.35\linewidth]{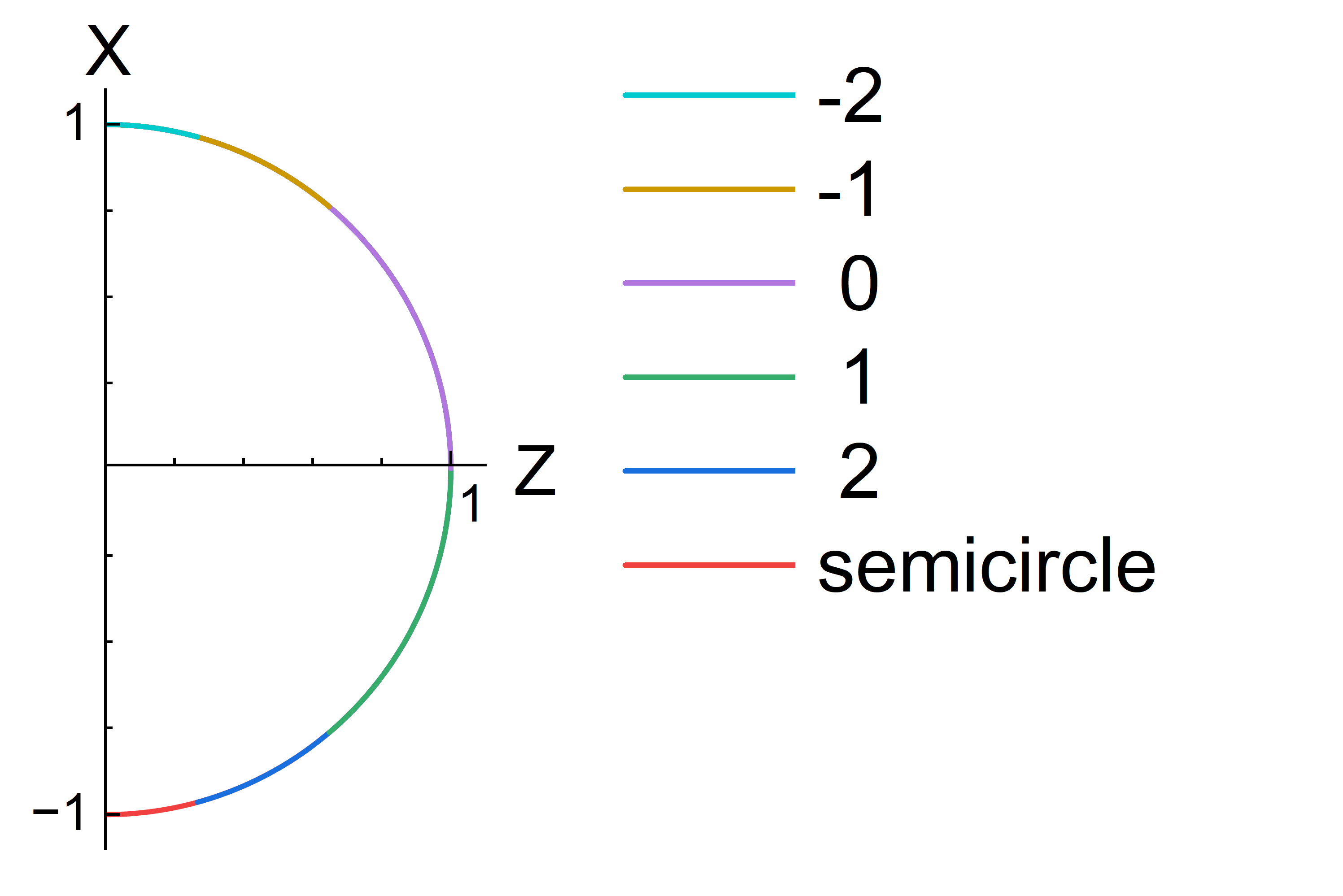}%
			\label{fig:Einstein_d=2_BCFT_RT_analytical_vs_numerical}}
		\caption{Comparison between analytical and numerical results for the HEE (a) and RT surfaces (b) of the one-branch BCFT in $(d+1=3)$ Einstein gravity. The legend in (b) indicates the values of $\rho$ for the corresponding numerical RT surfaces. Geometrically, the RT surface for a given value of $\rho$ completely overlaps with and extends beyond the RT surfaces corresponding to smaller values of $\rho$. The analytical RT surfaces correspond to segments of semicircles}
	\label{fig:Einstein_d=2_BCFT}
	\end{figure*}

\begin{figure*}[!tp]
		\centering
		\subfloat[one-branch $S^1_{\text{BCFT}}(\rho)$]{%
			\includegraphics[width=0.42\linewidth]{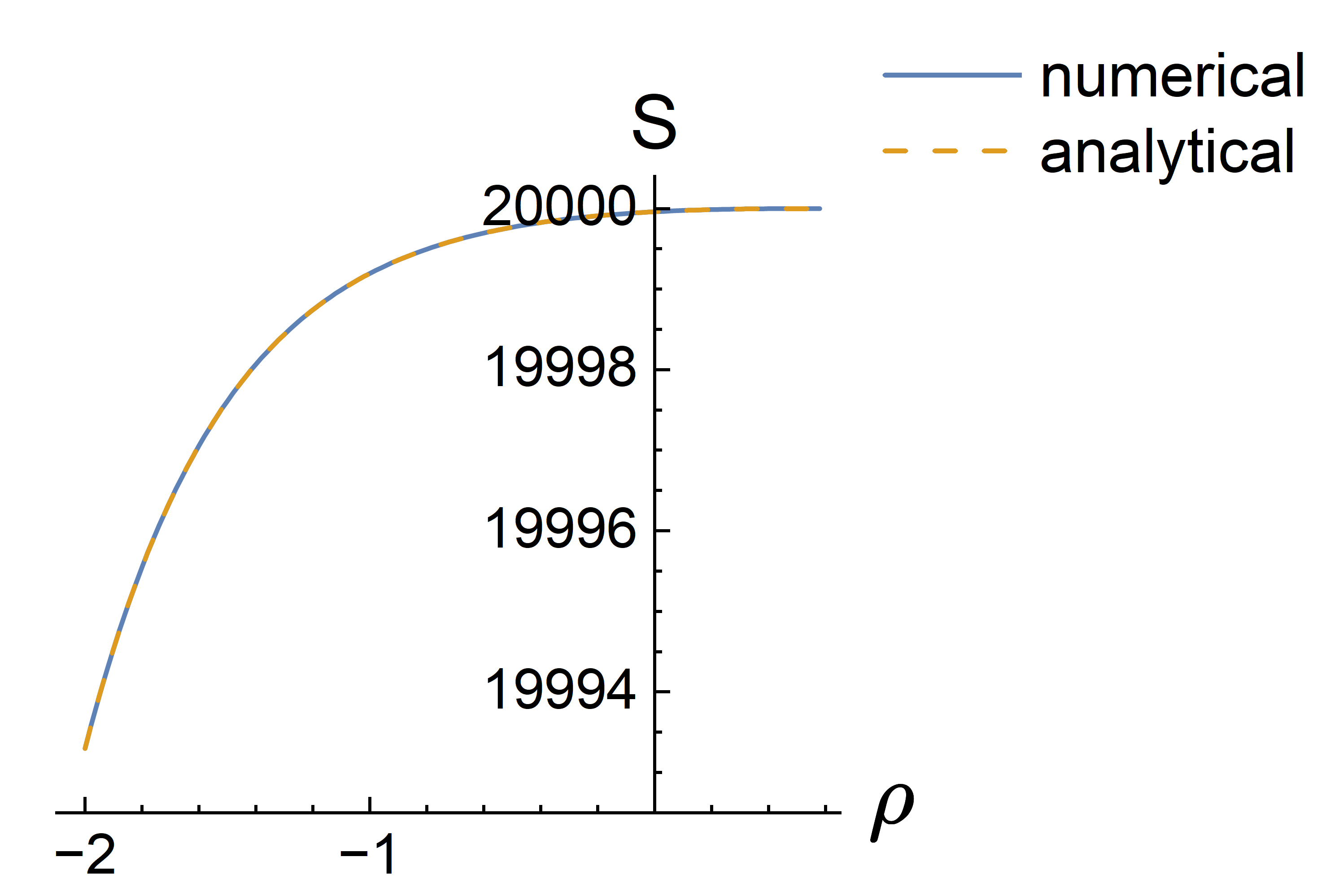}%
			\label{fig:Einstein_d=4_BCFT_analytical_vs_numerical}}\hfill
		\subfloat[RT surface, $\rho=-0.5$]{%
			\includegraphics[width=0.42\linewidth]{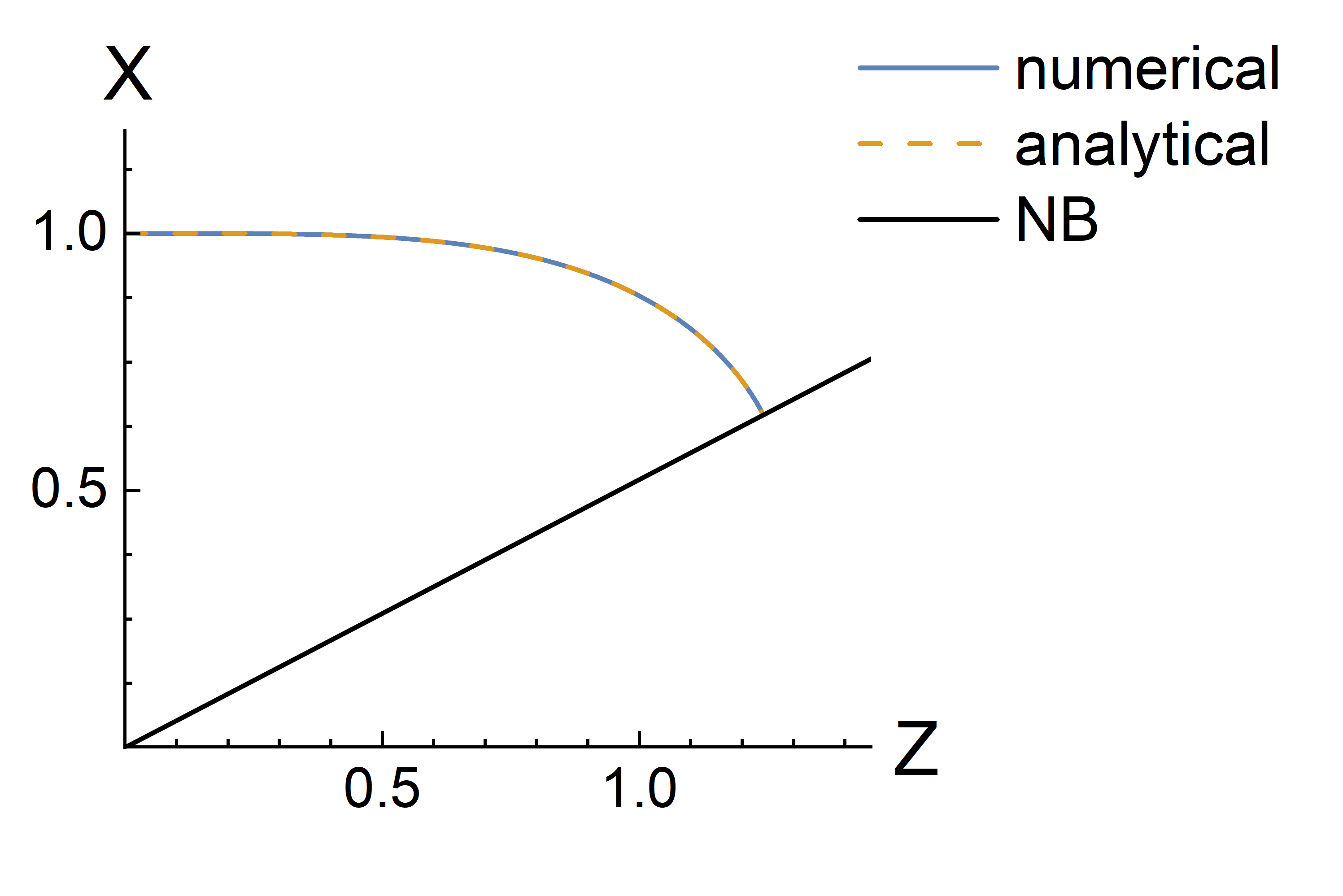}%
			\label{fig:Einstein_d=4_rho=-0.5_BCFT_RT_analytic_vs_numerical}}\par\smallskip
		\subfloat[RT surface, $\rho=0$]{%
			\includegraphics[width=0.42\linewidth]{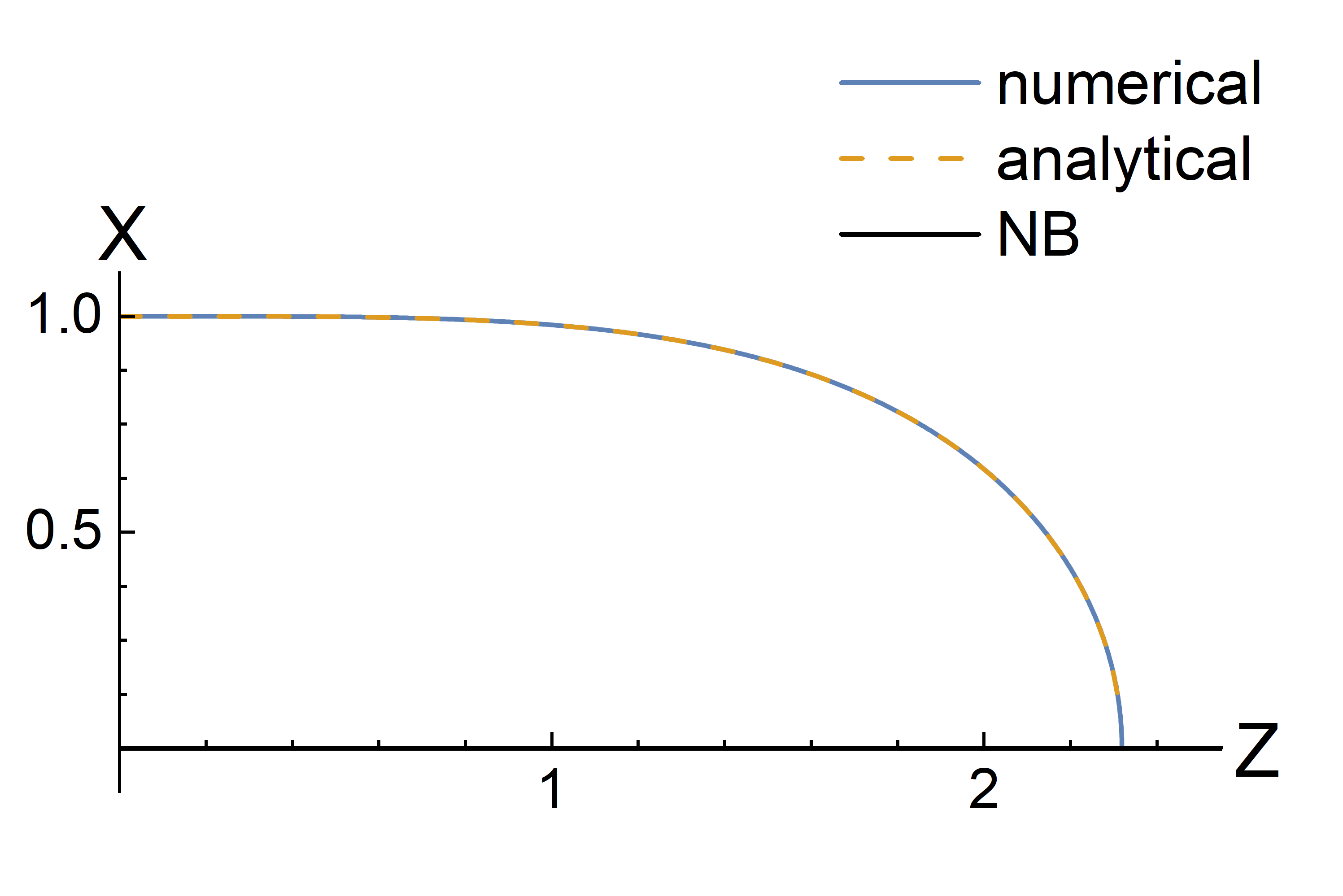}%
			\label{fig:Einstein_d=4_rho=0_BCFT_RT_analytic_vs_numerical}}\hfill
		\subfloat[RT surface, $\rho=0.5$]{%
			\includegraphics[width=0.42\linewidth]{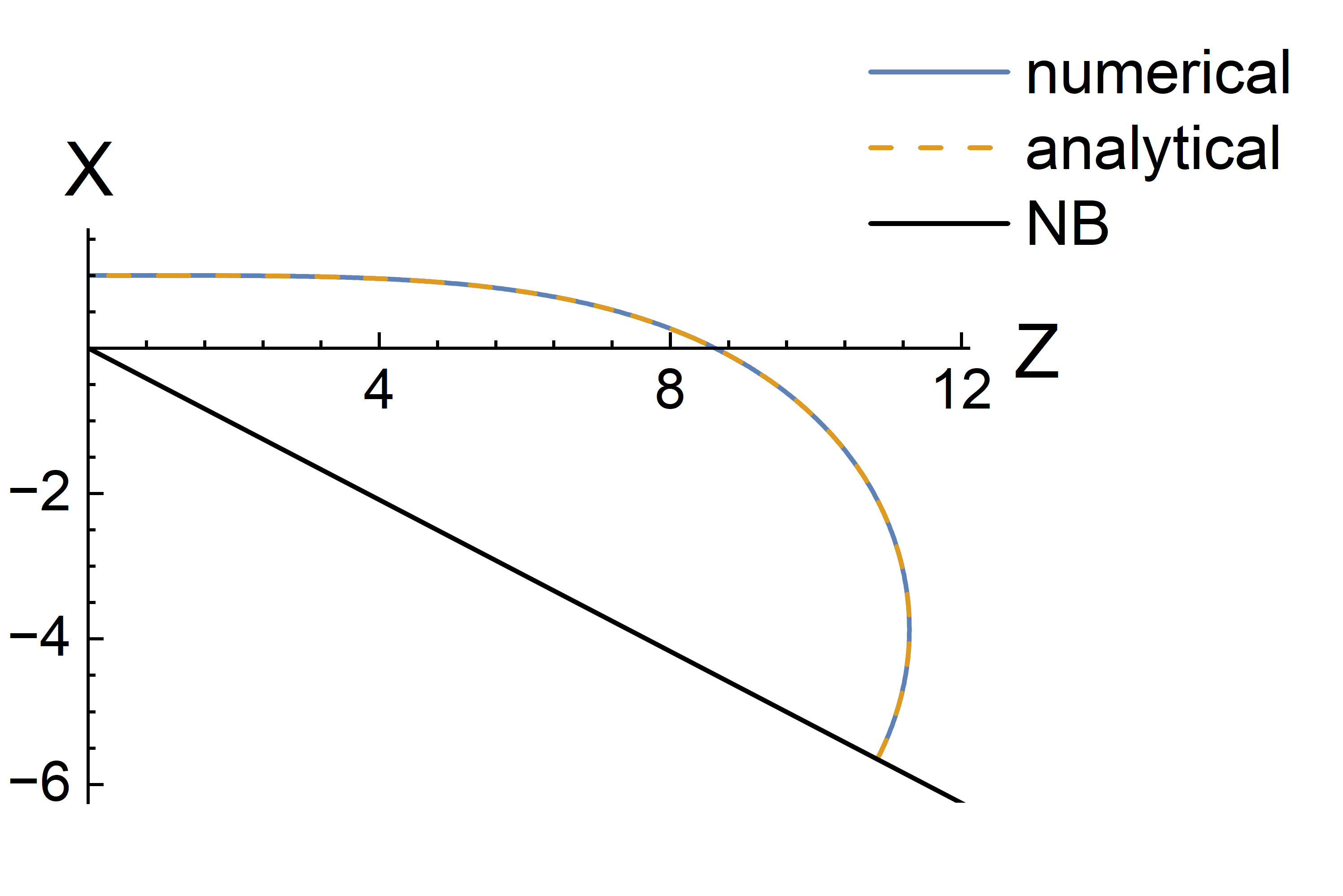}%
			\label{fig:Einstein_d=4_rho=0.5_BCFT_RT_analytic_vs_numerical}}
		\caption{Comparison between analytical and numerical results for the HEE (a) and RT surfaces (b--d) of the one-branch BCFT in $(d+1=5)$ Einstein gravity for various values of $\rho$}
	\label{fig:Einstein_d=4_BCFT}
	\end{figure*}

\begin{figure*}[!tp]
		\centering
		\subfloat[one-branch $S^1_{\text{BCFT}}(\rho)$]{%
			\includegraphics[width=0.42\linewidth]{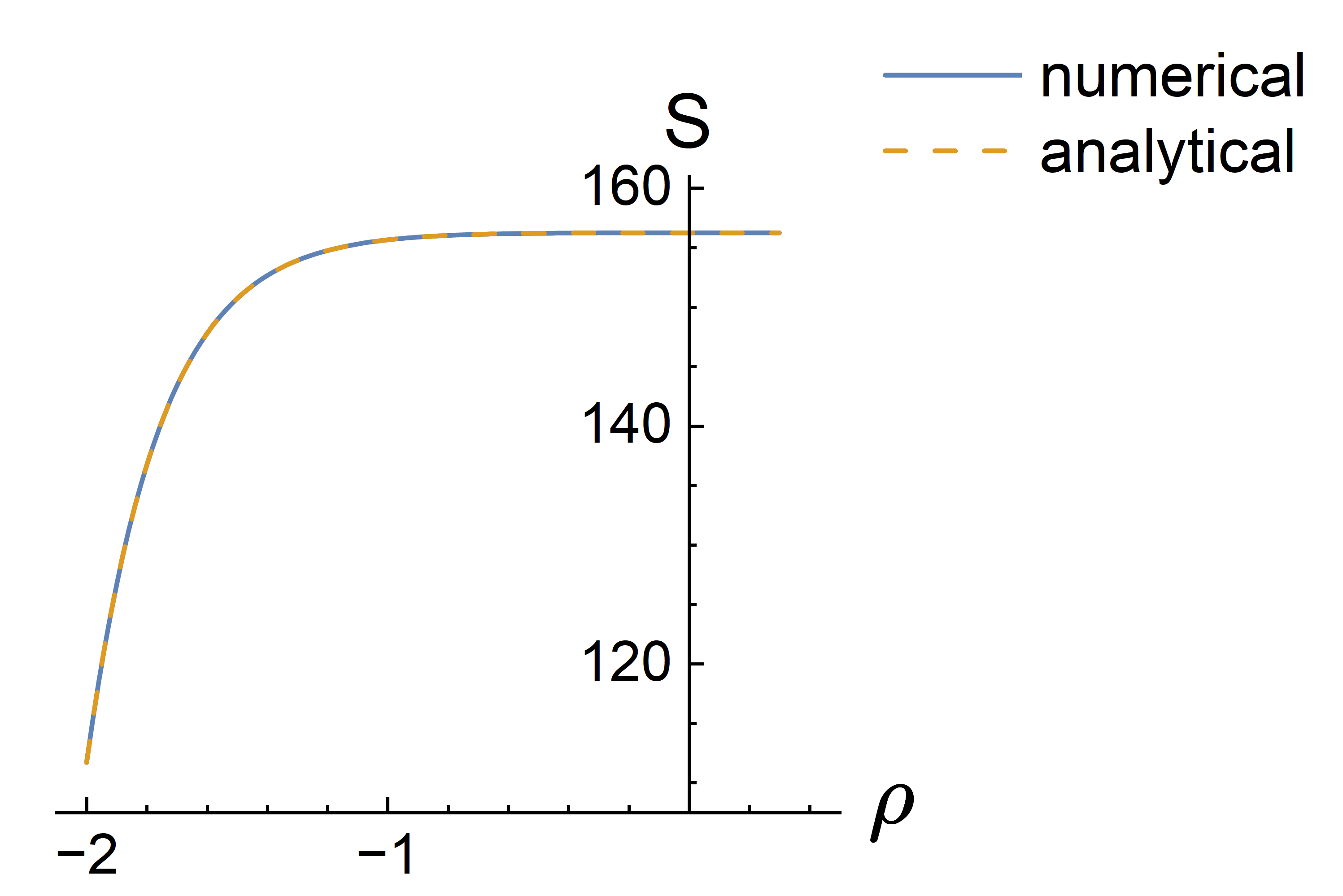}%
			\label{fig:Einstein_d=6_BCFT_analytical_vs_numerical}}\hfill
		\subfloat[$\rho=-0.3$]{%
			\includegraphics[width=0.42\linewidth]{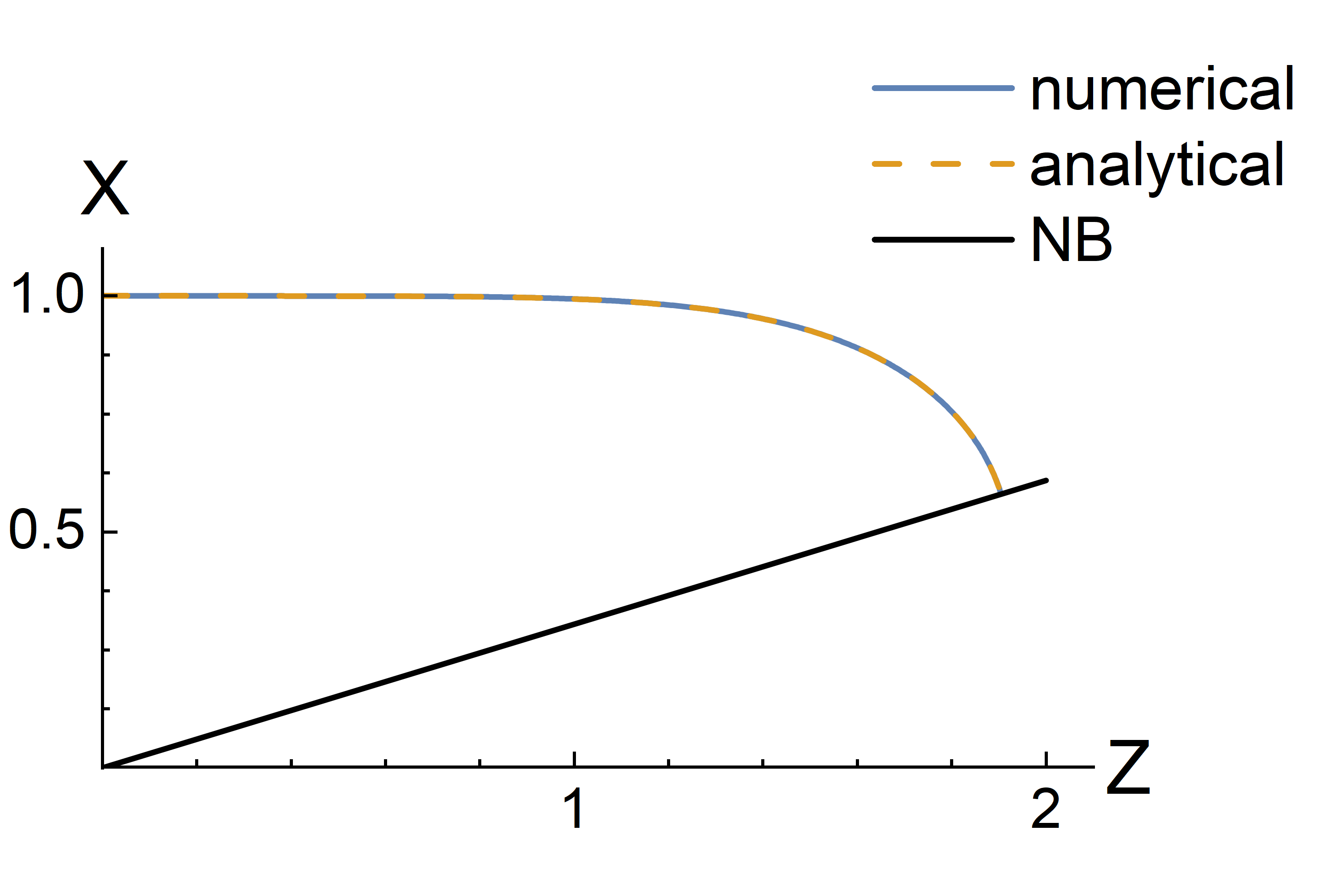}%
			\label{fig:Einstein_d=6_rho=-0.3_BCFT_RT_analytic_vs_numerical}}\par\smallskip
		\subfloat[$\rho=0$]{%
			\includegraphics[width=0.42\linewidth]{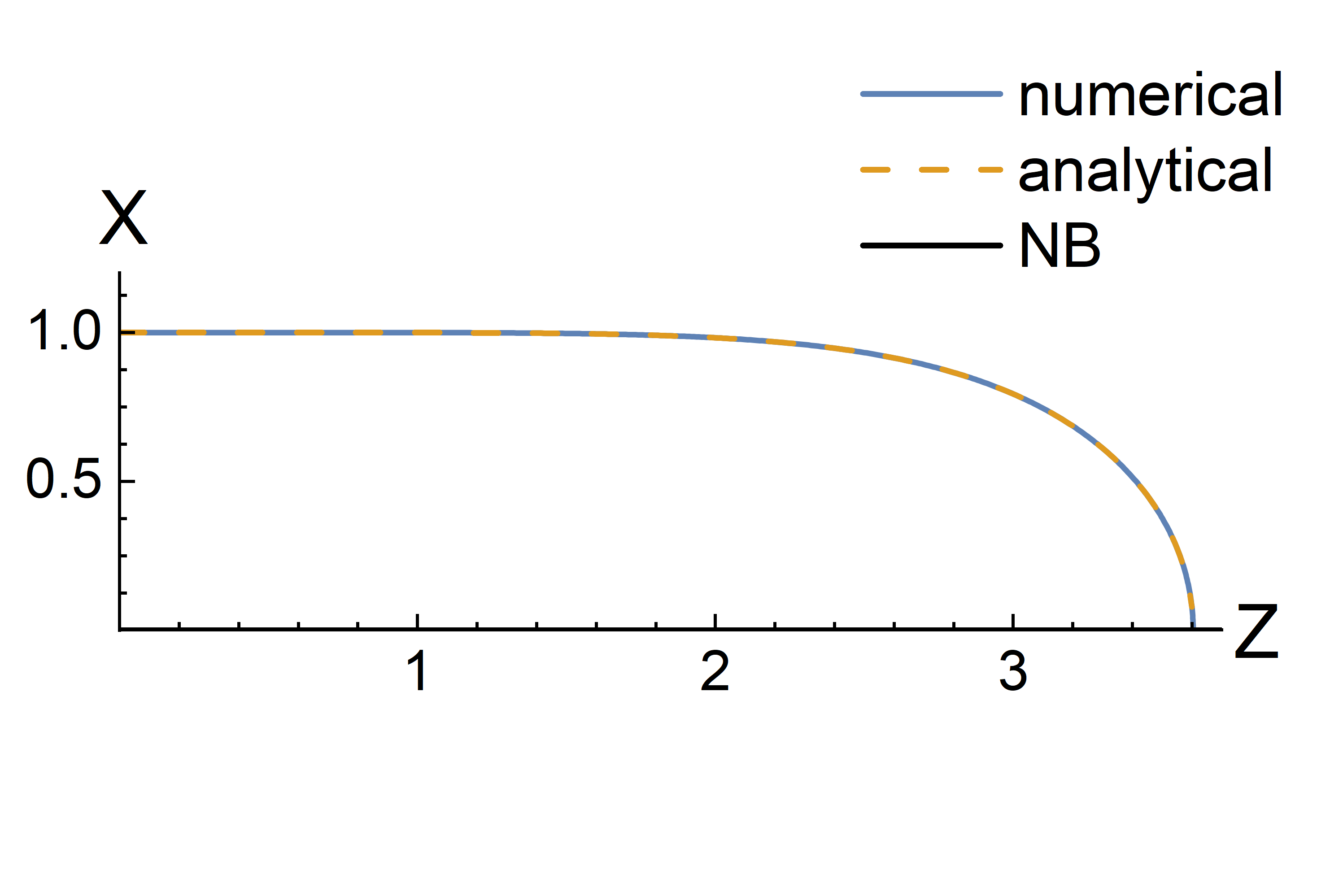}%
			\label{fig:Einstein_d=6_rho=0_BCFT_RT_analytic_vs_numerical}}\hfill
		\subfloat[$\rho=0.3$]{%
			\includegraphics[width=0.42\linewidth]{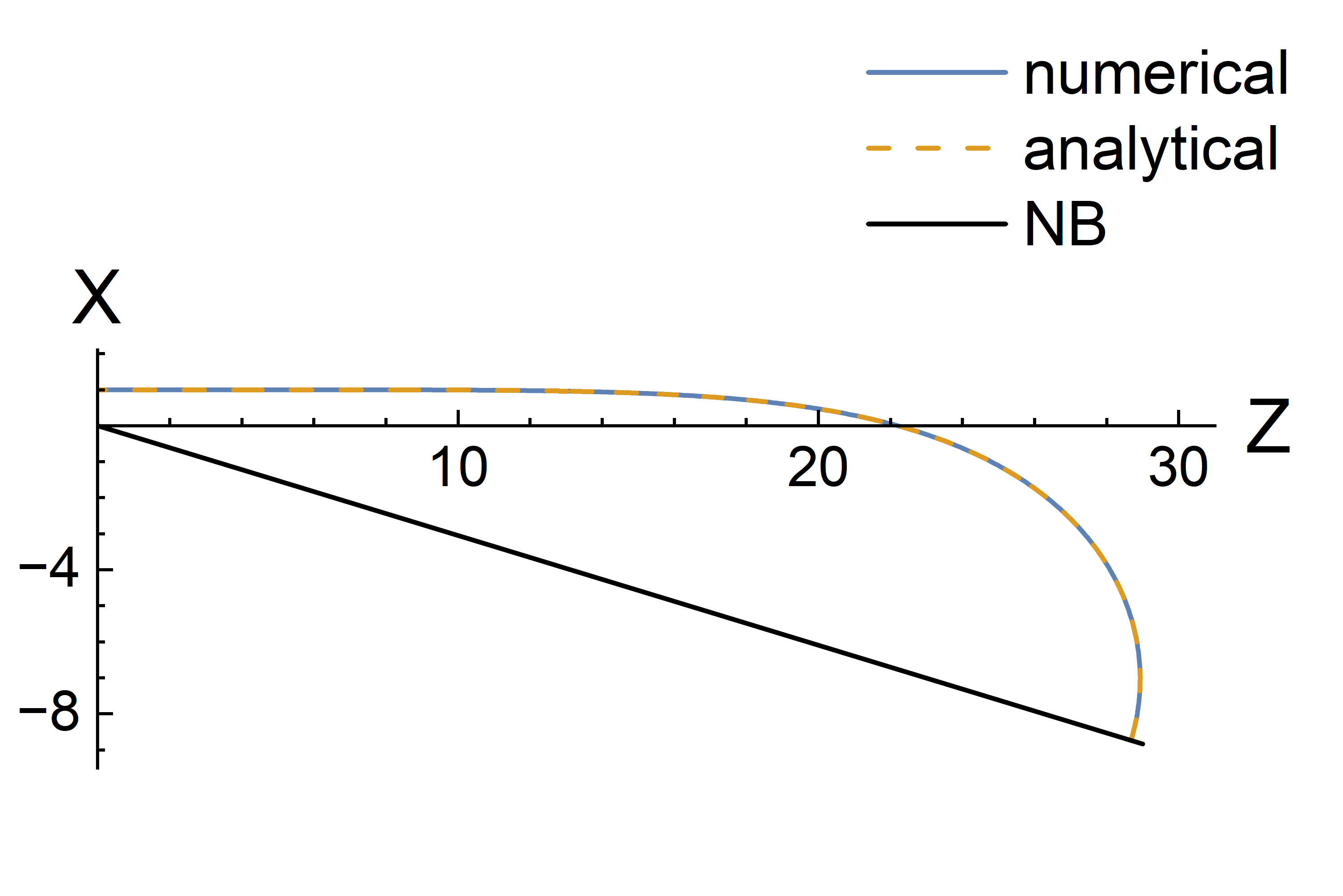}%
			\label{fig:Einstein_d=6_rho=0.3_BCFT_RT_analytic_vs_numerical}}
		\caption{Comparison between analytical and numerical results for the HEE (a) and RT surfaces (b--d) of the one-branch BCFT in $(d+1=7)$ Einstein gravity for various values of $\rho$}
	\label{fig:Einstein_d=6_BCFT}
	\end{figure*}

	We verify that our numerical approach correctly reproduces the exact HEE and the corresponding RT surface profiles for the one-branch AdS/BCFT, as illustrated in Figs.~\ref{fig:Einstein_d=2_BCFT}, \ref{fig:Einstein_d=4_BCFT}, and \ref{fig:Einstein_d=6_BCFT}. In these one-branch AdS/BCFT setups, we set $x_L=1$ and $4G=1$.

	Based on analytical calculations, the RT surface of the one-branch BCFT for $(d+1=3)$ Einstein gravity is a circular segment. The corresponding one-branch HEE, $S^1_{\text{BCFT}}(\rho)$, is given by
	
	\begin{equation}
		S^1_{\text{BCFT}}(\rho)=\frac{1}{4G}\bigg(\log\big(\frac{2\,x_L}{\varepsilon^S_z}\big)+\rho\bigg).
	\label{eq:Einstein_d=2_SBCFT_analytical}
	\end{equation}

	Analytical derivations for the RT surface and HEE of the one-branch BCFT in $(d+1\geq 4)$ Einstein gravity can be found in \cite{Chu:2017aab}. However, their conventions differ slightly from ours. Furthermore, they do not consider the case where the NB brane is positioned on the same side as the BCFT subregion. The refined analytical expressions for the HEE and RT surfaces, adapted to be compatible with our framework, along with their detailed derivations, are presented in the Appendix \ref{RT Surface and HEE of One-Branch BCFT in dgeq3 Einstein Gravity}.
	
	In our numerical calculations, we impose an upper bound on $z_b$, which corresponds to the maximum valid value of $\rho$ for physical solutions. As seen in Figs.~\ref{fig:Einstein_d=4_BCFT_analytical_vs_numerical} and \ref{fig:Einstein_d=6_BCFT_analytical_vs_numerical}, if the upper bound on $z_b$ is removed, the RT surface at sufficiently large $\rho$ degenerates into a straight line orthogonal to the $x$-axis, and the corresponding one-branch HEE, $S^1_{\text{BCFT}}(\rho)$, reaches a constant value. This behavior is evident from the plateaus shown in Figs.~\ref{fig:Einstein_d=4_BCFT_analytical_vs_numerical} and \ref{fig:Einstein_d=6_BCFT_analytical_vs_numerical}. We verify that these plateau values perfectly match the analytical results. Therefore, although our numerical framework cannot directly yield valid solutions for extremely large $\rho$, we can deduce the correct asymptotic behavior for large $\rho$ from the existing solutions at smaller $\rho$.

\begin{figure*}[!tp]
		\centering
		\subfloat[]{%
			\includegraphics[width=0.31\linewidth]{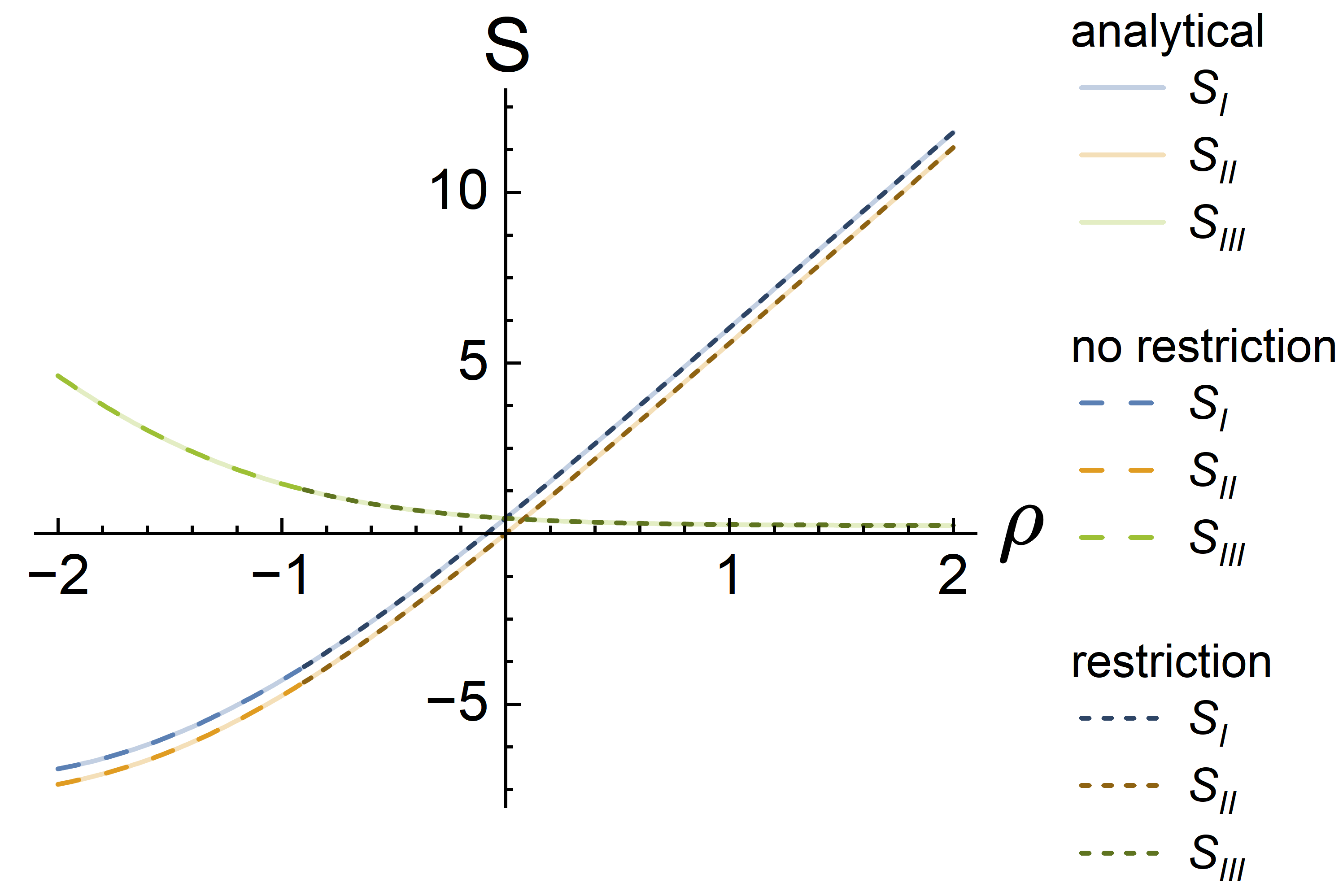}%
			\label{fig:Einstein_d=2_analytical_vs_numerical}}\hfill
		\subfloat[$\rho=-0.9$]{%
			\includegraphics[width=0.31\linewidth]{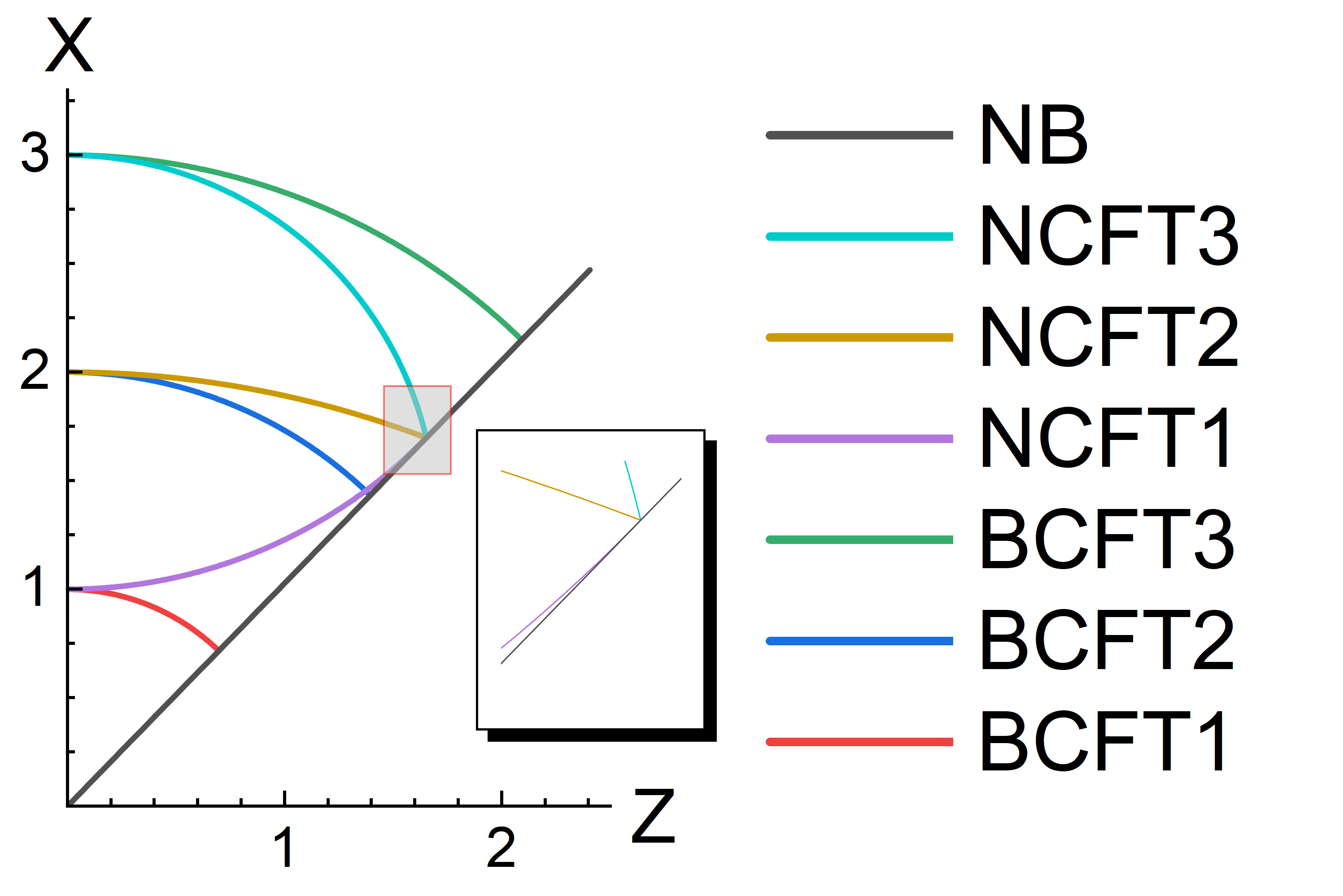}%
			\label{fig:Einstein_d=2_rho=-0.9}}\hfill
		\subfloat[$\rho=-1.0$]{%
			\includegraphics[width=0.31\linewidth]{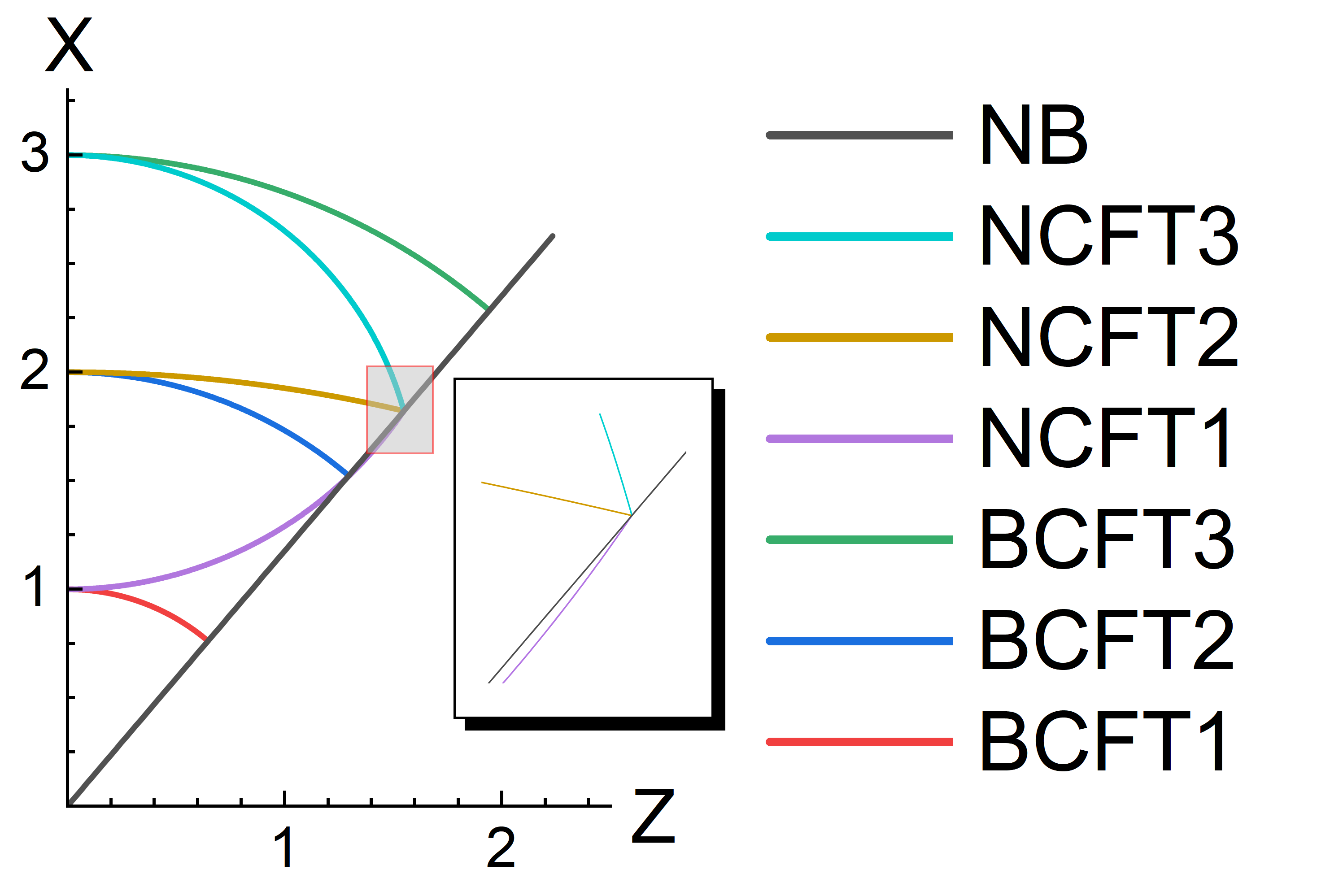}%
			\label{fig:Einstein_d=2_rho=-1.0}}
		\caption{Network HEE and RT surfaces for a one-node, three-branch configuration in $(d+1=3)$ Einstein gravity. We set $x_{(1)L}=1$, $x_{(2)L}=2$, $x_{(3)L}=3$, $4\pi G_{(1)}=3$, $4\pi G_{(2)}=2$, and $4\pi G_{(3)}=1$. (a) Solid lines represent the analytical results. The sparse dashed lines represent the unphysical numerical results that permit NB crossing, whereas the dense dashed lines represent the physically valid numerical results that prohibit NB crossing. (b) RT surfaces for $\rho=-0.9$, where NB crossing does not occur. (c) RT surfaces for $\rho=-1.0$, where NB crossing occurs}
	\label{fig:Einstein_d=2_NCFT example}
	\end{figure*}

	We also analyze the configuration provided in~\cite{Guo:2026prm}, with the results displayed in Fig.~\ref{fig:Einstein_d=2_NCFT example}. Our numerical results are nearly identical to their analytical results. The only distinction is that our results exhibit a narrower valid range for $\rho$. We identify an oversight in their analysis: the analytical results presented in~\cite{Guo:2026prm} do not account for whether the RT surface crosses the Net-brane (NB) into the unphysical region. If we artificially permit this crossing in our numerical framework, we recover the broader valid range of $\rho$ and the exact entropy values reported in~\cite{Guo:2026prm}. This crossing scenario is illustrated in Fig.~\ref{fig:Einstein_d=2_rho=-1.0}. As demonstrated in this figure, we consistently plot all $n$-branch RT surfaces together within a single panel.

	The analytical formulas for the entropies of a one-node, three-branch network in $(d+1=3)$ Einstein gravity are given by

	\begin{align}
		&S_{\text{BCFT}}(\rho)=\sum_m^n\frac{1}{4 G_{(m)}}
		\bigg(\log(\frac{2\,x_{(m)L}}{\varepsilon_z^S})+\rho\bigg)\nonumber\\
		&S_{\text{NCFT}}(\rho)=\sum_m^n\frac{1}{4 G_{(m)}}\nonumber\\*
		&\qquad\times\log\bigg(\frac{(x_{(m)L}+\sinh(\rho)z_b)^2+z_b^2}
		{z_b\,\varepsilon_z^S}\bigg),
	\label{eq:Einstein_d=2_NCFT}
	\end{align}
	where $z_b$ is determined by minimizing $S_{\text{NCFT}}(\rho)$.

	\subsection{New results}
	
	The figures in this subsection illustrate the functional dependence of $S_{\text{I}}(\rho)$, $S_{\text{II}}(\rho)$, $S_{\text{III}}(\rho)$, as well as $z_b(\rho)$, on the parameter $\rho$, across various sets of gravitational coupling coefficients and the dimensions $d$. The network configuration investigated in this subsection is consistently a one-node, three-branch network. Unless otherwise specified, we fix $x_{(m)L}=1$ and $4G_{(m)}=1$ for all branches; the only exception is the Einstein gravity case, where we assign distinct parameters to different branches.
	
	Overall, in all scenarios investigated, $S_{\text{I}}$ and $S_{\text{II}}$ increase monotonically with $\rho$, whereas $S_{\text{III}}$ decreases with $\rho$. Notably, $S_{\text{III}}$ remains non-negative throughout, implying that $S_{\text{NCFT}} \geq S_{\text{BCFT}}$. Physically, this indicates that the junction condition of the NCFT is generally more restrictive than the boundary conditions of the isolated BCFTs.
	
	\begin{figure*}[!tp]
		\centering
		\subfloat[$d=4$]{%
			\includegraphics[width=0.35\linewidth]{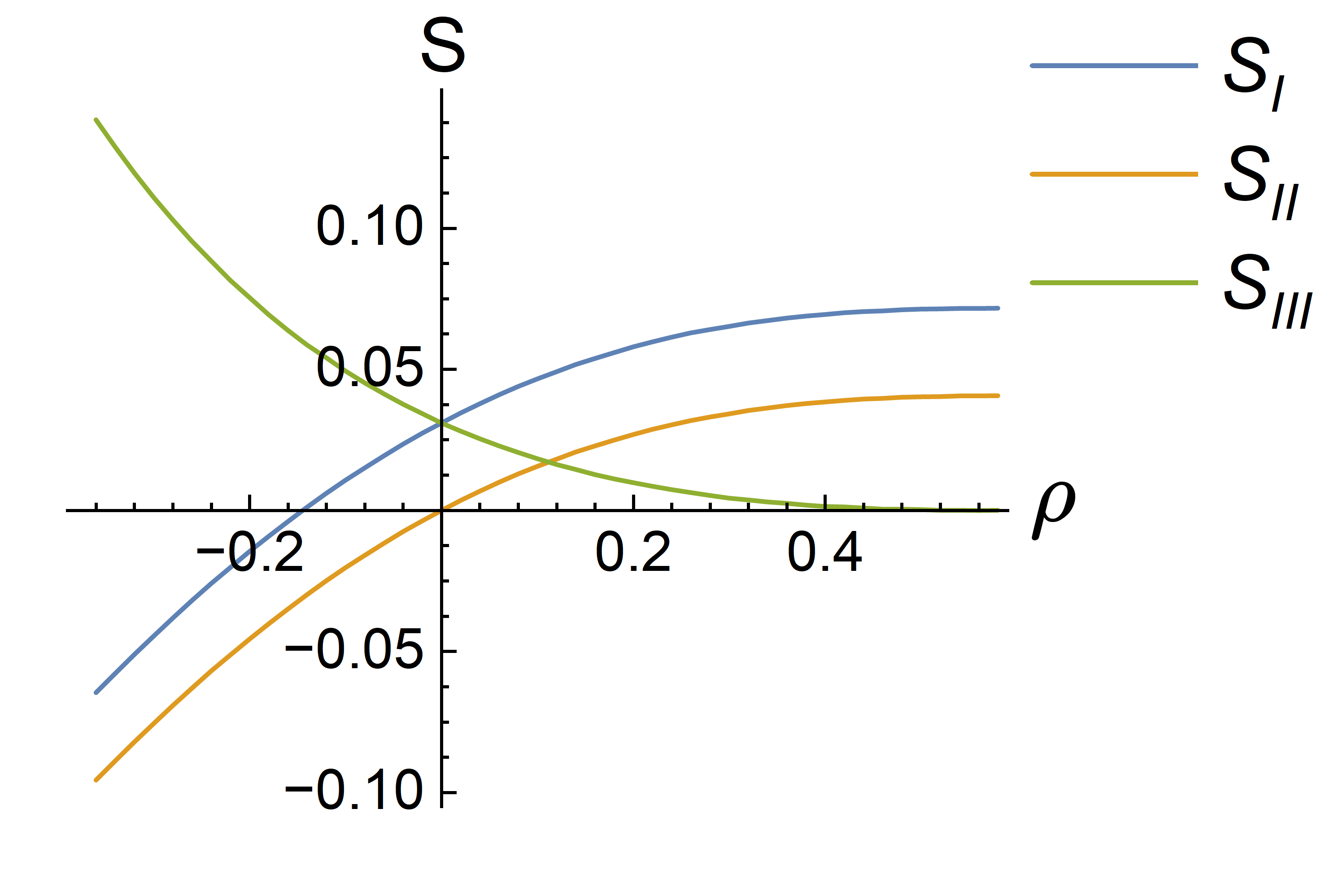}%
			\label{fig:Einstein_d=4_DW_SNCFT}}\hspace{0.06\linewidth}
		\subfloat[$d=6$]{%
			\includegraphics[width=0.35\linewidth]{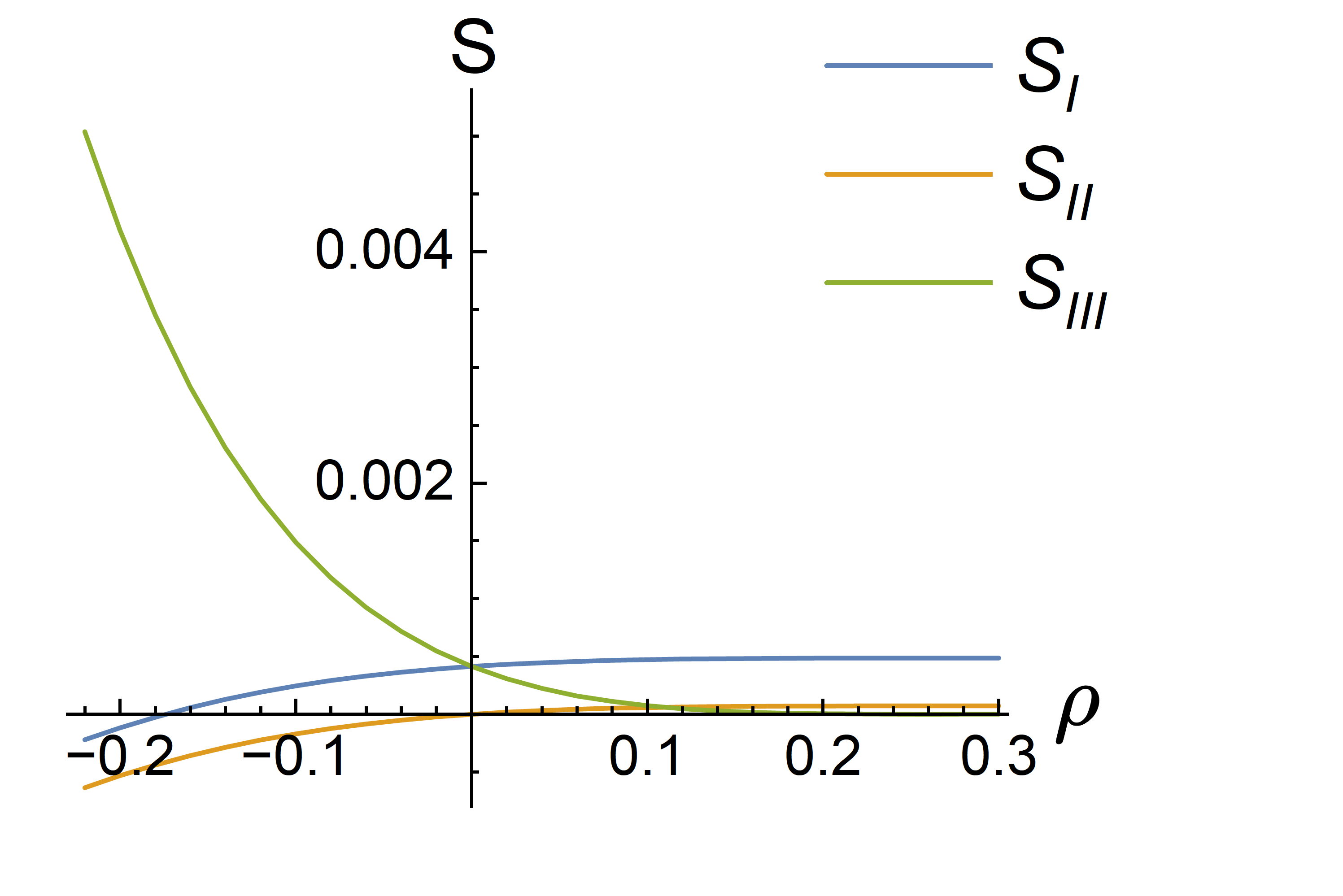}%
			\label{fig:Einstein_d=6_DW_SNCFT}}
		\caption{Network HEE for Einstein gravtiy in $(d+1=5,\,7)$. We set $x_{(1)L}=1$, $x_{(2)L}=2$, $x_{(3)L}=3$, $4\pi G_{(1)}=3$, $4\pi G_{(2)}=2$, and $4\pi G_{(3)}=1$}
		\par\medskip
		\centering
		\subfloat[]{%
			\includegraphics[width=0.42\linewidth]{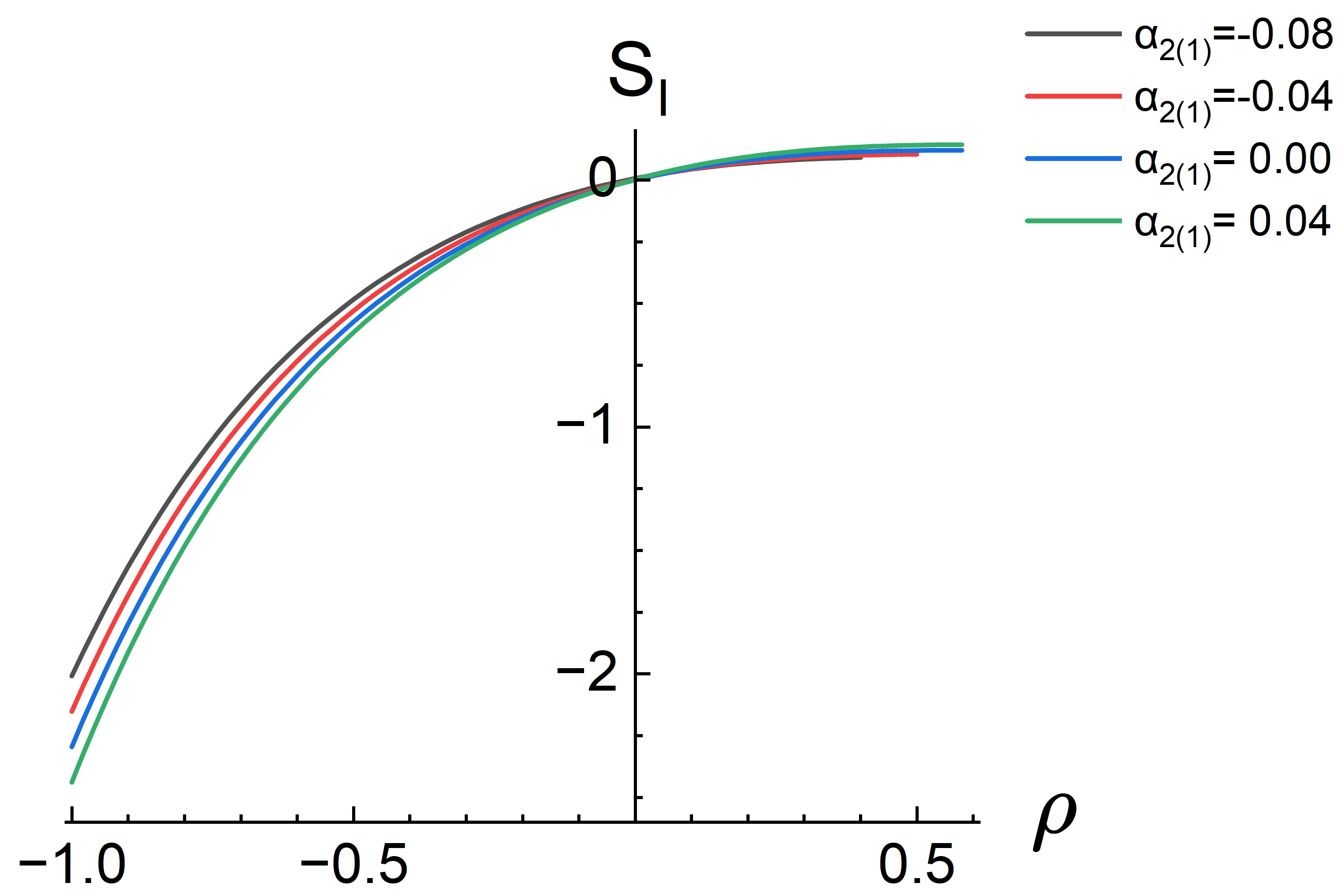}%
			\label{fig:d=4_GB_SI}}\hfill
		\subfloat[]{%
			\includegraphics[width=0.42\linewidth]{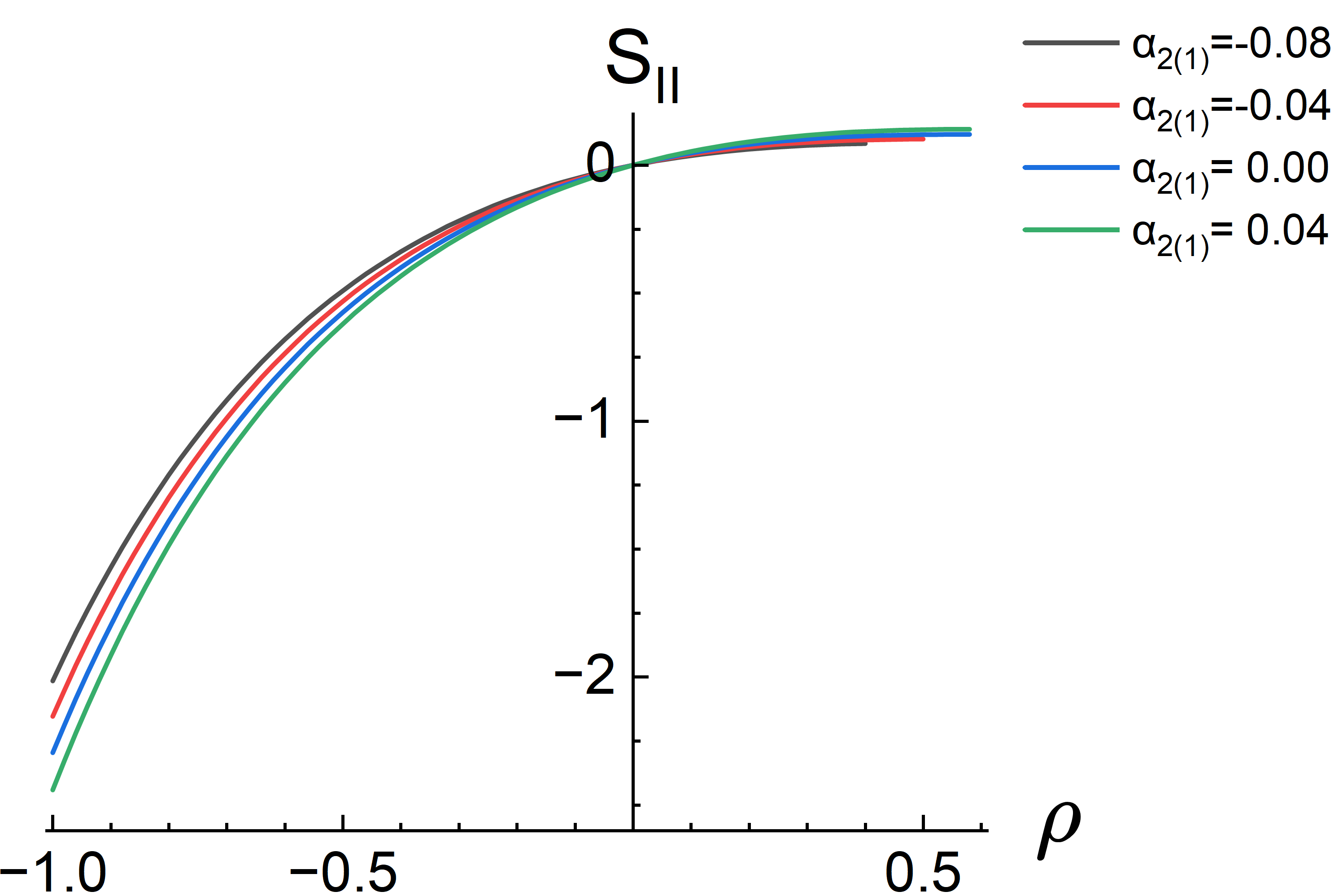}%
			\label{fig:d=4_GB_SII}}\par\smallskip
		\subfloat[]{%
			\includegraphics[width=0.42\linewidth]{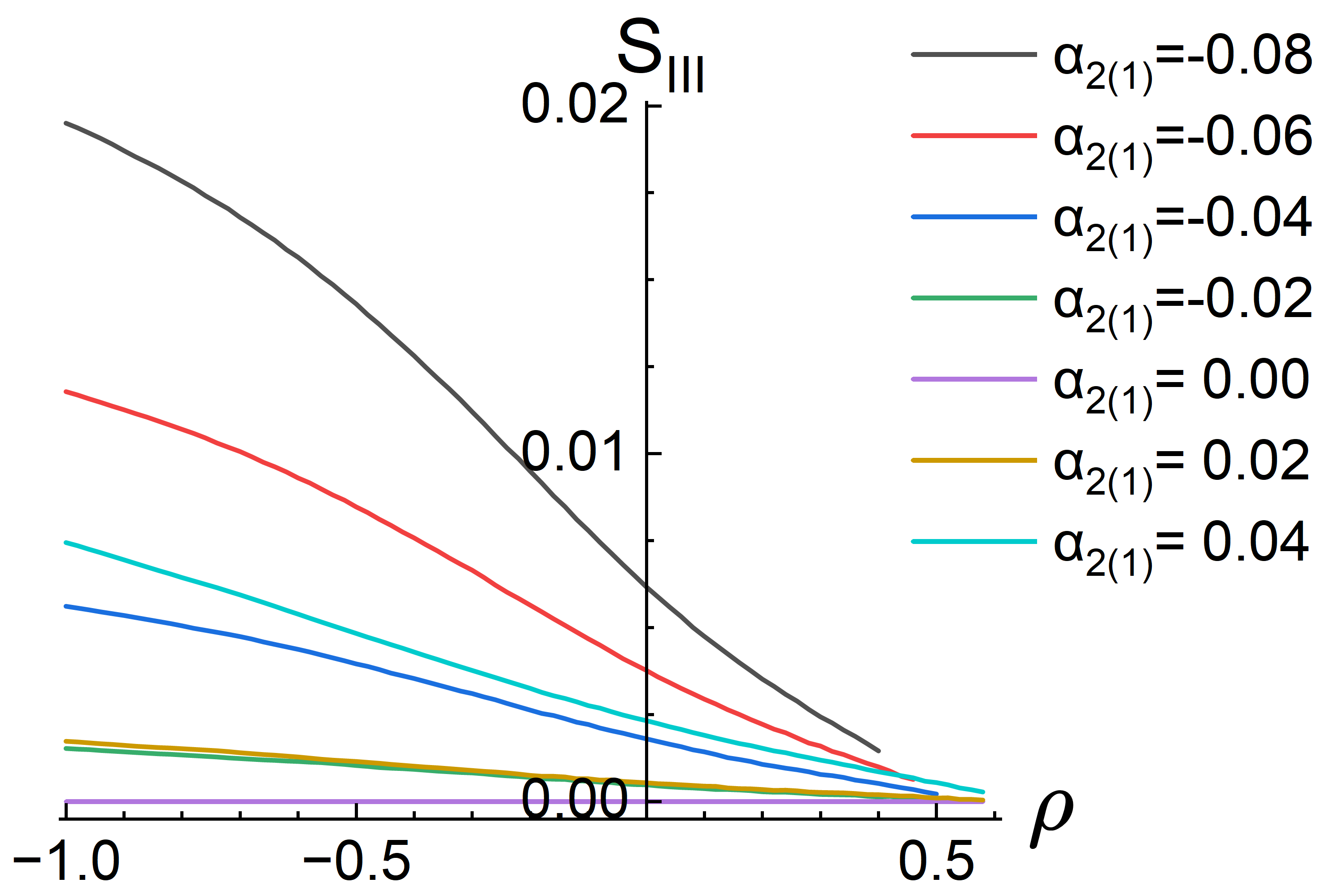}%
			\label{fig:d=4_GB_SIII}}\hfill
		\subfloat[]{%
			\includegraphics[width=0.42\linewidth]{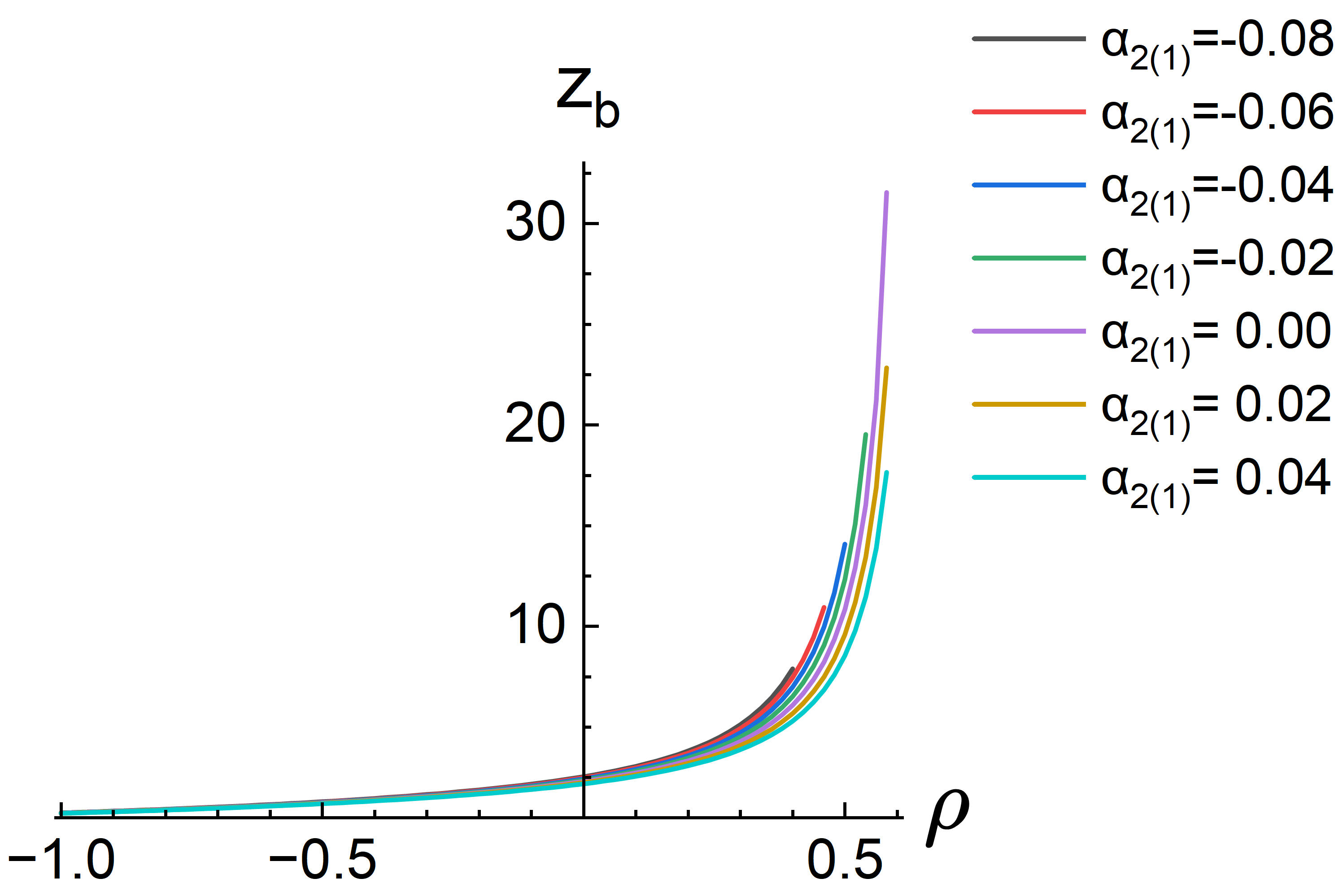}%
			\label{fig:d=4_GB_zb}}
		\caption{Network HEE and $z_b$ for Gauss-Bonnet gravtiy in $(d+1=5)$. We set $\alpha_{2(2)}=\alpha_{2(3)}=0$. Specifically for the calculation of $S_{\text{III}}$, we set $\varepsilon_z^S=1\times10^{-2}$ to reduce numerical errors}
	\label{fig:d=4_GB}
	\end{figure*}

	Furthermore, we observe that $S_{\text{I}}(\rho)$ and $S_{\text{II}}(\rho)$ are highly sensitive to $\rho$, but relatively insensitive to variations in the gravitational coupling coefficients. In contrast, the couplings exert a much more pronounced effect on $S_{\text{III}}$. Regardless of the spacetime dimension or the order of the Lovelock gravity, the behavior of the $S_{\text{III}}$ curves shares universal features: in the negative $\rho$ regime, a larger absolute value of the gravitational coupling leads to a greater $S_{\text{III}}$. In other words, as the couplings increase from negative values, the $S_{\text{III}}$ curves initially descend and subsequently ascend.

	Fig.~\ref{fig:d=4_GB} presents the results for $(d+1=5)$ Gauss-Bonnet (GB) gravity. We can see that as the GB coupling increases, $S_{\text{I}}$ and $S_{\text{II}}$ decrease in the negative $\rho$ regime, whereas they increase in the positive $\rho$ regime. Furthermore, as the coupling grows, the right-hand side of the $z_b$ curves develops a distinct peak profile. Specifically, for a fixed value of $\rho$, $z_b$ generally decreases with the coupling, while the maximum peak value of $z_b$ first increases and subsequently decreases.

\begin{figure*}[p]
		\centering
		\subfloat[]{%
			\includegraphics[width=0.389\linewidth]{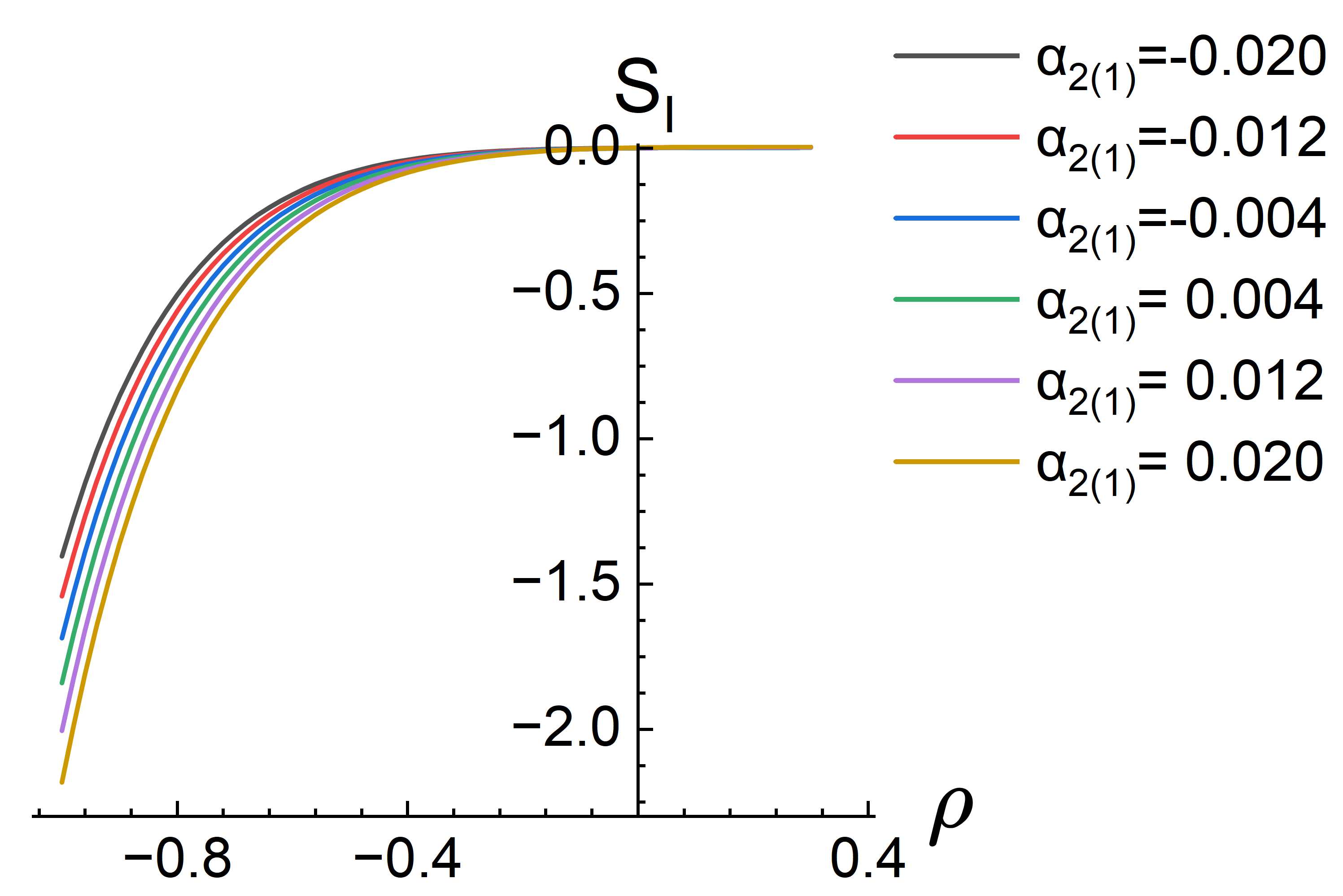}%
			\label{fig:d=6_GB_SI}}\hspace{0.07\linewidth}
		\subfloat[]{%
			\includegraphics[width=0.389\linewidth]{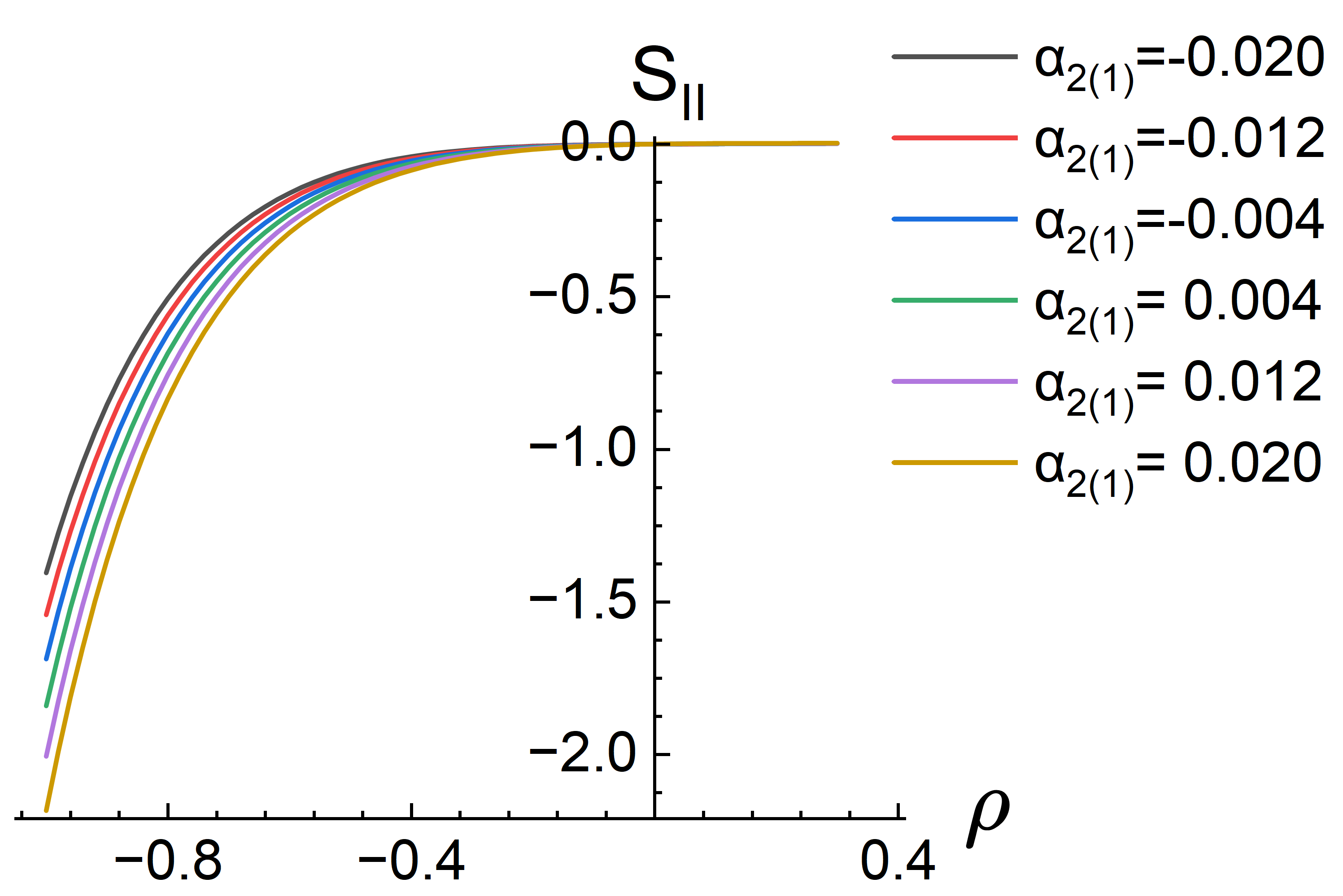}%
			\label{fig:d=6_GB_SII}}\par\smallskip
		\subfloat[]{%
			\includegraphics[width=0.389\linewidth]{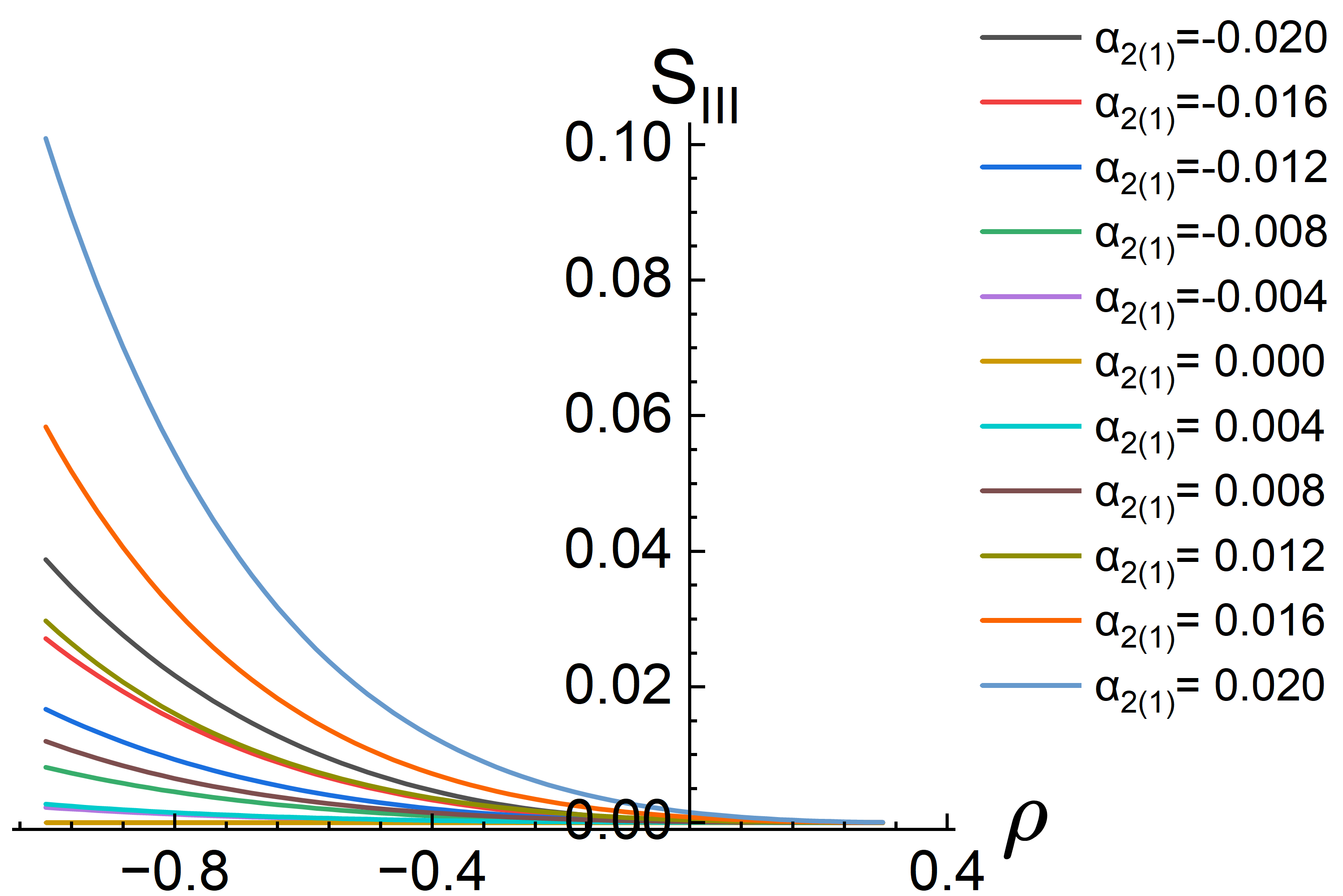}%
			\label{fig:d=6_GB_SIII}}\hspace{0.07\linewidth}
		\subfloat[]{%
			\includegraphics[width=0.389\linewidth]{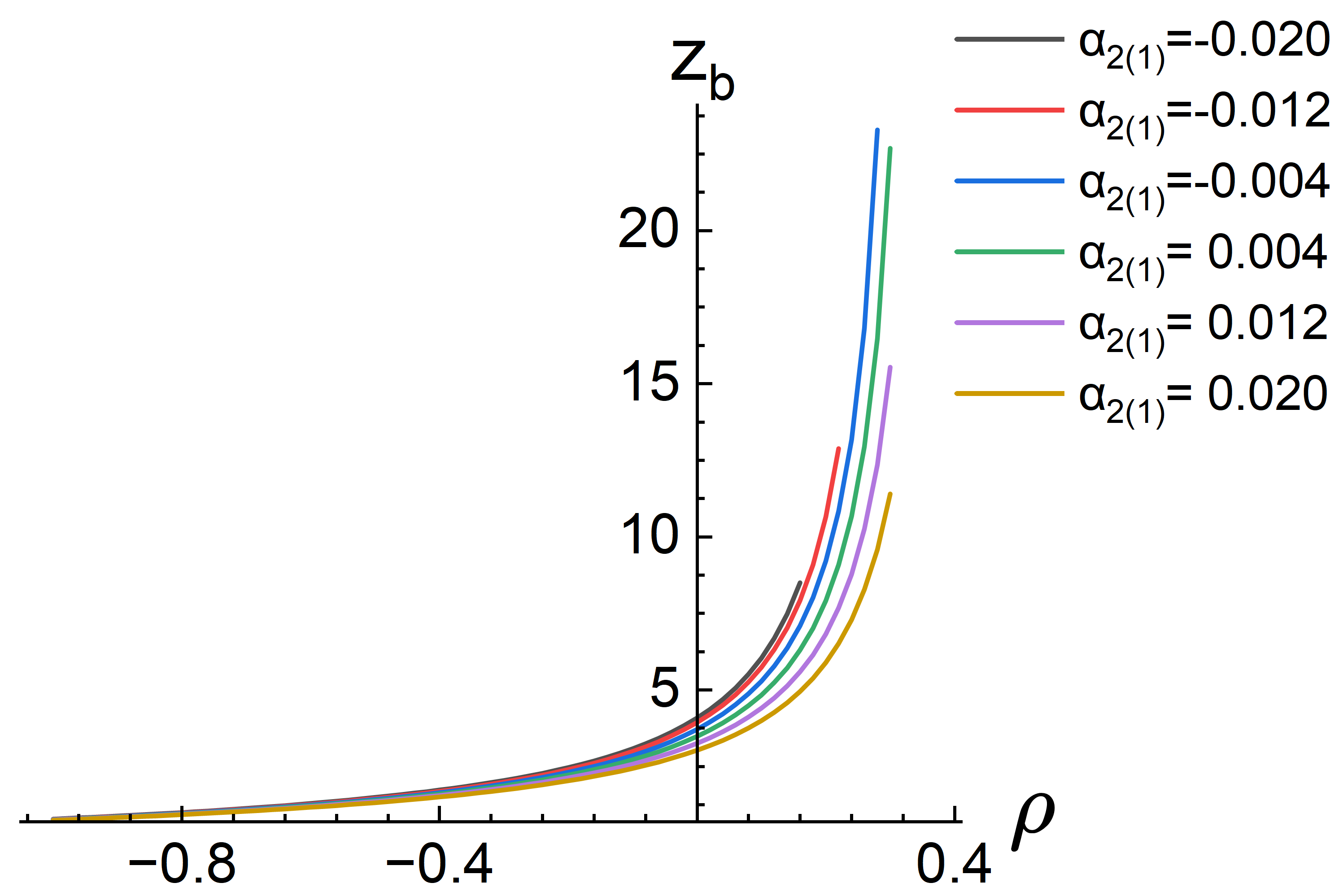}%
			\label{fig:d=6_GB_zb}}
		\caption{Network HEE and $z_b$ for Gauss-Bonnet gravtiy in $(d+1=7)$. We set $\alpha_{2(2)}=\alpha_{2(3)}=0$}
	\label{fig:d=6_GB}
	\par\medskip
		\centering
		\subfloat[]{%
			\includegraphics[width=0.389\linewidth]{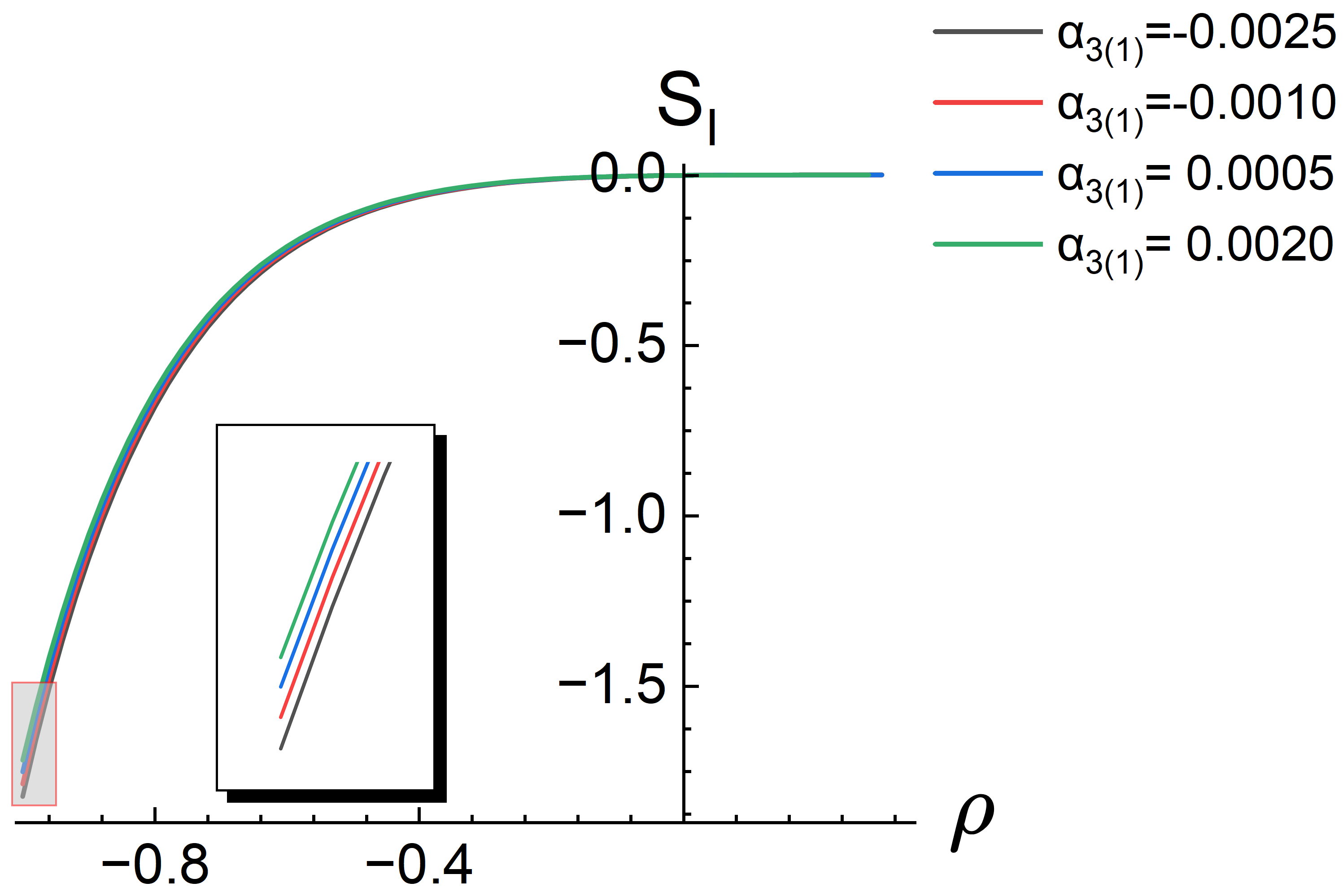}%
			\label{fig:d=6_cubicOnly_SI}}\hspace{0.07\linewidth}
		\subfloat[]{%
			\includegraphics[width=0.389\linewidth]{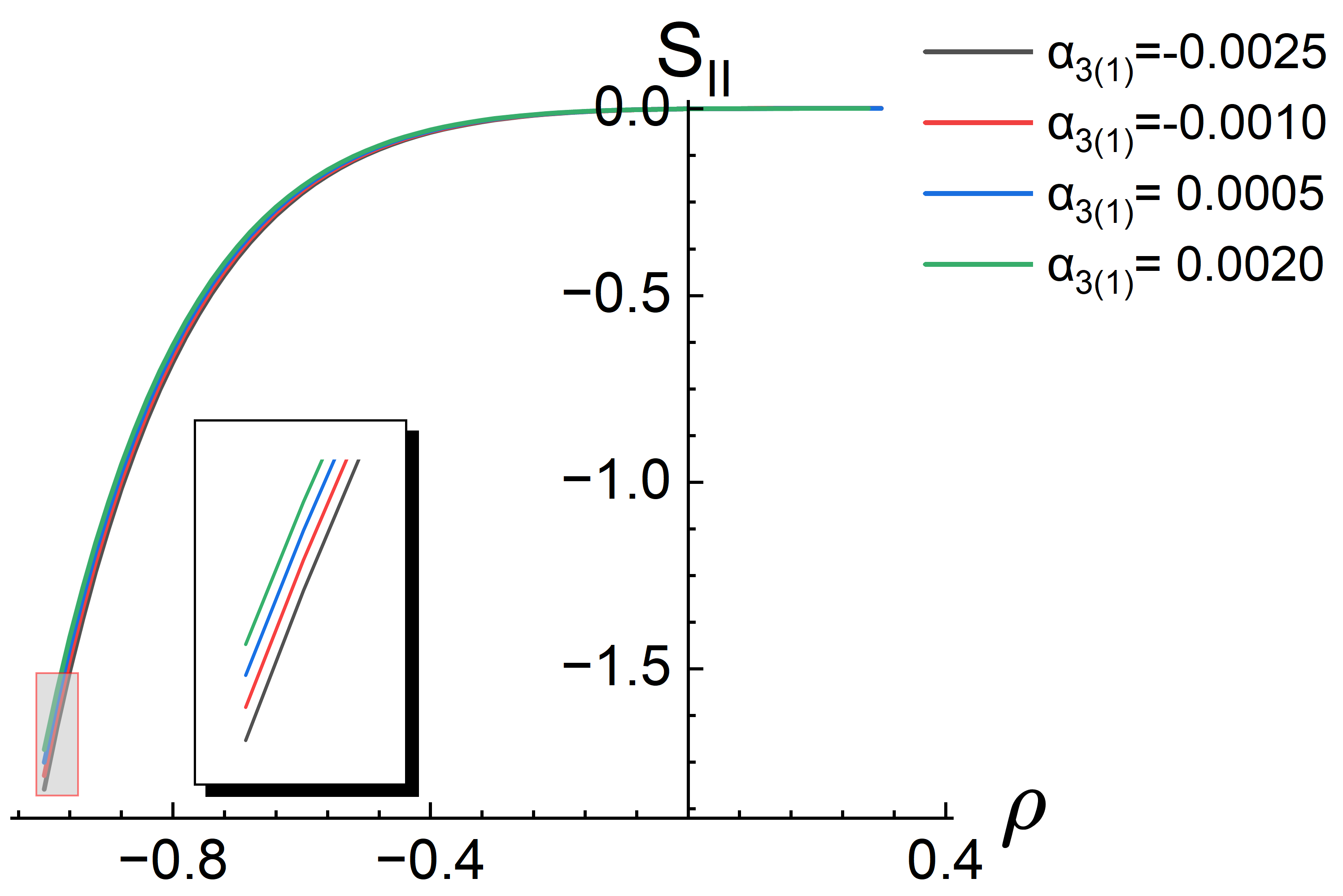}%
			\label{fig:d=6_cubicOnly_SII}}\par\smallskip
		\subfloat[]{%
			\includegraphics[width=0.389\linewidth]{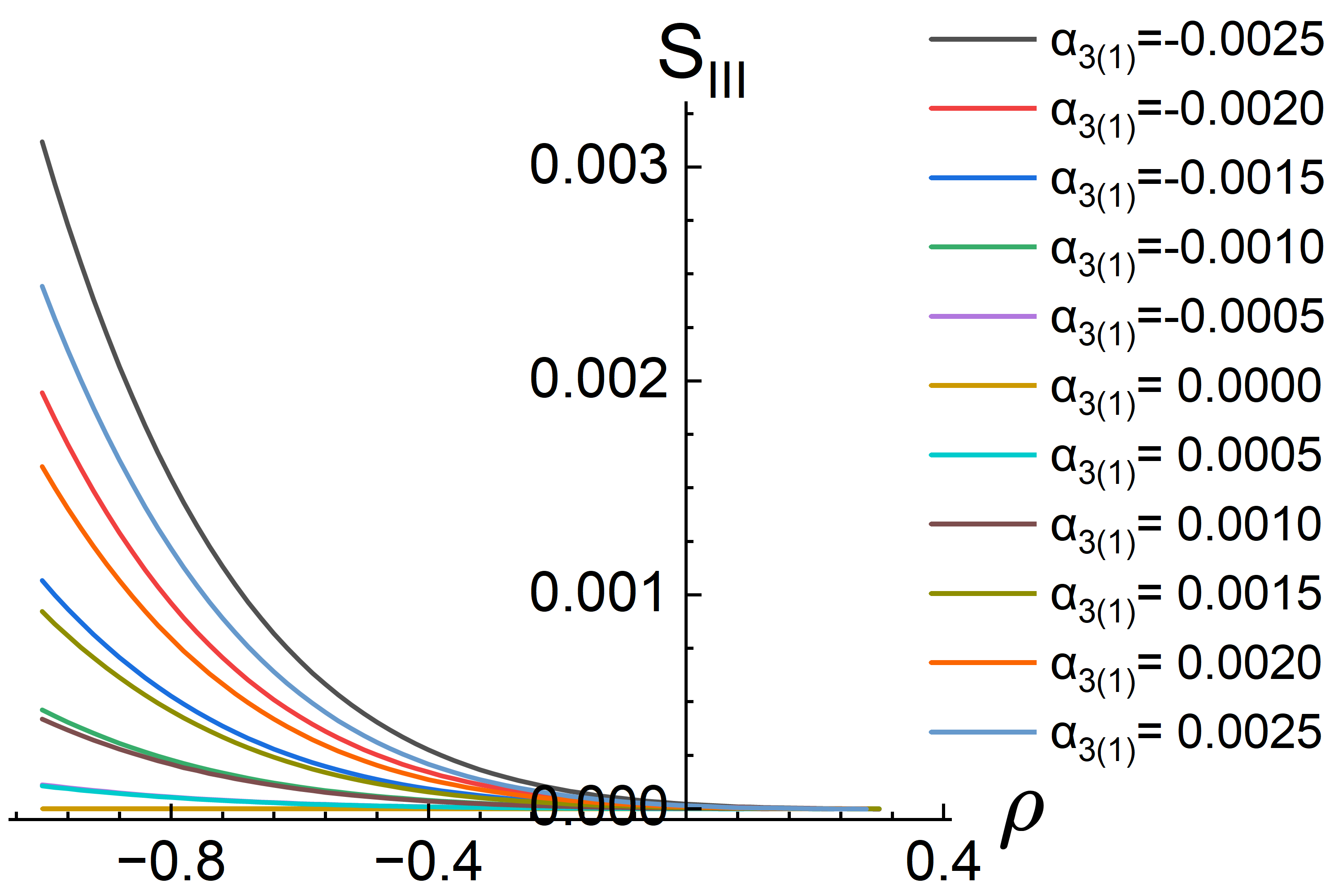}%
			\label{fig:d=6_cubicOnly_SIII}}\hspace{0.07\linewidth}
		\subfloat[]{%
			\includegraphics[width=0.389\linewidth]{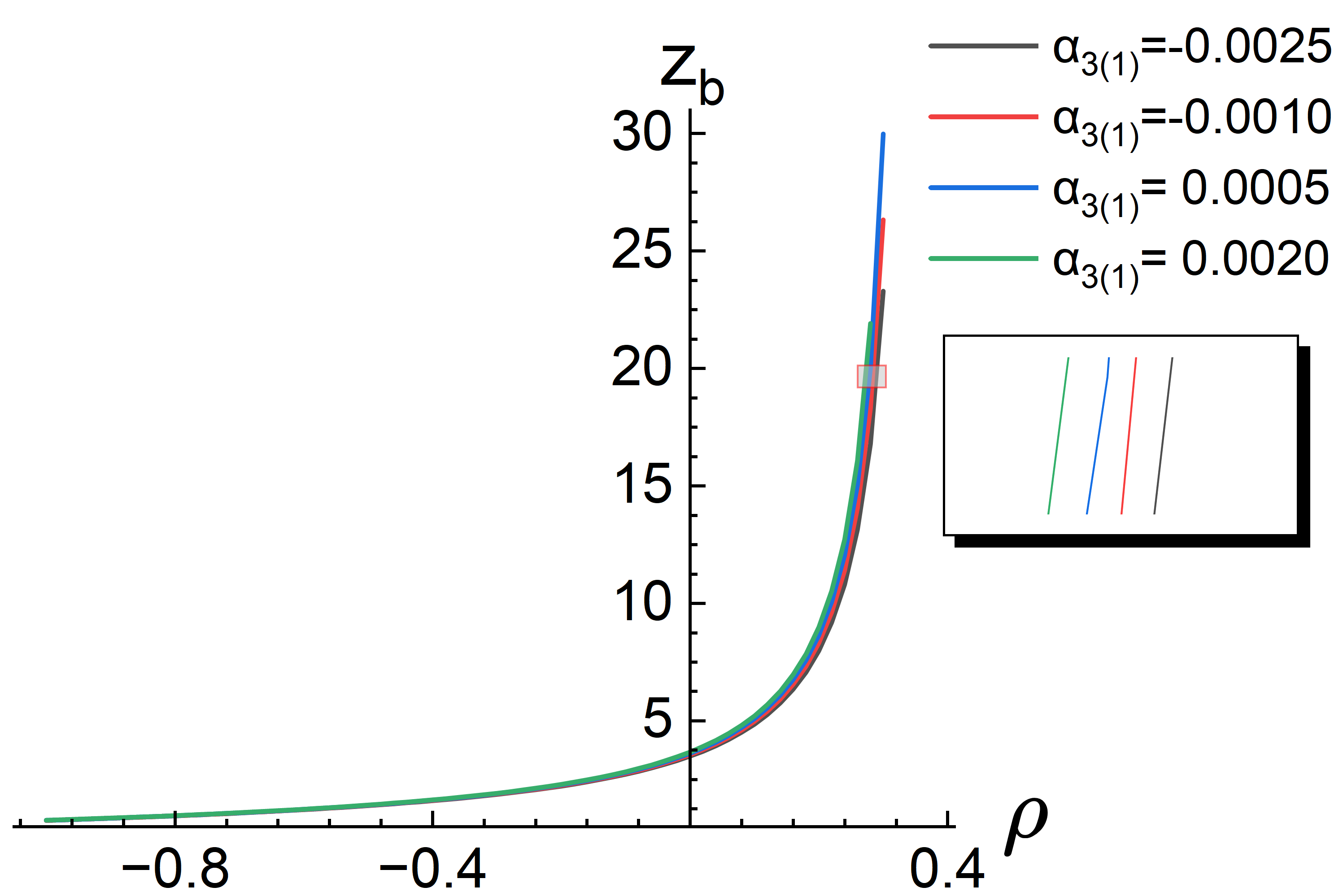}%
			\label{fig:d=6_cubicOnly_zb}}
		\caption{Network HEE and $z_b$ for purely cubic Lovelock gravtiy in $(d+1=7)$ (without the Gauss-Bonnet term). We set $\alpha_{3(2)}=\alpha_{3(3)}=0$}
	\label{fig:d=6_cubicOnly}
	\end{figure*}

	As shown in Fig.~\ref{fig:d=6_GB} for $(d+1=7)$ GB gravity, the network exhibits almost all the characteristic properties observed in the $(d+1=5)$ case. The only notable distinction is that the values of $S_{\text{I}}$ and $S_{\text{II}}$ in the positive $\rho$ regime are virtually unaffected by the GB couplings. Moving to Fig.~\ref{fig:d=6_cubicOnly}, which illustrates $(d+1=7)$ purely cubic Lovelock gravity (without the GB term), the overall influence of the couplings becomes weaker and the trend reverses. Specifically, as the cubic coupling decreases, $S_{\text{I}}$ and $S_{\text{II}}$ decrease in the negative $\rho$ regime, and the right-hand side of the $z_b$ curves again forms a peak profile. During this decrease, for a fixed value of $\rho$, $z_b$ generally decreases, while its maximum peak value first increases and subsequently decreases.

	Finally, Fig.~\ref{fig:d=6_ellipse} displays the results for $(d+1=7)$ Lovelock gravity up to the third order, featuring non-trivial values for both the GB and cubic couplings. We sample the couplings along an ellipse within the standard AdS/CFT bounds~\cite{Camanho:2009hu,Camanho:2010ru}. In this combined scenario, if we treat the two couplings collectively as a single parameter, the previously observed effects on network HEE and $z_b$ are successfully reproduced. It is worth noting that the cubic coupling exerts an opposite effect to that of the GB coupling.

	Although we only present the results where a single branch deviates from Einstein gravity, we have also computed the cases where all branches involve higher-order gravity. These additional scenarios yield similar outcomes: varying the gravitational coupling coefficients in one branch while keeping the other two fixed produces effects that are qualitatively the same as those described above.

	\twocolumn[{%
	\begin{minipage}{\textwidth}
	\widefiguretype
		\centering
		\subfloat[]{%
			\includegraphics[width=0.42\linewidth]{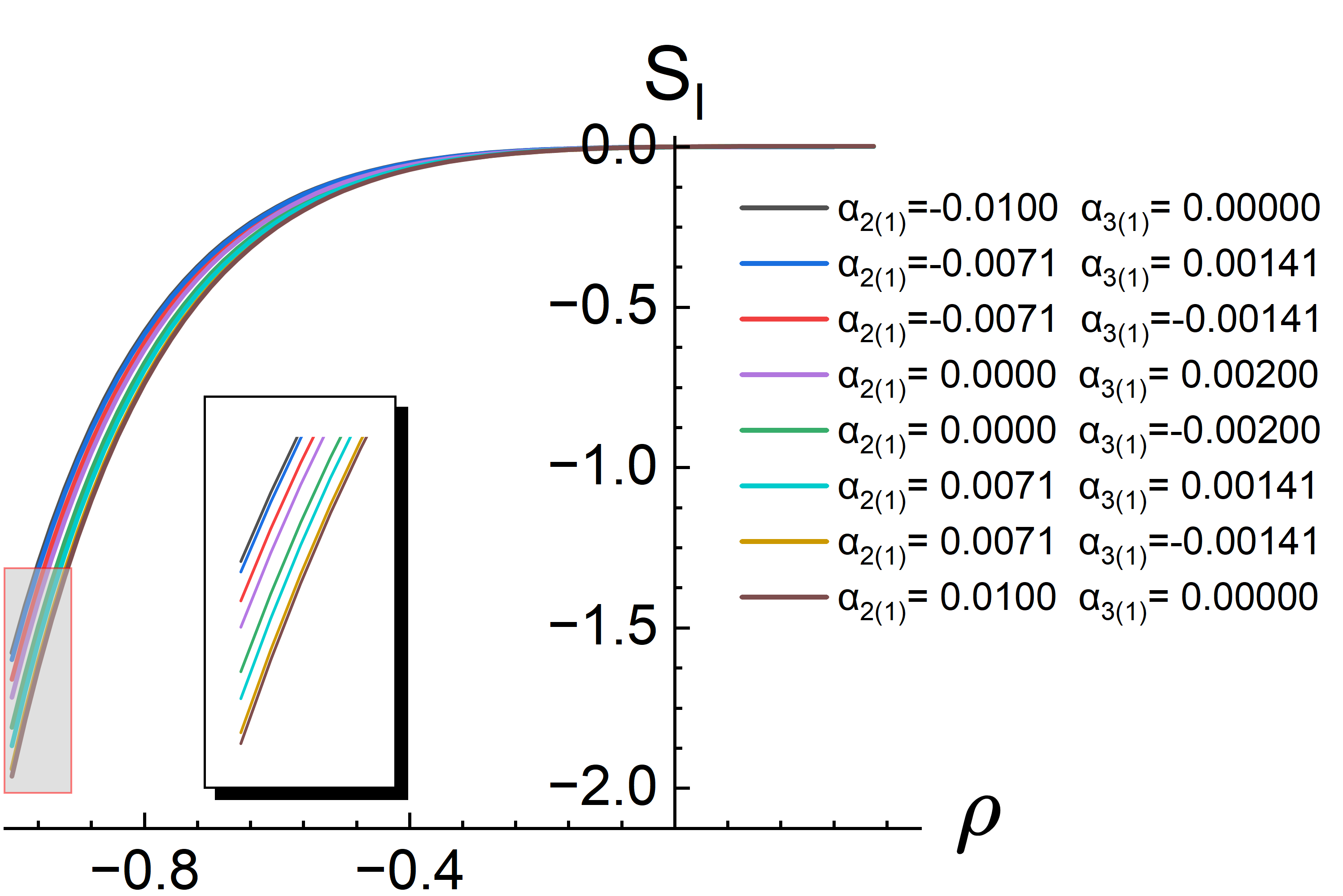}%
			\label{fig:d=6_ellipse_SI}}\hfill
		\subfloat[]{%
			\includegraphics[width=0.42\linewidth]{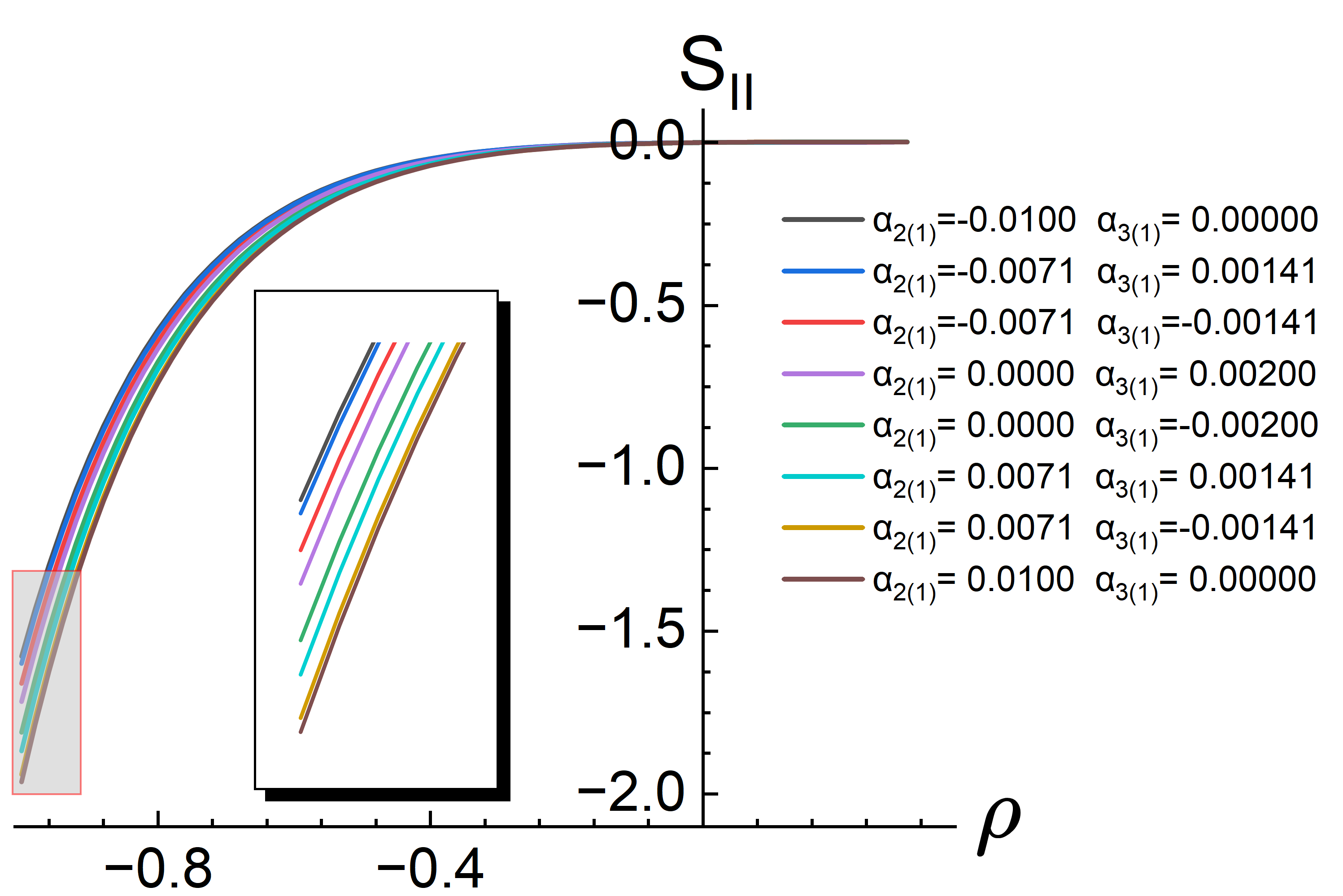}%
			\label{fig:d=6_ellipse_SII}}\par\smallskip
		\subfloat[]{%
			\includegraphics[width=0.42\linewidth]{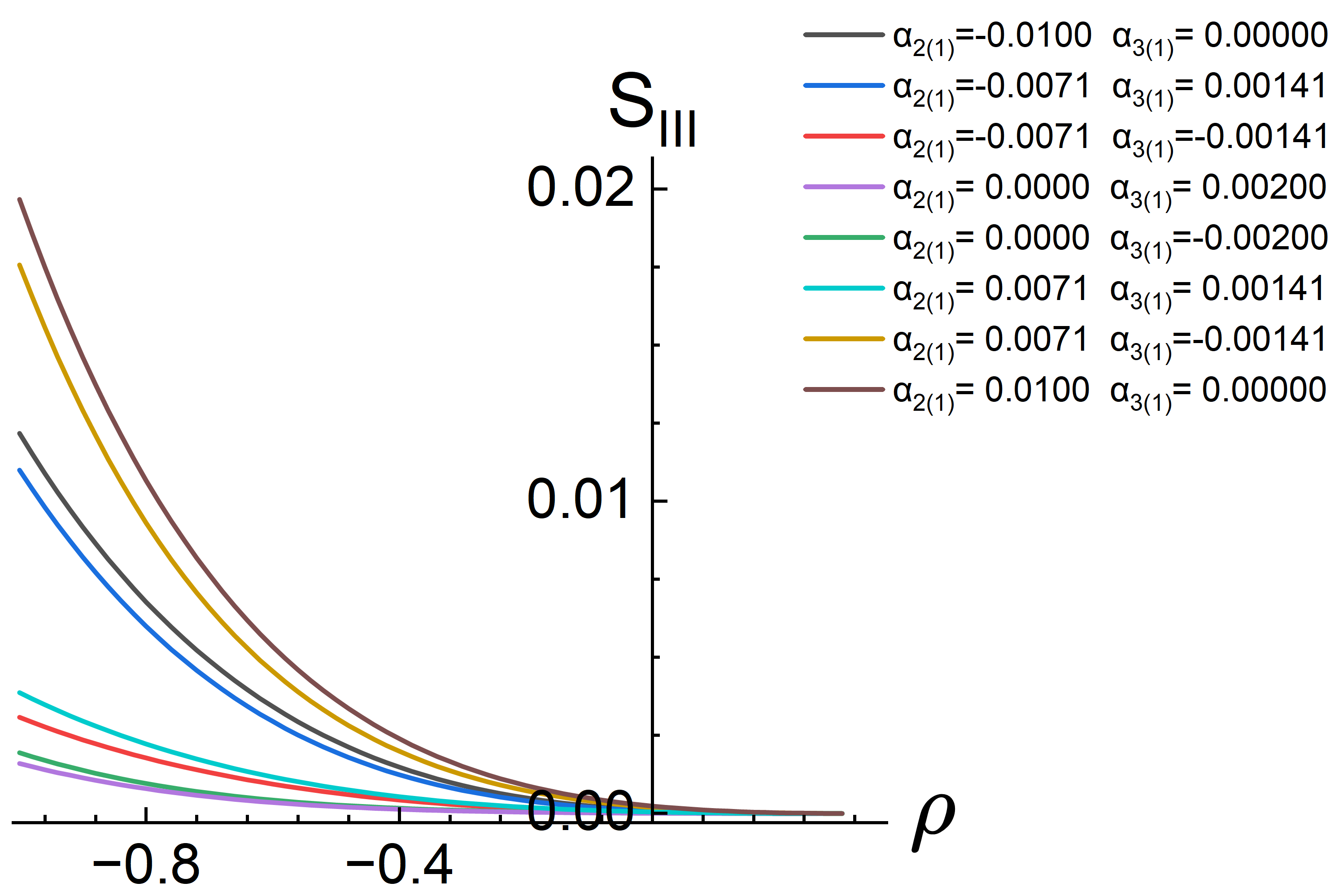}%
			\label{fig:d=6_ellipse_SIII}}\hfill
		\subfloat[]{%
			\includegraphics[width=0.42\linewidth]{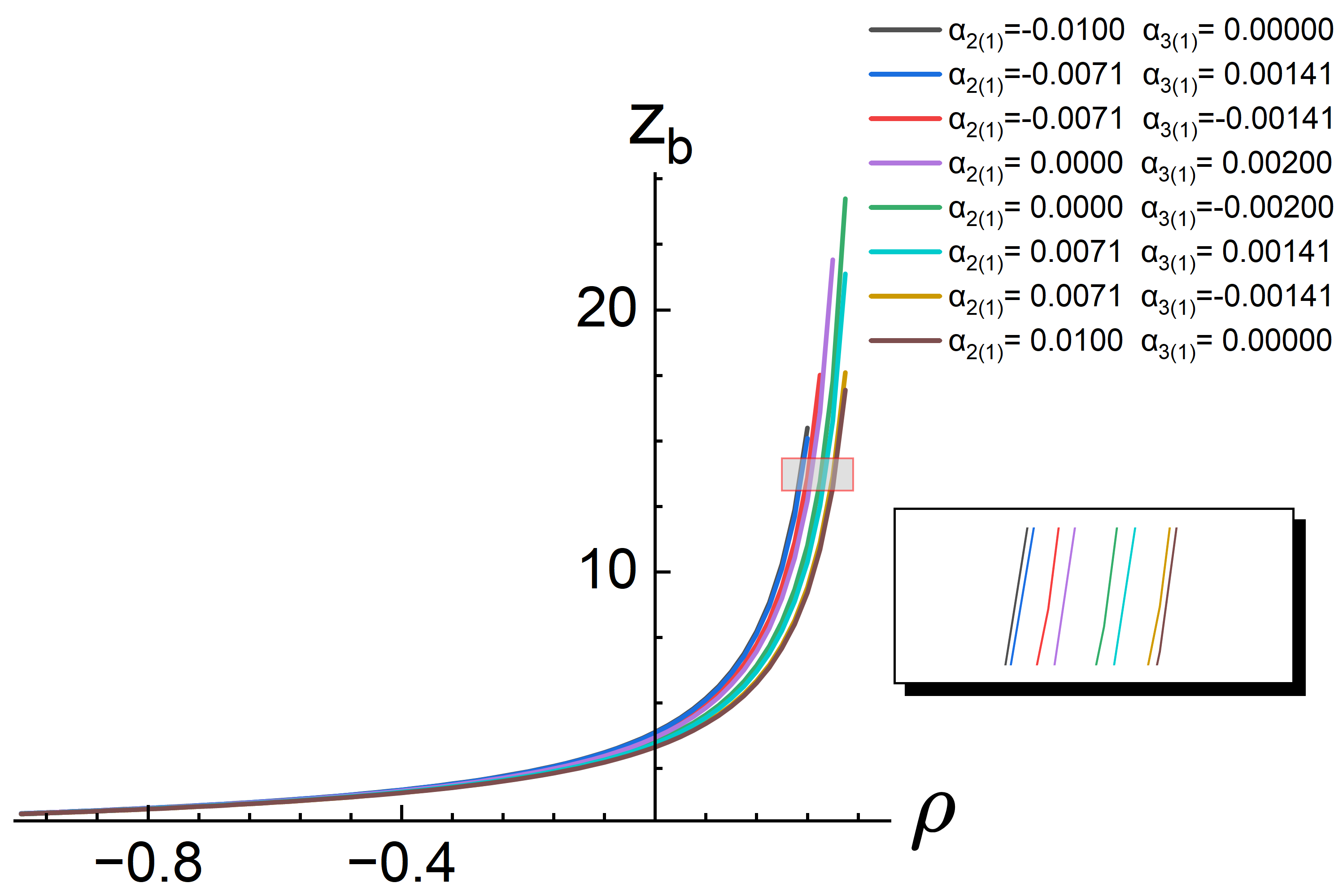}%
			\label{fig:d=6_ellipse_zb}}
		\caption{Network HEE and $z_b$ for Lovelock gravtiy up to the third order in $(d+1=7)$. We set $\alpha_{2(2)}=\alpha_{2(3)}=\alpha_{3(2)}=\alpha_{3(3)}=0$}
	\label{fig:d=6_ellipse}
	\end{minipage}\par\medskip
	}]
	\section{HEE in Generalized Network }\label{HEE in Generalized Network}

	In this section, we study the effect of network shape and size on the HEE. In the context of three-dimensional Einstein gravity ($\text{AdS}_3$), the RT surface on a constant-time slice reduces to a spacelike geodesic, and the HEE is strictly proportional to the minimum geodesic length. Since this length has a known analytical expression, we can readily investigate the network HEE by treating the nodal coordinates $z_b$ as explicit variational parameters. In the Poincar\'e patch of $\text{AdS}_3$, the geodesic distance $D$ between two points $(x_1,z_1)$ and $(x_2,z_2)$ on a constant-time slice is given by
\begin{equation}
	D=l\;\text{arccosh}\bigg(1+\frac{(x_1-x_2)^2+(z_1-z_2)^2}{2z_1z_2}\bigg).
\end{equation}
In our setup, the geodesic distance primarily involves two specific geometric configurations. First, for an external branch connecting $(L_E,\varepsilon_z)$ to a node at $(-\sinh(\rho)z_b,z_b)$, the distance is
\begin{equation}
	D=l\;\log\bigg(\frac{(L_E+\text{sinh}(\rho)z_b)^2+z_b^2}{z_b\,\varepsilon_z}\bigg).
\label{eq:external distance}
\end{equation}
Second, for an internal branch connecting two nodes at $(-\sinh(\rho)z_{b1},z_{b1})$ and $(L_I+\sinh(\rho)z_{b2},z_{b2})$, the distance becomes
\begin{equation}
	\begin{aligned}
	D&=l\;\text{arccosh}\biggl(1+\\
	 &\quad\frac{(L_I+\sinh(\rho)(z_{b1}+z_{b2}))^2+(z_{b2}-z_{b1})^2}
	 {2z_{b1}z_{b2}}\biggr).
	\end{aligned}
\label{eq:internal distance}
\end{equation}
Here, $L_I$ and $L_E$ denote the subregion lengths of the internal and external branches, respectively, while $p_I$ and $p_E$ represent the total number of internal and external branches. The Poincar\'e geometry has zero boundary stress~\cite{Guo:2025sbm}, whereas network vacua can carry edge-length-dependent Casimir stress~\cite{ZhaoMiao2025}. Furthermore, to satisfy the wedge inclusion condition, the parameter $\rho$ must be non-negative \cite{Guo:2026prm}.

	\begin{figure*}[!tp]
	\begin{minipage}[c]{0.44\linewidth}
		\caption{Schematic geometry for line-shaped networks. The number preceding ``L'' counts the links between adjacent nodes along the line-shaped networks, with each link comprising all internal branches connecting that pair of nodes. The numbers preceding ``W'' give the relative branch multiplicities of the successive links. $p_I$ and $p_E$ represent the total number of internal and external branches, respectively. We set $p_I=p_E=12$}
	\label{fig:line-shaped network}
	\end{minipage}\hfill
	\begin{minipage}[c]{0.52\linewidth}
		\centering
		\includegraphics[width=0.76\linewidth]{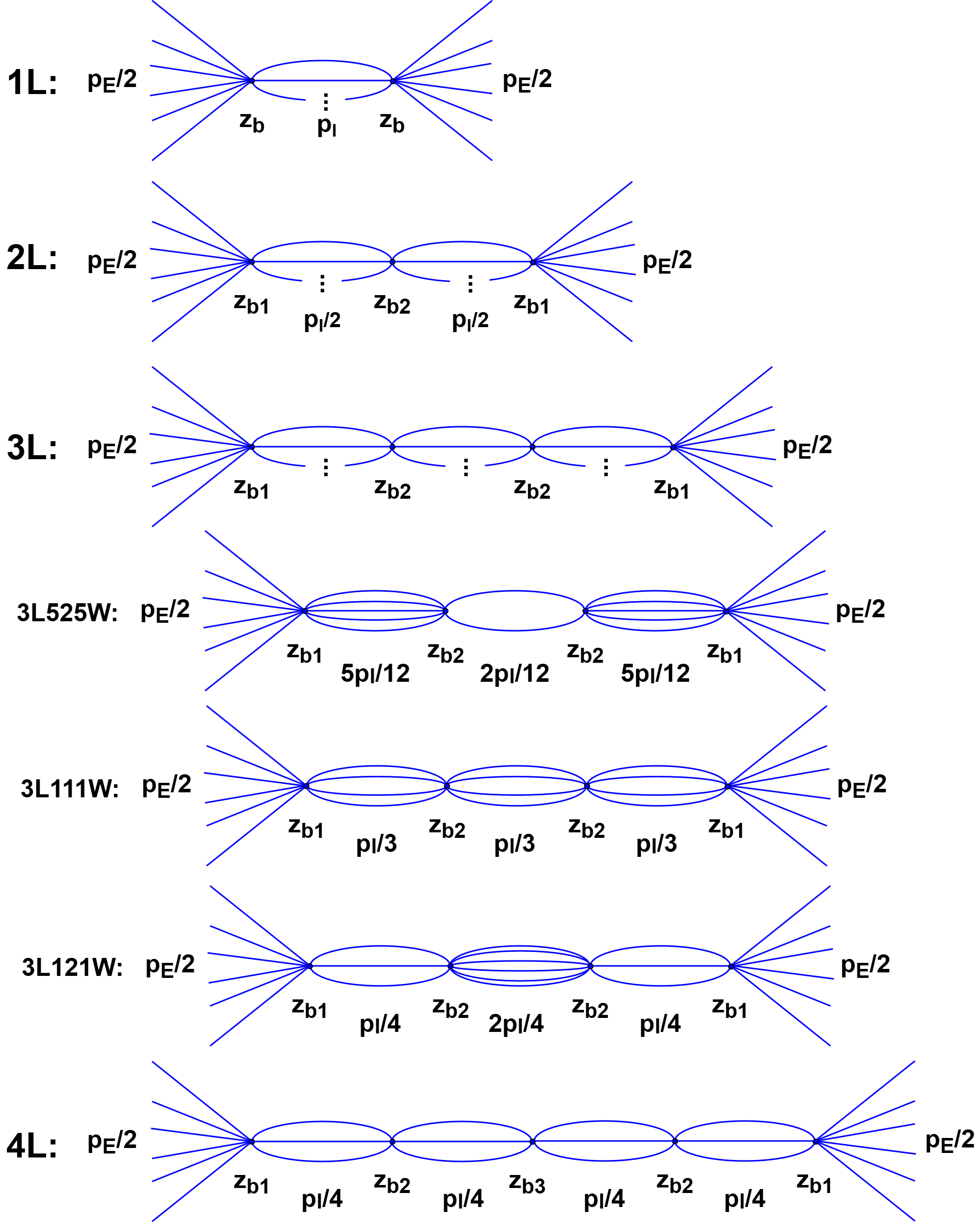}
	\end{minipage}\par\medskip
	\begin{minipage}{\linewidth}
		\centering
		\subfloat[]{%
			\includegraphics[width=0.43\linewidth]{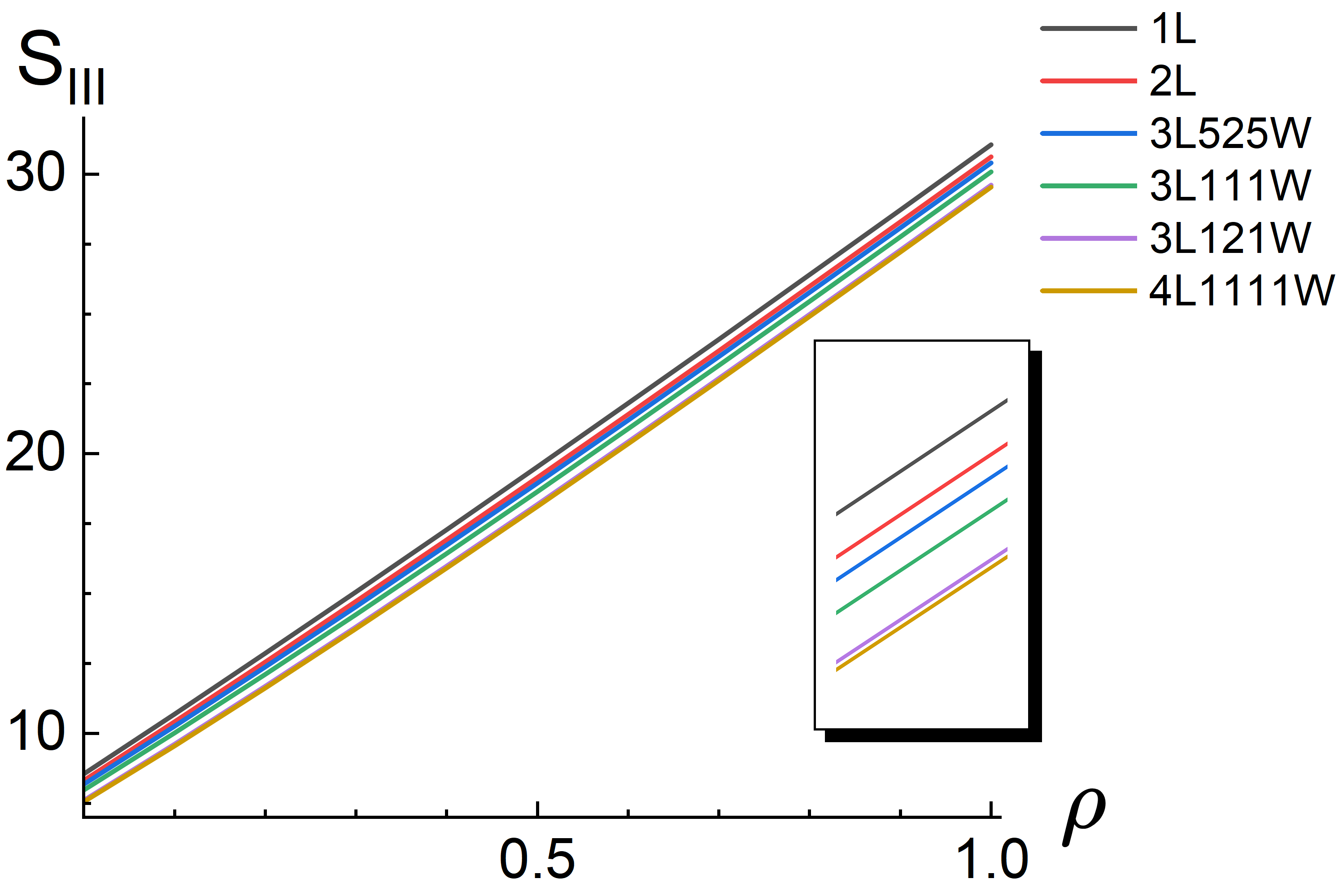}%
			\label{fig:line-shaped network entropy}}\hspace{0.06\linewidth}
		\subfloat[]{%
			\includegraphics[width=0.43\linewidth]{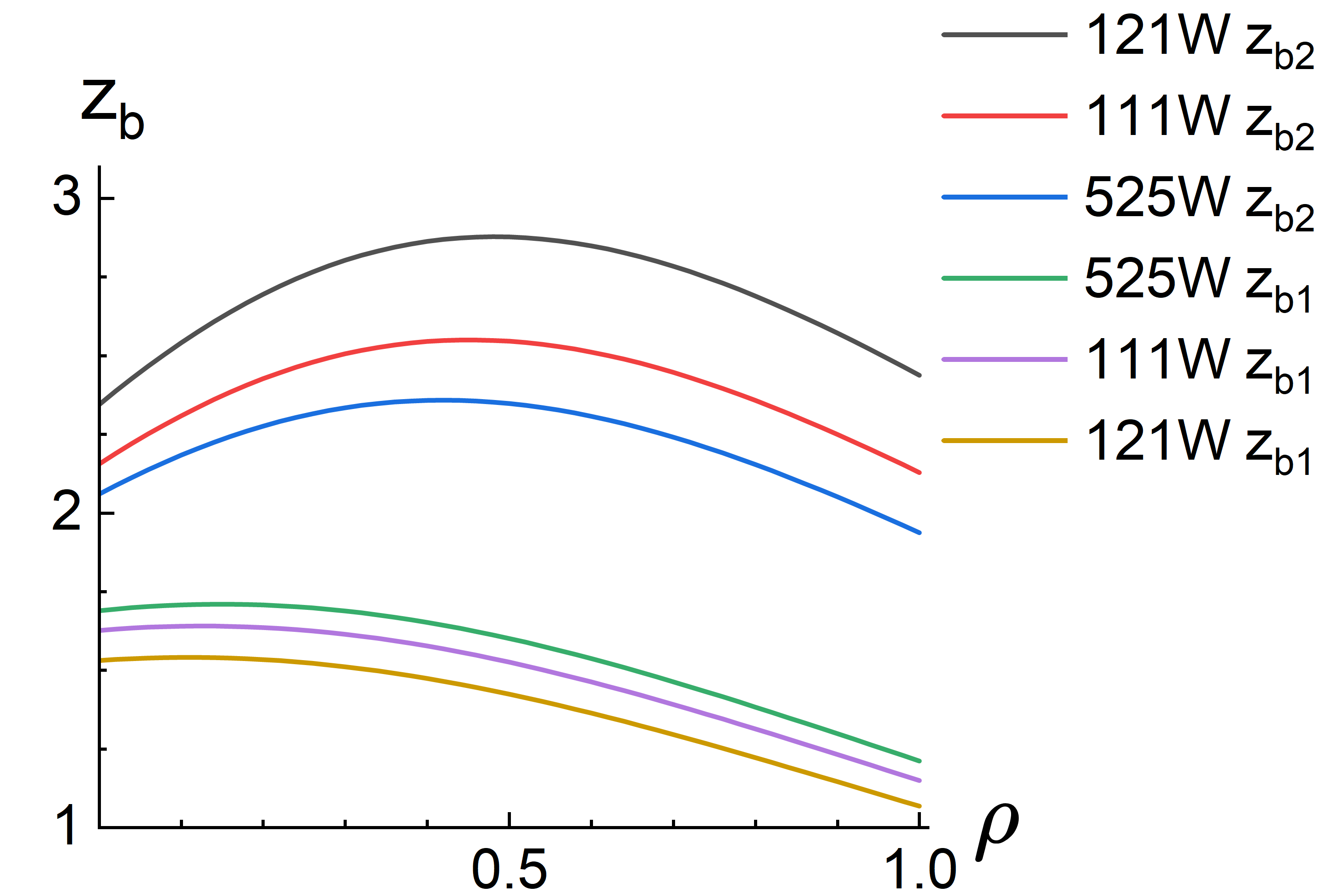}%
			\label{fig:line-shaped network zb}}
		\caption{(a) Network HEE for various line-shaped networks. (b) Nodal coordinate $z_b$ for 3L line-shaped networks. We set $L_I=L_E=1$ and $p_I=p_E=12$. $L_I$ and $L_E$ denote the subregion lengths of the internal and external branches, respectively}
	\label{fig:line-shaped network entropy and zb}
	\end{minipage}
	\end{figure*}

In the BCFT configuration, since the nodal coordinate $z_b$ is an independent variable for each branch, the minimal geodesic distance for internal branches vanishes, owing to the fact that the full BCFT has zero entropy. Consequently, we find that the BCFT entanglement entropy, $S_{\text{BCFT}}(\rho)$, is completely independent of internal parameters and can be expressed as
\begin{equation}
	S_\text{BCFT}(\rho)=\frac{l}{4G}\,p_E \bigg(\log\frac{2L_E}{\varepsilon_z}+\rho\bigg).
\end{equation}

The expression for $S_\text{NCFT}(\rho)$ follows directly from the distance formulas Eqs. (\ref{eq:external distance}) and (\ref{eq:internal distance}) for the external and internal branches. As an example, for a two-node network, it simply takes the form
\begin{equation}
	\begin{split}
	&S_\text{NCFT}(\rho) = \frac{l}{4G}\biggl[\\
	 &\quad p_E\log\bigg(\frac{(L_E+\text{sinh}(\rho)z_b)^2+z_b^2}{z_b\,\varepsilon_z}\bigg)\\
	 &\quad+p_I\;\text{arccosh}\bigg(1+\frac{(L_I+2\text{sinh}(\rho)z_b)^2}{2z_b^2}\bigg)\biggr].
	\end{split}
\end{equation}
The physical value of $S_\text{NCFT}(\rho)$ is determined by minimizing this expression with respect to the nodal variables $z_b$. This minimization process guarantees that the junction conditions for the RT surfaces are automatically satisfied at the extremal points. Using $S_\text{BCFT}(\rho)$ and $S_\text{NCFT}(\rho)$, we can then explicitly compute the three types of network HEE defined previously.

	We observe that the type-III network HEE ($S_{\text{III}}$) captures the effects of network shape and size more effectively than types I and II ($S_{\text{I}}$ and $S_{\text{II}}$). This is because $S_{\text{I}}$ and $S_{\text{II}}$ vary rapidly with $\rho$, which tends to overshadow the subtle effects arising from the network's geometry. Moreover, we find that highly symmetric network topologies—such as a simple two-node internal line, a triangle, or a square—exhibit identical minimal behaviors. Specifically, their total geodesic distances are minimized when all nodal variables are equal ($z_{b1}=z_{b2}=\cdots$). Consequently, non-trivial variations in the network HEE primarily arise from structural asymmetries. To explore this, we focus on how line-shaped networks affect the network HEE. 

	The geometry for line-shaped networks is illustrated in Fig. \ref{fig:line-shaped network}. In our notation, the number preceding ``L'' counts the links between adjacent nodes along the line-shaped networks, with each link comprising all internal branches connecting that pair of nodes. The numbers preceding ``W'' give the relative branch multiplicities of the successive links. As shown in Fig. \ref{fig:line-shaped network entropy and zb}(a), the network HEE decreases as the number or depth of the internal branches increases. Furthermore, as depicted in Fig. \ref{fig:line-shaped network entropy and zb}(b), in networks with deeper internal structures, the nodal coordinate $z_b$ increases at deeper positions and slightly decreases at shallower positions—a spatial readjustment that ultimately minimizes the total network HEE.

	\section{Conclusion}\label{Conclusion}

   	In this work, we have investigated the HEE of network NCFTs. We derive a more general expression for the Net-brane (NB) tension, as well as the extremal and junction conditions for RT surfaces up to cubic Lovelock gravity. Notably, these conditions can be readily generalized to higher-order Lovelock gravity theories.

	In our numerical computations, We employ the shooting method. Our numerical computations accurately recover previous analytical results for Einstein gravity. Our numerical results for the BCFT are in excellent agreement with the analytical expressions. The only minor distinction is that the RT surface solution orthogonal to the $x$-axis cannot be directly generated by our numerical algorithm; however, it can be unambiguously deduced from the asymptotic behavior of the numerical solutions. Furthermore, we present numerical results focusing on a one-node, three-branch network configuration, exploring the effects of gravitational coupling coefficients on both the network HEE and the profile of the RT surface. We find that the couplings exert the most pronounced effect on the type-III network HEE, $S_{\text{III}}$, exhibiting a non-monotonic behavior. Conversely, the effects of the couplings on $S_{\text{I}}$, $S_{\text{II}}$, and the nodal coordinate $z_b$ are monotonic and relatively weak. These latter effects become even weaker as the spacetime dimension or the order of the gravity theory increases.

	Furthermore, we analyze the influence of the network configuration on the HEE and $z_b$ within $(d+1=3)$ Einstein gravity. As previously discussed, $S_{\text{III}}$ is the HEE candidate that is least dependent on the NB parameter $\rho$ and can capture the intrinsic properties of the network configuration. We observe that the total HEE of the isolated BCFTs is independent of the parameters of the internal branches. Additionally, the overall network HEE remains unaffected by symmetric network configurations; rather, it is the asymmetric structures that primarily dictate the network HEE. Consequently, we investigate line-shaped structures where the internal links are separated by multiple nodes. By adjusting the number of internal links and the multiplicity of internal branches, we find that the network HEE decreases as these parameters increase. This reduction in entropy is driven by the fact that $z_b$ is pushed deeper into the bulk at the inner nodes. 

	While we have restricted our configuration analysis to $(d+1=3)$ Einstein gravity due to the availability of simple analytical expressions, extending this analysis to higher spacetime dimensions or higher-order gravity theories constitutes a promising avenue for future research. Exploring the effects of more complex network configurations and topologies is also a highly future direction.
	
	Finally, we note that in our current numerical computations for higher-order Lovelock gravity, the valid ranges of the gravitational coupling coefficients are adopted from \cite{Camanho:2009hu, Camanho:2010ru}, which were originally derived for standard AdS/CFT setups. More precise bounds on these couplings, specifically tailored to NCFT, could be determined through a detailed Kaluza-Klein mode analysis for specific network configurations, spacetime dimensions, and gravity orders. 

	Moreover, analogous to the well-known g-theorem in AdS/BCFT, the network HEE in AdS/NCFT might lead to a corresponding network g-theorem. The primary open challenges in establishing this include: identifying the precise analogue of the boundary central charge from the numerical network HEE, determining the correct form of the g-function, and pinpointing the specific parameter that represents the renormalization group (RG) flow. Resolving these questions remains a compelling task for future investigations.

	\backmatter
	%

	\begin{appendices}
	\counterwithin*{equation}{section}
	\renewcommand*{\theHequation}{appendix.\Alph{section}.\arabic{equation}}
	\compactappendixheadings
	\section{Explicit Tensor Expressions in Lovelock Theories}\label{Explicit Tensor Expressions in Lovelock Theories}

	The explicit expressions for $\mathcal{L}^{(p)}$ and $\mathcal{Q}^{(p)}$ in the action of Lovelock gravity, Eq. (\ref{eq:Lovelock action}), read
	\begin{equation}
    	\mathcal{L}^{(p)}=\frac{1}{2^p}g^{\mu_1\mu_2\cdots\mu_{2p}}_{\nu_1\nu_2\cdots\nu_{2p}}R^{\nu_1\nu_2}_{\mu_1\mu_2}\cdots R^{\nu_{2p-1}\nu_{2p}}_{\mu_{2p-1}\mu_{2p}},
	\label{eq:Lp}
	\end{equation}
	\begin{equation}
	\begin{split}
	&\mathcal{Q}^{(p)} = \frac{4p}{2^p}
	 h^{i_1 i_2\cdots i_{2p-1}}_{j_1 j_2\cdots j_{2p-1}}
	 \sum_{k=0}^{p-1}\binom{p-1}{k}\frac{(-2)^k}{2k+1}\\
	 &\quad\times(\prescript{b}{}{K})^{j_1}_{i_1}
	 (\prescript{b}{}{K})^{j_2}_{i_2}\cdots
	 (\prescript{b}{}{K})^{j_{2k-1}}_{i_{2k-1}}
	 (\prescript{b}{}{K})^{j_{2k}}_{i_{2k}}\\
	 &\times(\prescript{b}{}{K})^{j_{2p-1}}_{i_{2p-1}}
	 (\prescript{b}{}{R})^{j_{2k+1} j_{2k+2}}_{i_{2k+1} i_{2k+2}}
	 \cdots(\prescript{b}{}{R})^{j_{2p-3} j_{2p-2}}_{i_{2p-3} i_{2p-2}},
	\end{split}
	\label{eq:Qp}
	\end{equation}
	where $g^{\mu_1\mu_2\cdots\mu_{2p}}_{\nu_1\nu_2\cdots\nu_{2p}}$ and $h^{i_1 i_2\cdots i_{2p-1}}_{j_1 j_2\cdots j_{2p-1}}$ are the generalized Kronecker deltas defined by the bulk metric $g_{\mu\nu}$ and the induced boundary metric $h_{ij}$, respectively. Note that we use $\mu, \nu, \rho, \dots$ to denote bulk tensor indices and $i, j, k, \dots$ to denote boundary indices. Additional notations are summarized in Table \ref{tab:notation}. Due to the complete antisymmetry among all upper or lower indices, $\mathcal{L}^{(p)}$ automatically vanishes if $2p > d+1$. This explains why the maximum order of the Lovelock gravity terms is $\frac{d}{2}$ for odd $(d+1)$ dimensions (for even $(d+1)$ dimensions, $\mathcal{L}^{(\frac{d+1}{2})}$ is a topological term). Here, $\binom{p-1}{k}$ represents the binomial coefficient. The term $(\prescript{b}{}{K})$ denotes the extrinsic curvature evaluated from the bulk to the boundary, and $(\prescript{b}{}{R})$ represents the intrinsic curvature on the boundary (assuming the boundary is a timelike hypersurface). In the model in this paper, since the boundary is a Net-brane (NB), we denote $(\prescript{b}{}{K})$ and $(\prescript{b}{}{R})$ as $(\prescript{NB}{}{K})$ and $(\prescript{NB}{}{R})$. For completeness, we provide the formal definition of the generalized Kronecker delta:
	\begin{equation}
	\begin{gathered}
	g^{\mu_1\mu_2\cdots\mu_{2p}}_{\nu_1\nu_2\cdots\nu_{2p}}=\\
		\begin{vmatrix}
		g^{\mu_1}_{\nu_1} & g^{\mu_2}_{\nu_1} & \cdots & g^{\mu_{2p}}_{\nu_1} \\
		g^{\mu_1}_{\nu_2} & g^{\mu_2}_{\nu_2} & \cdots & g^{\mu_{2p}}_{\nu_2} \\
		\vdots & \vdots & \ddots & \vdots \\
		g^{\mu_1}_{\nu_{2p}} & g^{\mu_2}_{\nu_{2p}} & \cdots & g^{\mu_{2p}}_{\nu_{2p}}
		\end{vmatrix}.
	\end{gathered}
	\label{eq:generalized Kronecker delta}
	\end{equation}
	
	As an example, the first three terms of $\mathcal{L}^{(p)}$ are given by: $\mathcal{L}^{(0)}=1$ (the cosmological constant term), $\mathcal{L}^{(1)}=R$ (the Einstein term), and $\mathcal{L}^{(2)}=R^2-4R_{\mu\nu}R^{\mu\nu}+R_{\mu\nu\rho\sigma}R^{\mu\nu\rho\sigma}$ (the Gauss-Bonnet term). Note that $\mathcal{Q}^{(0)}=0$.
	
	In the Lovelock HEE, Eq. (\ref{eq:Lovelock HEE}), the quantities $\tilde{\mathcal{L}}^{(p)}$ and $\tilde{\mathcal{Q}}^{(p)}$ satisfy $\tilde{\mathcal{L}}^{(p)}=p\mathcal{L}^{(p-1)}$ and $\tilde{\mathcal{Q}}^{(p)}=p\mathcal{Q}^{(p-1)}$. Their explicit expressions are given by:
	
	\begin{equation}
	\begin{aligned}
	\tilde{\mathcal{L}}^{(p)}&=\frac{p}{2^{p-1}}
	 (h_\Gamma)^{\alpha_1\alpha_2\cdots\alpha_{2(p-1)}}_{\beta_1\beta_2\cdots\beta_{2(p-1)}}\\
	 &\quad\times(\prescript{\Gamma}{}{R})^{\beta_1\beta_2}_{\alpha_1\alpha_2}
	 \cdots(\prescript{\Gamma}{}{R})^{\beta_{2(p-1)-1}\beta_{2(p-1)}}_{\alpha_{2(p-1)-1}\alpha_{2(p-1)}},
	\end{aligned}
	\label{eq:tLp}
	\end{equation}
	\begin{equation}
	\begin{split}
	&\tilde{\mathcal{Q}}^{(p)} = \frac{4p(p-1)}{2^{p-1}}
	 (h_\gamma)^{a_1 a_2\cdots a_{2(p-1)-1}}_{b_1 b_2\cdots b_{2(p-1)-1}}
	 \\[-0.3ex]
	 &\quad\times\sum_{k=0}^{p-2}\binom{p-2}{k}\frac{(-2)^k}{2k+1}
	 (\prescript{\gamma}{}{K})^{b_1}_{a_1}
	 (\prescript{\gamma}{}{K})^{b_2}_{a_2}\cdots
	 \\[-0.3ex]
	 &\qquad(\prescript{\gamma}{}{K})^{b_{2k-1}}_{a_{2k-1}}
	 (\prescript{\gamma}{}{K})^{b_{2k}}_{a_{2k}}
	 (\prescript{\gamma}{}{K})^{b_{2(p-1)-1}}_{a_{2(p-1)-1}}\\
	 &\times(\prescript{\gamma}{}{R})^{b_{2k+1} b_{2k+2}}_{a_{2k+1} a_{2k+2}}
	 \cdots(\prescript{\gamma}{}{R})^{b_{2(p-1)-3} b_{2(p-1)-2}}_{a_{2(p-1)-3} a_{2(p-1)-2}}.
	\end{split}
	\label{eq:tQp}
	\end{equation}
	
	Similarly, $(\prescript{\gamma}{}{K})$ represents the extrinsic curvature evaluated from the generalized RT surface $\Gamma$ to the intersection $\gamma$.
	
	The quantities $\mathcal{E}^{(p)}_{\mu\nu}$ and $\mathcal{J}^{(p)}_{ij}$ in the varied Lovelock action, Eq. (\ref{eq:action variation}), take the following explicit forms:

	\begin{equation}
	\begin{aligned}
	 \mathcal{E}^{(p)}_{\mu\nu}&=-\frac{1}{2^{p+1}}g_{\sigma\mu}g^{\sigma\mu_1\mu_2\cdots\mu_{2p}}_{\nu\nu_1\nu_2\cdots\nu_{2p}}\\
	 &\quad\times R^{\nu_1\nu_2}_{\mu_1\mu_2}\cdots R^{\nu_{2p-1}\nu_{2p}}_{\mu_{2p-1}\mu_{2p}},
	\end{aligned}
	\label{eq:Ep}
	\end{equation}
	\begin{align}
    	\mathcal{J}^{(p)}_{ij}&=\frac{2p}{2^p}h^{i_1 i_2\cdots i_{2p-1}}_{j_1 j_2\cdots j_{2p-1}}\nonumber\\*
		&\quad\times\sum_{k=0}^{p-1}\binom{p-1}{k}\frac{(-2)^k}{2k+1}\nonumber\\*
		&\times\Biggl[(2k+1)(\prescript{NB}{}{K})^{j_1}_{i_1}(\prescript{NB}{}{K})^{j_2}_{i_2}\cdots\nonumber\\*
		&\qquad (\prescript{NB}{}{K})^{j_{2k-1}}_{i_{2k-1}}(\prescript{NB}{}{K})^{j_{2k}}_{i_{2k}}\nonumber\\*
		&\qquad\times(\prescript{NB}{}{K})^{j_{2p-1}}_i h_{i_{2p-1}j}\nonumber\\*
		&\times(\prescript{NB}{}{R})^{j_{2k+1}j_{2k+2}}_{i_{2k+1}i_{2k+2}}\nonumber\\*
		&\qquad\cdots(\prescript{NB}{}{R})^{j_{2p-3}j_{2p-2}}_{i_{2p-3}i_{2p-2}}\nonumber\\*
		&+2(p-k-1)(\prescript{NB}{}{K})^{j_1}_{i_1}(\prescript{NB}{}{K})^{j_2}_{i_2}\cdots\nonumber\\*
		&\qquad (\prescript{NB}{}{K})^{j_{2k-1}}_{i_{2k-1}}(\prescript{NB}{}{K})^{j_{2k}}_{i_{2k}}\nonumber\\*
		&\qquad\times(\prescript{NB}{}{K})^{j_{2p-1}}_{i_{2p-1}}\nonumber\\*
		&\times(\prescript{NB}{}{R})^{j_{2k+1}j_{2k+2}}_{i_{2k+1}i_{2k+2}}\nonumber\\*
		&\qquad\cdots(\prescript{NB}{}{R})^{j_{2p-5}j_{2p-4}}_{i_{2p-5}i_{2p-4}}\nonumber\\*
		&\qquad\times(\prescript{NB}{}{R})^{j_{2p-3}l}_{i_{2p-3}i_{2p-2}}h_{li} h^{j_{2p-2}}_j\Biggr]\nonumber\\*
		&\quad-\frac{1}{2}Q^{(p)}h_{ij}.
	\label{eq:Jp}
	\end{align}

	The quantities $\tilde{\mathcal{E}}^{(p)}_{\alpha\beta}$ and $\tilde{\mathcal{J}}^{(p)}_{ab}$ in the varied Lovelock HEE, Eq. (\ref{eq:HEE variation}), take the following explicit forms:

	\begin{equation}
	\begin{aligned}
	 \tilde{\mathcal{E}}^{(p)}_{\alpha\beta}&=-\frac{p}{2^{p}}(h_\Gamma)_{\gamma\alpha}(h_\Gamma)^{\gamma\alpha_1\alpha_2\cdots\alpha_{2(p-1)}}_{\beta\beta_1\beta_2\cdots\beta_{2(p-1)}}\\
	 &\quad\times(\prescript{\Gamma}{}{R})^{\beta_1\beta_2}_{\alpha_1\alpha_2}\cdots (\prescript{\Gamma}{}{R})^{\beta_{2(p-1)-1}\beta_{2(p-1)}}_{\alpha_{2(p-1)-1}\alpha_{2(p-1)}},
	\end{aligned}
	\label{eq:tEp}
	\end{equation}
	\begin{align}
    	\tilde{\mathcal{J}}^{(p)}_{ab}&=\frac{2p(p-1)}{2^{p-1}}(h_\gamma)^{a_1 a_2\cdots a_{2(p-1)-1}}_{b_1 b_2\cdots b_{2(p-1)-1}}\nonumber\\*
		&\quad\times\sum_{k=0}^{p-2}\binom{p-2}{k}\frac{(-2)^k}{2k+1}\nonumber\\*
		&\times\Biggl[(2k+1)(\prescript{\gamma}{}{K})^{b_1}_{a_1}(\prescript{\gamma}{}{K})^{b_2}_{a_2}\cdots\nonumber\\*
		&\qquad (\prescript{\gamma}{}{K})^{b_{2k-1}}_{a_{2k-1}}(\prescript{\gamma}{}{K})^{b_{2k}}_{a_{2k}}\nonumber\\*
		&\qquad\times(\prescript{\gamma}{}{K})^{b_{2(p-1)-1}}_a (h_\gamma)_{a_{2(p-1)-1}b}\nonumber\\*
		&\times(\prescript{\gamma}{}{R})^{b_{2k+1}b_{2k+2}}_{a_{2k+1}a_{2k+2}}\nonumber\\*
		&\qquad\cdots(\prescript{\gamma}{}{R})^{b_{2(p-1)-3}b_{2(p-1)-2}}_{a_{2(p-1)-3}a_{2(p-1)-2}}\nonumber\\*
		&+2(p-k-2)(\prescript{\gamma}{}{K})^{b_1}_{a_1}(\prescript{\gamma}{}{K})^{b_2}_{a_2}\cdots\nonumber\\*
		&\qquad (\prescript{\gamma}{}{K})^{b_{2k-1}}_{a_{2k-1}}(\prescript{\gamma}{}{K})^{b_{2k}}_{a_{2k}}\nonumber\\*
		&\qquad\times(\prescript{\gamma}{}{K})^{b_{2(p-1)-1}}_{a_{2(p-1)-1}}\nonumber\\*
		&\times(\prescript{\gamma}{}{R})^{b_{2k+1}b_{2k+2}}_{a_{2k+1}a_{2k+2}}\nonumber\\*
		&\qquad\cdots(\prescript{\gamma}{}{R})^{b_{2(p-1)-5}b_{2(p-1)-4}}_{a_{2(p-1)-5}a_{2(p-1)-4}}
\nonumber\\*
		&\qquad\times(\prescript{\gamma}{}{R})^{b_{2(p-1)-3}c}_{a_{2(p-1)-3}a_{2(p-1)-2}}\nonumber\\*
		&\qquad\times(h_\gamma)_{ca}(h_\gamma)^{b_{2(p-1)-2}}_b\Biggr]\nonumber\\*
		&\quad-\frac{1}{2}\tilde{Q}^{(p)}(h_\gamma)_{ab}.
	\label{eq:tJp}
	\end{align}

	\begin{table*}[t]
		\centering
		\caption{Notations of AdS/NCFT}
		\label{tab:notation}
		\setlength{\tabcolsep}{4pt}
		\begin{tabular}{@{}lcccc@{}}
			\toprule
	    	& Bulk branch $B_{(m)}$ & Net-brane $NB$ & \shortstack{RT surface branch\\$\Gamma_{(m)}$} & Intersection $\gamma$  \\
			\midrule
		indices & $\mu$, $\nu$, $\rho$, $\cdots$ & $i$, $j$, $k$, $\cdots$ & $\alpha$, $\beta$, $\gamma$, $\cdots$ & $a$, $b$, $c$, $\cdots$ \\
 		coordinate & $x^\mu_{(m)}=(t,z_{(m)},x_{(m)},\eta^a)$ & $y^i=(t,z,\eta^a)$ & $\xi^\alpha_{(m)}=(s_{(m)},\eta^a)$ & $\eta^a$ \\
 		metric & $g_{(m)\mu\nu}$ & $h_{ij}$ & $(h_{\Gamma_{(m)}})_{\alpha\beta}$ & $(h_\gamma)_{ab}$\\
		curvature & $R_{(m)}$ & $(\prescript{NB}{}{R})$ & $(\prescript{\Gamma_{(m)}}{}{R})$ & $(\prescript{\gamma}{}{R})$ \\
		extrinsic curvature & $\varnothing$ & $(\prescript{NB}{}{K})_{(m)}$ & $\varnothing$ & $(\prescript{\gamma}{}{K})_{(m)}$ \\
 			\bottomrule
 		\end{tabular}
	\end{table*}

	\section{Derivation of Extremal and Junction Conditions for RT Surfaces}\label{Derivation of Extremal and Junction Conditions for RT Surfaces}

	This appendix derives the extremal and junction conditions for RT surfaces. The derivation closely follows \cite{Guo:2026prm}. For convenience, we restate the variation of the HEE Eq. (\ref{eq:HEE variation}) below

	\begin{align}
		&\delta S = \sum_{m}^{n}\frac{1}{4 G_{(m)}}\int_{\Gamma_{(m)}}
		  d^{d-1}\xi_{(m)}\sqrt{h_{\Gamma_{(m)}}}\nonumber\\*
		  &\quad\times\bigl(\alpha_{1(m)}\tilde{\mathcal{E}}^{(1)}_{(m)\alpha\beta}
		  +\alpha_{2(m)}\tilde{\mathcal{E}}^{(2)}_{(m)\alpha\beta}
		  \nonumber\\*
		&\qquad+\alpha_{3(m)}\tilde{\mathcal{E}}^{(3)}_{(m)\alpha\beta}\bigr)
		  \delta (h_{\Gamma_{(m)}})^{\alpha\beta}\nonumber\\*
		  &\quad+\int_{\gamma}d^{d-2}\eta\sqrt{h_{\gamma}}
		  \biggl[\sum_{m}^{n}\frac{1}{4 G_{(m)}}\nonumber\\*
		  &\scalebox{0.86}{$\displaystyle
		  {}\times\bigl(\alpha_{1(m)}\tilde{\mathcal{J}}^{(1)}_{(m)ab}
		  +\alpha_{2(m)}\tilde{\mathcal{J}}^{(2)}_{(m)ab}
		  +\alpha_{3(m)}\tilde{\mathcal{J}}^{(3)}_{(m)ab}\bigr)
		  \biggr]\delta (h_{\gamma})^{ab}.$}
	\label{Appendix eq:HEE variation}
	\end{align}
	
	The RT surfaces $\Gamma_{(m)}$ and their intersection $\gamma$ can be parameterized by their embedding functions $x_{(m)}^{\mu}(\xi_{(m)}^{\alpha})$ and $x_{(m)}^{\mu}(\eta^a)$, respectively. The induced metrics on $\Gamma_{(m)}$ and $\gamma$ are then expressed as
	\begin{equation}
		\begin{aligned}
		(h_{\Gamma_{(m)}})_{\alpha\beta}&=\frac{\partial x_{(m)}^{\mu}}{\partial \xi_{(m)}^{\alpha}}
		\frac{\partial x_{(m)}^{\nu}}{\partial \xi_{(m)}^{\beta}}g_{(m)\mu\nu},\\
		(h_{\gamma})_{ab}&=\frac{\partial x_{(m)}^{\mu}}{\partial \eta^a}
		\frac{\partial x_{(m)}^{\nu}}{\partial \eta^b}g_{(m)\mu\nu}.
		\end{aligned}
	\label{eq:embedding function}
	\end{equation}
	
	Consequently, the variations of the induced metrics are given by
	\begin{align}
		\delta (h_{\Gamma_{(m)}})_{\alpha\beta}
		  &=\partial_{(m)\alpha}\delta x_{(m)}^{\mu}
		  \partial_{(m)\beta} x_{(m)}^{\nu}g_{(m)\mu\nu}\nonumber\\*
		&\quad+\partial_{(m)\alpha}x_{(m)}^{\mu}
		  \partial_{(m)\beta}\delta x_{(m)}^{\nu}g_{(m)\mu\nu}\nonumber\\*
		&\quad+\partial_{(m)\alpha}x_{(m)}^{\mu}\partial_{(m)\beta}x_{(m)}^{\nu}\nonumber\\*
		&\qquad\times\partial_{(m)\rho}g_{(m)\mu\nu}\delta x_{(m)}^\rho,
	\label{eq:RT metric variation}
	\end{align}
	\begin{equation}
	\begin{split}
	\delta (h_{\gamma})_{ab}
	  &= \partial_a\delta x_{(m)}^{\mu}\partial_b x_{(m)}^{\nu}g_{(m)\mu\nu}\\
	  &\quad+\partial_a x_{(m)}^{\mu}\partial_b\delta x_{(m)}^{\nu}g_{(m)\mu\nu}\\
	  &\quad+\partial_a x_{(m)}^{\mu}\partial_b x_{(m)}^{\nu}\\
	  &\qquad\times\partial_{(m)\rho}g_{(m)\mu\nu}\delta x_{(m)}^\rho.
	\end{split}
	\label{eq:gamma metric variation}
	\end{equation}
	
	Substituting Eqs. (\ref{eq:RT metric variation}) and (\ref{eq:gamma metric variation}) into Eq. (\ref{Appendix eq:HEE variation}) and integrating by parts, we obtain
	\begin{align}
		\delta S&=\sum_m^n-\frac{1}{2G_{(m)}}\int_{\Gamma_{(m)}}d^{d-1}\xi_{(m)}\sqrt{h_{\Gamma_{(m)}}}\nonumber\\*
		&\quad\left(\mathcal{A}_{(m)}^{\alpha\beta}(K_\Gamma)_{(m)\mu\alpha\beta}\right)\delta x_{(m)}^\mu(\xi_{(m)}^\alpha)\nonumber\\
		&+\int_\gamma d^{d-2}\eta\sqrt{h_\gamma}\biggl[\nonumber\\*
		&\quad\sum_m^n\frac{1}{2 G_{(m)}}\bigl(n_{{(m)}\alpha}\mathcal{A}_{(m)}^{\alpha\beta}\partial_{(m)\beta}x_{(m)}^\nu g_{(m)\mu\nu}\nonumber\\
		&\quad-\hat{D}_a \mathcal{B}_{(m)}^{ab}\partial_b x_{(m)}^\nu g_{(m)\mu\nu}\nonumber\\*
		&\quad-\mathcal{B}_{(m)}^{ab}(K_\gamma)_{(m)\mu ab}\bigr)\frac{\partial x_{(m)}^\mu}{\partial y^i(\eta^a)}\biggr]\delta y^i(\eta^a),
	\label{eq:HEE variation for condition}
	\end{align}
	where $n_{(m)\alpha}$ represents the normal vector pointing from the RT surface $\Gamma_{(m)}$ to the intersection $\gamma$. The covariant derivative on $\gamma$ is denoted by $\hat{D}_a$. The extrinsic curvatures $(K_\Gamma)_{(m)}$ and $(K_\gamma)_{(m)}$, evaluated from the bulk branch $B_{(m)}$ to the RT surface $\Gamma_{(m)}$ and the intersection $\gamma$ respectively, are defined as
	\begin{align}
		&(K_\Gamma)^{\quad\;\,\mu}_{(m)\alpha\beta}=\partial_{(m)\alpha}\partial_{(m)\beta}x_{(m)}^\mu\nonumber\\*
		&\qquad-(\prescript{\Gamma_{(m)}}{}{\Gamma})^\gamma_{\alpha\beta}\partial_{(m)\gamma}x_{(m)}^\mu\nonumber\\*
		&\qquad+(\prescript{B_{(m)}}{}{\Gamma})^\mu_{\sigma\rho}\partial_{(m)\alpha}x_{(m)}^\sigma\partial_{(m)\beta}x_{(m)}^\rho,\nonumber\\
		&(K_\gamma)^{\quad\;\,\mu}_{(m)ab}=\partial_a\partial_bx_{(m)}^\mu\nonumber\\*
		&\qquad-(\prescript{\gamma}{}{\Gamma})^{\quad\;\,c}_{(m)ab}\partial_c x_{(m)}^\mu\nonumber\\*
		&\qquad+(\prescript{B_{(m)}}{}{\Gamma})^\mu_{\sigma\rho}\partial_a x_{(m)}^\sigma\partial_b x_{(m)}^\rho,
	\label{eq:KG Kg extrinsic curvatures}
	\end{align}
	where $(\prescript{B_{(m)}}{}{\Gamma})^\mu_{\sigma\rho}$, $(\prescript{\Gamma_{(m)}}{}{\Gamma})^\gamma_{\alpha\beta}$ and $(\prescript{\gamma}{}{\Gamma})^{\quad\;\,c}_{(m)ab}$ represent the Levi-Civita connections in bulk branch $B_{(m)}$, on the RT surface $\Gamma_{(m)}$, and at the intersection $\gamma$, repsectively. Furthermore, $\mathcal{A}_{(m)}^{\alpha\beta}$ and $\mathcal{B}_{(m)}^{ab}$ are defined as:
	\begin{align}
		&\mathcal{A}_{(m)}^{\alpha\beta}=-\bigl(\alpha_{1(m)}\tilde{\mathcal{E}}^{(1)\alpha\beta}_{(m)}+\alpha_{2(m)}\tilde{\mathcal{E}}^{(2)\alpha\beta}_{(m)}\nonumber\\*
		&\qquad+\alpha_{3(m)}\tilde{\mathcal{E}}^{(3)\alpha\beta}_{(m)}\bigr),\nonumber\\
		&\mathcal{B}_{(m)}^{ab}=-\bigl(\alpha_{1(m)}\tilde{\mathcal{J}}^{(1)ab}_{(m)}+\alpha_{2(m)}\tilde{\mathcal{J}}^{(2)ab}_{(m)}\nonumber\\*
		&\qquad+\alpha_{3(m)}\tilde{\mathcal{J}}^{(3)ab}_{(m)}\bigr).
	\label{eq:Am Bm}
	\end{align}
	
	Eq. (\ref{eq:HEE variation for condition}) dictates the extremal and junction conditions:

	\begin{equation}
		\text{Extremal Condition:}\quad\mathcal{A}_{(m)}^{\alpha\beta}(K_\Gamma)_{(m)\mu\alpha\beta}=0,
	\label{Appendix eq:raw extremal condition}
	\end{equation}
	\begin{align}
		&\text{Junction Condition:}\nonumber\\
		&\sum_m^n\frac{1}{G_{(m)}}\biggl(
		 n_{{(m)}\alpha}\mathcal{A}_{(m)}^{\alpha\beta}
		 \partial_{(m)\beta}x_{(m)}^\nu g_{(m)\mu\nu}\nonumber\\*
		&\qquad-\hat{D}_a\mathcal{B}_{(m)}^{ab}
		 \partial_b x_{(m)}^\nu g_{(m)\mu\nu}\nonumber\\*
		&\qquad-\mathcal{B}_{(m)}^{ab}(K_\gamma)_{(m)\mu ab}
		 \biggr)\frac{\partial x_{(m)}^\mu}{\partial y^i(\eta^a)}=0.
	\label{Appendix eq:raw junction condition}
	\end{align}

	As indicated by Eq. (\ref{eq:Am Bm}), higher-order Lovelock gravity terms can be directly appended to this equation.

	Next, upon substituting the coordinate ansatz from the main text into Eqs. (\ref{Appendix eq:raw extremal condition}) and (\ref{Appendix eq:raw junction condition}), the explicit extremal and junction conditions are obtained:
	\begin{align}
		&\text{Extremal condition:}\nonumber\\
 		&a_{1(m)}=l^4\alpha_{1(m)}-2(d-3)(d-2)l^2\alpha_{2(m)}\nonumber\\*
		&\qquad+3(d-5)(d-4)(d-3)(d-2)\alpha_{3(m)}\nonumber\\[0.5ex]
		&a_{2(m)}=(d-3)(d-2)\bigg[l^2\alpha_{2(m)}\nonumber\\*
		&\qquad-3(d-5)(d-4)\alpha_{3(m)}\bigg]\nonumber\\[0.5ex]
		&a_{3(m)}=(d-5)(d-4)(d-3)(d-2)\alpha_{3(m)}\nonumber\\[0.5ex]
		&\theta_{(m)}'(s_{(m)})=\frac{(d-1)\sin(\theta_{(m)}(s_{(m)}))}{z_{(m)}(s_{(m)})}\nonumber\\*
		&\times\scalebox{0.78}{$\displaystyle
		  \frac{a_{1(m)}+2a_{2(m)}\sin^2(\theta_{(m)}(s_{(m)}))
		        +3a_{3(m)}\sin^4(\theta_{(m)}(s_{(m)}))}
		       {a_{1(m)}+6a_{2(m)}\sin^2(\theta_{(m)}(s_{(m)}))
		        +15a_{3(m)}\sin^4(\theta_{(m)}(s_{(m)}))}$},
	\label{eq:specific extremal condition}
	\end{align}
	\begin{align}
		&\text{AdS/NCFT junction condition:}\nonumber\\
 		&b_{1(m)}=l^4\alpha_{1(m)}\nonumber\\[1.5ex]
		&b_{2(m)}=(d-3)(d-2)l^2\alpha_{2(m)}\nonumber\\[1.5ex]
		&b_{3(m)}=(d-5)(d-4)(d-3)(d-2)\alpha_{3(m)}\nonumber\\[1.5ex]
		&\sum_m^n\frac{1}{G_{(m)}}\bigg\{\bigg[b_{1(m)}-2b_{2(m)}\cos^2(\theta_{(m)0})\nonumber\\*
		&\qquad+3b_{3(m)}\cos^4(\theta_{(m)0})\bigg]\sin(\theta_{(m)0})\sinh(\rho)\nonumber\\[1.5ex]
		&\quad-\bigg[b_{1(m)}-2b_{2(m)}
		  \bigg(\cos^2(\theta_{(m)0})-2\bigg)\nonumber\\*
		&\qquad+b_{3(m)}\cos^2(\theta_{(m)0})
		\nonumber\\*
		&\qquad\times\bigg(3\cos^2(\theta_{(m)0})-4\bigg)\bigg]\cos(\theta_{(m)0})\bigg\}=0,
	\label{eq:specific NCFT junction condition}
	\end{align}
	where $\theta_{(m)0} \equiv \theta_{(m)}(s_{(m)}=0)$ represents the value of $\theta_{(m)}$ at the intersection $\gamma$.

	\begin{samepage}
	Note that if the weighted sum over branches is removed, Eq. (\ref{eq:specific NCFT junction condition}) cleanly reduces to the standard AdS/BCFT junction condition for a single branch:
	\begin{align}
		&\text{AdS/BCFT junction condition:}\nonumber\\
		&\bigg[b_{1(m)}-2b_{2(m)}\cos^2(\theta_{(m)0})\nonumber\\*
		&\qquad+3b_{3(m)}\cos^4(\theta_{(m)0})\bigg]\sin(\theta_{(m)0})\sinh(\rho)\nonumber\\[1.5ex]
		&\quad-\bigg[b_{1(m)}-2b_{2(m)}
		  \bigg(\cos^2(\theta_{(m)0})-2\bigg)\nonumber\\*
		&\qquad+b_{3(m)}\cos^2(\theta_{(m)0})
		\nonumber\\*
		&\qquad\times\bigg(3\cos^2(\theta_{(m)0})-4\bigg)\bigg]\cos(\theta_{(m)0})=0.
	\label{eq:specific BCFT junction condition}
	\end{align}
	\end{samepage}

	Notably, from Eq. (\ref{eq:specific BCFT junction condition}), one can see that in pure Einstein gravity, the RT surface is strictly normal to the Net-brane at the intersection $\gamma$, perfectly recovering the well-known AdS/BCFT junction condition. 

	\section{Specific Expressions for Boundary and Bulk HEE}\label{Specific Expressions for Boundary and Bulk HEE}

	Under the coordinate ansatz introduced in the main text, the explicit expressions for the boundary and bulk entropy terms are given as follows:

	\begin{equation}
	\begin{split}
		&S_{\text{Bdy}} = V\times\sum_m^n\frac{1}{4G_{(m)}}\biggl[\\
				&\quad 4(d-2)\frac{(z_b)^{d-2}}{l^{d-5}}\cos(\theta_{(m)0})\\
		&\quad\times\bigg(l^2\alpha_{2(m)}\\
		&\qquad-(d-4)(d-3)\alpha_{3(m)}\cos^2(\theta_{(m)0})\bigg)\biggr],
	\end{split}
	\label{eq:SBdy}
	\end{equation}
	\begin{align}
		&c_{1(m)}=l^4\alpha_{1(m)}\nonumber\\[1.5ex]
		&c_{2(m)}=(d-2)l^2\alpha_{2(m)}\nonumber\\[1.5ex]
		&c_{3(m)}=(d-4)(d-3)(d-2)\alpha_{3(m)}\nonumber\\[1.5ex]
		&S_{\text{Bulk}}=V\times\sum_m^n\frac{1}{4G_{(m)}}\int ds_{(m)}\nonumber\\*
		&\quad
		\frac{l^{d-5}}{z_{(m)}(s_{(m)})^{d-1}}\nonumber\\*
		&\qquad\times\bigg[\bigg(c_{1(m)}-2(d-1)c_{2(m)}\cos^2(\theta_{(m)}(s_{(m)}))
		\nonumber\\*
		&\qquad+3(d-1)c_{3(m)}\cos^4(\theta_{(m)}(s_{(m)}))\bigg)\nonumber\\*
		&\qquad\quad-4z_{(m)}(s_{(m)})\sin(\theta_{(m)}(s_{(m)}))
		\nonumber\\*
		&\qquad\times\partial_{s_{(m)}}\theta_{(m)}(s_{(m)})\nonumber\\*
		&\qquad\quad\times\bigg(c_{2(m)}-3c_{3(m)}\cos^2(\theta_{(m)}(s_{(m)}))\bigg)\bigg],
	\label{eq:SBulk}
	\end{align}
	where $V$ represents the integral over the transverse $\eta^a$ directions, which corresponds to the volume of the transverse $(d-2)$-dimensional space:
	\begin{equation}
		V=\int d^{d-2}\eta.
	\label{eq:V}
	\end{equation}

	\section{RT Surface and HEE of One-Branch BCFT in \texorpdfstring{$(d+1 \geq 4)$}{(d+1 >= 4)} Einstein Gravity}\label{RT Surface and HEE of One-Branch BCFT in dgeq3 Einstein Gravity}

	Let RT surface be described by $x=x(z)$ with the boundary condition
	\begin{equation}
		x(0)=x_L.
	\end{equation}

	The induced metric on the RT surface is given by
	\begin{equation}
		d(\prescript{\Gamma}{}{s})^2=\frac{l^2}{z^2}\bigg[\bigg(1+x'^2(z)\bigg)dz^2+\delta_{ab}d\eta^a d\eta^b\bigg].
	\label{eq:RT x(z) coordinate}
	\end{equation}

	The equation of motion of RT surface is given by
	\begin{equation}
		\frac{x'(z)}{z^{d-1}\sqrt{1+x'^2(z)}}=C,\;C=\text{constant}.
	\label{eq:RT EOM}
	\end{equation}

	We denote the angle of NB brane by $\varphi=-\arctan\big(\sinh(\rho)\big)$. If $\varphi\leq 0$, a possible solution to Eq.~(\ref{eq:RT EOM}) corresponds to $C=0$, which implies that the equation of the RT surface is simply $x=x_L$. However, for a general value of $\varphi$, there exists a more general solution corresponding to $C\neq 0$. When both solutions are physically admissible, the true RT surface is the one that yields the smaller HEE, in accordance with the minimal area prescription. In the $C\neq 0$ case, we assume that $x'(z_*)=\infty$ at the turning point $z=z_*$. Let $(x_b,z_b)$ denote the coordinates of the intersection between the RT surface and the NB brane, satisfying $x_b=-\sinh(\rho)z_b$. In Einstein gravity, the RT surface is strictly normal to the NB brane at the intersection. This orthogonality imposes the boundary condition
	\begin{equation}
		x'(z_b)=-\cot(\varphi).
	\label{eq:RT BC}
	\end{equation}

	Applying the condition $x'(z_*)=\infty$ and the boundary condition in Eq.~(\ref{eq:RT BC}), we obtain

	\begin{align}
		&x'(z)=\frac{(\frac{z}{z_*})^{d-1}}{\sqrt{1-(\frac{z}{z_*})^{2(d-1)}}},\\
		&z_b=z_*\cos^\frac{1}{d-1}(\varphi).
	\end{align}

	Furthermore, the geometric configuration yields the relation

	\begin{equation}
		x_L-x_b=\int^{z_*}_0 x'(z)dz-\text{Sign}(\varphi)\int^{z_*}_{z_b} x'(z)dz,
	\label{eq:zs eq}
	\end{equation}
	where $\text{Sign}(\varphi)$ is the signum function.

	This allow us to solve for $z_*$,
	\begin{equation}
		z_{*}=\frac{x_L}{F(\varphi)},
	\label{eq:zs}
	\end{equation}
	where
	\begin{equation}
	\begin{split}
		F(\varphi)&=\text{Sign}(\varphi)
		\frac{B\big(\cos^2(\varphi);\frac{d}{2(d-1)},\frac{1}{2}\big)}{2(d-1)}\\
		&\quad+\big(1-\text{Sign}(\varphi)\big)
		\frac{\sqrt{\pi}\,\Gamma\big(\frac{d}{2(d-1)}\big)}{\Gamma\big(\frac{1}{2(d-1)}\big)}\\
		&\quad+\tan(\varphi)\cos^\frac{1}{d-1}(\varphi),
	\end{split}
	\label{eq:F of varphi}
	\end{equation}
	Here, $\Gamma(a)$ represents the gamma function, and $B(x;a,b)$ denotes the incomplete beta function.

	Consequently, the HEE and the profile of the RT surface can be readily determined. The one-branch HEE is expressed as
	\begin{align}
	&S^1_{\text{BCFT}}=V\times\frac{1}{4G}\big(\frac{1}{z_*}\big)^{d-2}\frac{l^{d-1}}{2(d-1)}\nonumber\\*
		&\quad\bigg[-B\bigg(\big(\frac{\varepsilon^S_z}{z_*}\big)^{2(d-1)};\frac{2-d}{2(d-1)},\frac{1}{2}\bigg)\nonumber\\[1.5ex]
		&\quad+\big(1-\text{Sign}(\varphi)\big)\frac{\sqrt{\pi}\,\Gamma\big(\frac{2-d}{2(d-1)}\big)}{\Gamma\big(\frac{1}{2(d-1)}\big)}\nonumber\\*
		&\quad+\text{Sign}(\varphi)B\bigg(\cos^2(\varphi);\frac{2-d}{2(d-1)},\frac{1}{2}\bigg)\bigg],
	\label{eq:Einstein one-branch HEE}
	\end{align}
	where $V$ is defined in Eq.~(\ref{eq:V}).

	The explicit profile of the RT surface is given by
	\begin{equation}
		\begin{gathered}
		x(z)=\\[-0.4ex] 
		{ \renewcommand{\arraystretch}{2.4}
		\left\{\begin{array}{@{}l@{\hspace{2pt}}l@{}}
    			\begin{aligned}[t]&\int_z^{z_b}\frac{\big(\frac{w}{z_*}\big)^{d-1}}{\sqrt{1-\big(\frac{w}{z_*}\big)^{2(d-1)}}}\,dw\\[-0.3ex]&\quad-\sinh(\rho)z_b\end{aligned} & \varphi \geq 0 \\
    			\left\{\begin{array}{@{}l@{\,}l@{}}
        			\begin{aligned}[t]&\int^z_{z_b}\frac{\big(\frac{w}{z_*}\big)^{d-1}}{\sqrt{1-\big(\frac{w}{z_*}\big)^{2(d-1)}}}\,dw\\[-0.3ex]&\quad-\sinh(\rho)z_b\end{aligned} & \substack{\text{lower segment,}\\ z_b\leq z\leq z_*} \\
        			\begin{aligned}[t]&\int_z^{z_*}\frac{\big(\frac{w}{z_*}\big)^{d-1}}{\sqrt{1-\big(\frac{w}{z_*}\big)^{2(d-1)}}}\,dw\\[-0.3ex]&\quad+x_*\end{aligned} & \substack{\text{upper segment,}\\ z<z_*} \\ 
    			\end{array}\right. & \varphi < 0
		\end{array}\right.
		},
	\end{gathered}
	\label{eq:Einstein one-branch RT}
	\end{equation}
	where
	\begin{equation}
		x_*=\int^{z_*}_{z_b}\frac{\big(\frac{w}{z_*}\big)^{d-1}}{\sqrt{1-\big(\frac{w}{z_*}\big)^{2(d-1)}}}\,dw-\sinh(\rho)z_b.
	\end{equation}
	
	The sign of $\varphi$ leads to two geometric cases for the RT surface. Specifically, a negative $\varphi$ results in a backbending RT surface that comprises both lower and upper segments (as illustrated in Figs.~\ref{fig:Einstein_d=4_rho=0.5_BCFT_RT_analytic_vs_numerical} and \ref{fig:Einstein_d=6_rho=0.3_BCFT_RT_analytic_vs_numerical}). Conversely, a positive $\varphi$ yields a non-backbending RT surface (as illustrated in Figs.~\ref{fig:Einstein_d=4_rho=-0.5_BCFT_RT_analytic_vs_numerical} and \ref{fig:Einstein_d=6_rho=-0.3_BCFT_RT_analytic_vs_numerical}).

	\end{appendices}
	
	\renewcommand{\bibfont}{\fontsize{9}{10.3}\selectfont\interlinepenalty=10000\relax}
	\setlength{\bibsep}{0.16em}
	\begingroup
	\def\calabresearticleid{06002}
	\def\bfpage#1{%
	  \def\currentarticleid{#1}%
	  \ifx\currentarticleid\calabresearticleid P\fi
	  #1%
	}
	\bibliography{ref2}
	\endgroup
\end{document}